\documentclass[12pt]{article}
\pdfoutput=1 

\usepackage{jheppub}

\author[a]{Xenia de la Ossa,}
\author[b]{Mateo Galdeano,}
\author[a,c]{and Enrico Marchetto}
\affiliation[a]{Mathematical Institute, University of Oxford, Andrew Wiles Building, Woodstock Road,\\Oxford, OX2 6GG, U.K.}
\affiliation[b]{Fachbereich Mathematik, Universit\"at Hamburg, Bundesstr. 55, 20146, Hamburg, Germany}
\affiliation[c]{Deutsches Elektronen-Synchrotron DESY, Notkestr. 85, 22607 Hamburg, Germany}
\emailAdd{delaossa@maths.ox.ac.uk}
\emailAdd{mateo.galdeano@uni-hamburg.de}
\emailAdd{enrico.marchetto@desy.de}
\preprint{DESY-24-197}

\usepackage{amsmath}
\usepackage{amssymb}
\usepackage[dvipsnames*,svgnames]{xcolor}
\usepackage{titlesec}
\usepackage{enumitem} 
\titleformat{\chapter}[display]
  {\centering \Huge \bfseries \color{black}}{\thechapter}{10pt}{}
\usepackage{hyperref}
\usepackage{cleveref}
\usepackage[T1]{fontenc}
\usepackage{caption}
\usepackage{subcaption}
\usepackage{pgfplots}
\usetikzlibrary{datavisualization}
\usetikzlibrary{datavisualization.formats.functions}
\usepackage{braket}
\usepackage{tensor}
\usepackage{dsfont}
\usepackage{tcolorbox}
\usepackage{tikz}
\usetikzlibrary{decorations.pathreplacing,calligraphy}
\usetikzlibrary{decorations.pathreplacing}
\usetikzlibrary{decorations.pathmorphing, patterns,shapes}
\usepackage{bm}
\usepackage{upgreek}
\usepackage{mathtools}
\usepackage{cancel}
\usepackage{float, array, xspace, amscd, amsthm, amssymb}
\let\amslrcorner\lrcorner
\let\lrcorner\amslrcorner
\usepackage{fancyhdr}
\usepackage{longtable}
\usetikzlibrary{matrix}
\usepackage{comment}
\usepackage{bbold}
\usepackage[page]{appendix}
\usepackage{slashed}
\usepackage{mathrsfs}
\usepackage{xfrac}
\usepackage{scalerel,stackengine}
\usepackage{multirow}
\usepackage{ragged2e}
\DeclareCaptionJustification{justified}{\justifying}
\usepackage{amsthm}

\usepackage{textcomp}

\usepackage{yhmath}

\usepackage{lmodern}

\usepackage{anyfontsize}

\newcommand\reallywidehat[1]{%
\savestack{\tmpbox}{\stretchto{%
  \scaleto{%
    \scalerel*[\widthof{\ensuremath{#1}}]{\kern-.6pt\bigwedge\kern-.6pt}%
    {\rule[-\textheight/2]{1ex}{\textheight}}
  }{\textheight}%
}{0.5ex}}%
\stackon[1pt]{#1}{\tmpbox}%
}

\newcommand{\SVseven}{\mathrm{SV}^{\mathrm{G}_2}}
\newcommand{\SVeight}{\mathrm{SV}^\mathrm{Spin(7)}}
\newcommand{\FGseven}{\mathrm{FG}} 

\renewcommand{\Re}{\operatorname{Re}}
\renewcommand{\Im}{\operatorname{Im}}

\newcommand{\op}[1]{\boldsymbol{#1}}
\newcommand{\de}{\mathrm d}

\newcommand{\phu}{\varphi}

\newcommand{\be}{\begin{equation}}
\newcommand{\ee}{\end{equation}}

\def\veps{\varepsilon}

\newcommand{\del}{\partial}
\newcommand{\bdel}{\Bar{\partial}}

\newcommand{\tr}{\text{Tr}}
\newcommand{\nn}{\nonumber}
\newcommand{\dd}{{\rm d}}

\def\bea#1\eea{\begin{align}#1\end{align}}

\newif\ifcomments
\commentstrue
\newif\ifdetails
\detailstrue
\definecolor{GreenXD}{RGB}{28, 112, 46}
\definecolor{GreenXDdetail}{RGB}{80, 117, 88}
\definecolor{BlueMG}{RGB}{0, 0, 255}
\definecolor{BlueMGdetail}{RGB}{50, 0, 100}
\definecolor{detail}{RGB}{110,110,110}

\newcommand{\cn}{\mathcal{N}}

\newcommand{\ct}{\mathcal{T}}
\newcommand{\cj}{\mathcal{J}}
\newcommand{\co}{\mathcal{O}}

\DeclareFontFamily{U}{bbold}{}
\DeclareFontShape{U}{bbold}{m}{n}
{  <-5.5> s*[1.05] bbold5
	<5.5-6.5> s*[1.05] bbold6
	<6.5-7.5> s*[1.05] bbold7
	<7.5-8.5> s*[1.05] bbold8
	<8.5-9.5> s*[1.05] bbold9
	<9.5-11.5> s*[1.05] bbold10
	<11.5-16> s*[1.05] bbold12
	<16-> s*[1.05] bbold17
}{}

\makeatletter
\newcommand{\extp}{\@ifnextchar^\@extp{\@extp^{\,}}}
\def\@extp^#1{\mathop{\bigwedge\nolimits^{\!#1}}}
\makeatother

\theoremstyle{definition}

\begin{document}

\title{$\mathcal{SW}$-algebras and strings with torsion}

\abstract{We explore the connection between super $\mathcal{W}$-algebras ($\mathcal{SW}$-algebras) and $\mathrm{G}$-structures with torsion. The former are realised as symmetry algebras of strings with $\mathcal{N}=(1,0)$ supersymmetry on the worldsheet, while the latter are associated with generic string backgrounds with non-trivial Neveu--Schwarz flux $H$. In particular, we focus on manifolds featuring $\mathrm{Spin}(7)$, $\mathrm{G}_2$, $\mathrm{SU}(2)$, and $\mathrm{SU}(3)$-structures. We compare the full quantum algebras with their classical limits, obtained by studying the commutators of superconformal and $\mathcal{W}$-symmetry transformations---which preserve the action of the $(1,0)$ non-linear $\sigma$-model. We show that, at first order in the string length scale $\ell_s$, the torsion deforms some of the OPE coefficients corresponding to special holonomy through a scalar torsion class.
}

\maketitle

\section{Introduction}
\label{sec:Introduction}

It is well-known that string compactifications with no fluxes on manifolds of special holonomy are associated with extended chiral symmetry algebras 
on the worldsheet \cite{Zumino:1979et, Hull:1985jv, Banks:1987cy, Odake:1988bh, Eguchi:1988vra, Shatashvili:1994zw, Figueroa-OFarrill:1996tnk}. This provides an intriguing connection between differential geometry and two-dimensional conformal field theories (CFTs) that can be exploited to gain insight into both subjects. The prime example of this synergy is the use of spectral flow \cite{Schwimmer:1986mf} to unveil mirror symmetry for Calabi--Yau manifolds \cite{Lerche:1989uy}. More recently, generalisations of these techniques have been applied to describe mirror symmetry for manifolds with $\mathrm{G}_2$ or $\mathrm{Spin}(7)$ holonomy \cite{Shatashvili:1994zw, Gaberdiel:2004vx, Chuang:2004th, Fiset:2018huv, Braun:2019lnn, Fiset:2021ruv, Braun:2023zcm}.

Perhaps surprisingly, outside the realm of special holonomy manifolds---that is, manifolds equipped with a \emph{torsion-free} $\mathrm{G}$-structure---our knowledge of the underlying chiral algebra remains quite limited, even though recent efforts have been made in this direction \cite{Alvarez-Consul:2020hbl, Fiset:2021azq, Alvarez-Consul:2023zon}. In this work we take steps to identify the worldsheet algebras associated with string backgrounds involving a $\mathrm{G}$-structure \emph{with torsion}. In particular, we are interested in critical strings compactified on a $d$-dimensional internal Riemannian manifold $\mathcal{M}$, equipped with a (torsionful) $\mathrm{G}$-structure. Furthermore, we will assume that the torsion is given by a non-zero Neveu--Schwarz (NS) flux $H$ on $\mathcal{M}$.

The fundamental object of study for us are \emph{$\mathcal{W}$-algebras}: extensions of the Virasoro algebra for which the operator product expansions (OPEs) can be non-linear in the generators \cite{Belavin:1984vu, Zamolodchikov:1985wn}.\footnote{$\mathcal{W}$-algebras are often studied following the formalism of Vertex Operator Algebras (VOAs) proposed in \cite{Borcherds:1983sq, Frenkel:1986hu} and reviewed in \cite{Kac:1996wd}. We adopt a different approach, following for example \cite{Thielemans:1994er}.} More concretely, we focus on algebras preserving $\mathcal{N}=1$ supersymmetry, which we call \emph{super} $\mathcal{W}$-algebras or \emph{$\mathcal{SW}$-algebras} for short. These have been extensively studied and classified according to the number and the conformal weights of their generators \cite{Figueroa-OFarrill:1990mzn, Figueroa-OFarrill:1990tqt, Figueroa-OFarrill:1990ldq, Blumenhagen:1990jv, Blumenhagen:1991nm, Romans:1991wi, Blumenhagen:1992vr, Blumenhagen:1992sa, Bouwknegt:1992wg, Blumenhagen:1994qe, Blumenhagen:1994wn}.

Unfortunately, the geometric meaning (if any) of these $\mathcal{SW}$-algebras is obscured in the purely algebraic classifications: a more elaborate procedure is needed to identify the algebra underlying a given string background. An approach that has been extensively used in the literature is to focus on geometries such that the corresponding $\mathcal{SW}$-algebra admits a straightforward description, for example in terms of a free field realisation \cite{Shatashvili:1994zw, Howe:1994tv, Gaberdiel:2004vx, Chuang:2004th, deBoer:2005pt, Braun:2019lnn, Fiset:2021ruv, Galdeano:2023ide}, Gepner models \cite{Gepner:1987vz, Gepner:1987qi, Roiban:2001cp, Eguchi:2001xa, Blumenhagen:2001jb, Blumenhagen:2001qx, Eguchi:2001ip, Noyvert:2002mc, Braun:2023zcm} or WZW models and coset realisations \cite{Sugiyama:2001qh, Sugiyama:2002ag, Naka:2002xs, Eguchi:2003yy, Eguchi:2004yi, Sriharsha:2006zc, Melnikov:2017yvz, Fiset:2021azq}. {Another way---found in the mathematical literature---to provide a geometric interpretation of a $\mathcal{W}$-algebra consists in finding an embedding into the sections of the \emph{chiral de Rham complex} \cite{Malikov:1998dw}, which is a sheaf of vertex algebras over a manifold. Some examples of this are \cite{Heluani:2006pk, Ben-Zvi:2008sma, Ekstrand:2009zd, Ekstrand:2010wu, Heluani:2014uaa, Diaz:2016feg, Alvarez-Consul:2020hbl, Alvarez-Consul:2023zon}.} 

We follow an alternative strategy and consider a non-linear $\sigma$-model with target a critical string background as above. Howe and Papadopoulos discovered that, for every covariantly constant differential form on $\mathcal{M}$, the $\sigma$-model enjoys an additional (non-linear) classical symmetry \cite{Bilal:1990dm, Howe:1991im, Howe:1991vs, Howe:1991ic}. These are called \emph{$\mathcal{W}$-symmetries} and hold even in the presence of a non-zero NS flux \cite{Howe:2010az} (see also \cite{Stojevic:2006pq}) and for heterotic $\sigma$-models that include a gauge bundle sector \cite{delaOssa:2018azc, Grimanellis:2023nav, Papadopoulos:2023exe}.

\begin{figure}
\centering
\includegraphics[width=0.8\textwidth]{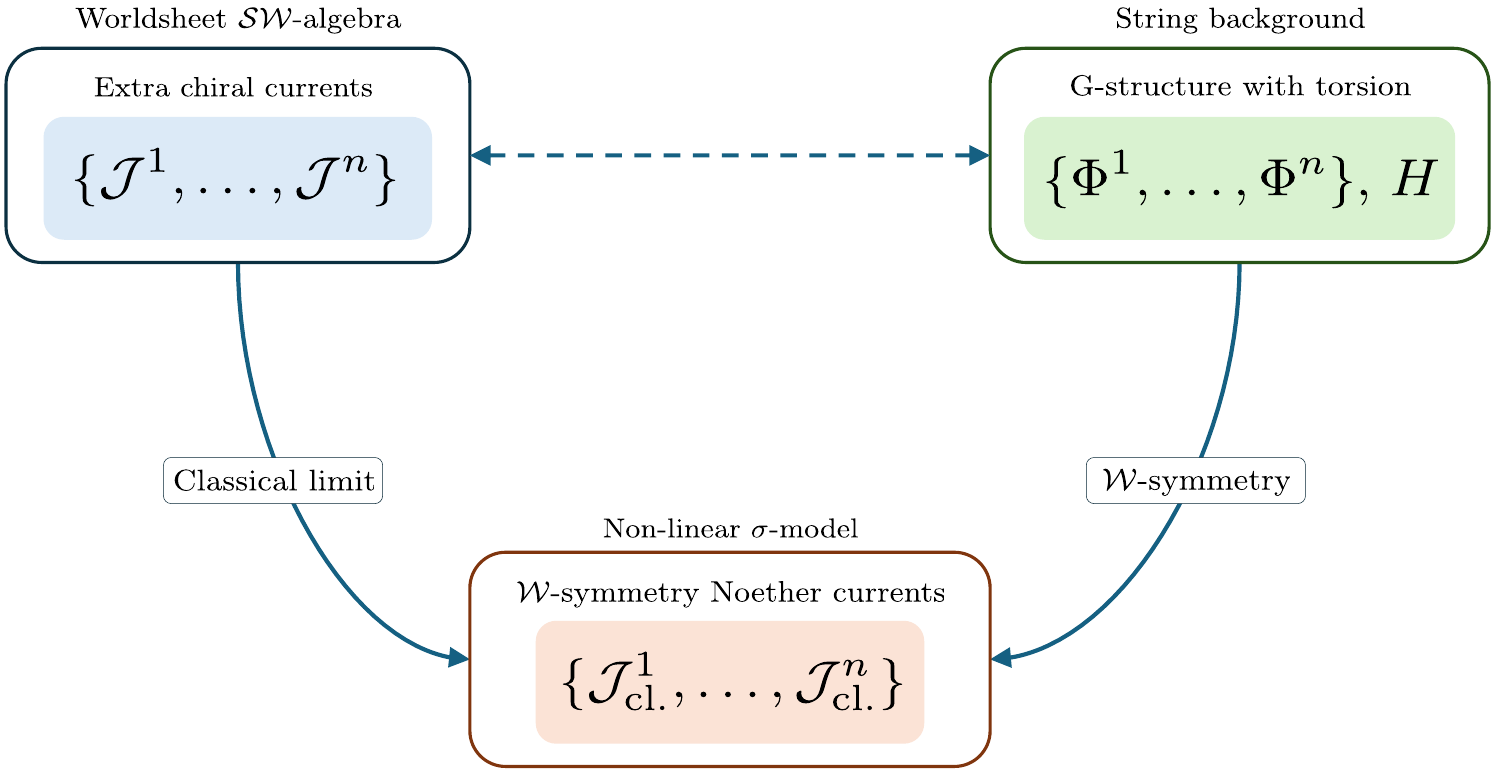}
\caption{\emph{Triangle of relationships between the geometry of the string background, the $\mathcal{SW}$-algebra of the worldsheet CFT and the non-linear $\sigma$-model. The dashed line represents the correspondence we want to study, which can be worked out explicitly (perturbatively in the string length scale $\ell_s$) at the level of the classical non-linear $\sigma$-model.}}
\label{fig: scheme}
\end{figure}

A $\mathrm{G}$-structure with torsion on $\mathcal{M}$ can be equivalently described in terms of a collection of differential forms $\lbrace \Phi^1, \dots, \Phi^n \rbrace$, known as \emph{characteristic forms}, that are covariantly constant with respect to a connection with torsion. This torsion is identified with the NS flux $H$ at the level of the $\sigma$-model. We thus find that the $\sigma$-model enjoys a classical algebra of $\mathcal{W}$-symmetries associated with $\mathcal{M}$, with a classical Noether (super) current $\cj_{\text{cl.}}^i$ for each covariantly constant form $\Phi^i$
\begin{equation*}
     \Phi^i \overset{\mathcal{W}}{\longrightarrow} \cj_{\text{cl.}}^i \, .
\end{equation*}
In the torsion-free setting, this algebra is usually regarded as a \emph{classical limit} of the algebra of worldsheet chiral symmetries, where the Noether currents are classical versions of the quantum chiral currents $\lbrace \cj^1, \dots, \cj^n \rbrace$,
\begin{equation*}
     \cj^i \overset{\text{cl.}}{\longrightarrow} \cj_{\text{cl.}}^i \, .
\end{equation*}
These relationships are pictorially summarised in Figure \ref{fig: scheme}.
Nevertheless, to the best of our knowledge this correspondence has not been made fully explicit in the literature so far. In this paper, we propose a procedure to compare the chiral algebra with its classical limit, simultaneously identifying the correct $\mathcal{SW}$-algebra and providing a geometric interpretation for its couplings. We do so for $(1,0)$ $\sigma$-models with NS flux $H$, a setup that includes $\mathrm{G}$-structures with torsion.\footnote{Since we focus on the holomorphic sector, the gauge bundle will not play an important role and one could argue that the same result should hold for the holomorphic sector of $(1,1)$ $\sigma$-models as well.} On the one hand, from the number (and degree) of the differential forms describing the $\mathrm{G}$-structure on $\mathcal{M}$---and using the classification results for $\mathcal{SW}$-algebras---we can find a family of consistent quantum algebras that are candidates to describe the string background. These families depend on parameters---such as the central charge $c$---that lack any geometrical meaning at this stage, but are present in the OPEs of the operators $\mathcal{J}^i$.

On the other hand,  from the classical algebra of $\mathcal{W}$-symmetries we are able to produce a set of \emph{classical OPEs} for the chiral currents $\mathcal{J}_{\text{cl.}}^i$. The attribute ``classical'' indicates that these OPEs encode the same information of a set of commutators $[\delta^1_{\epsilon_{1}}, \delta^2_{\epsilon_{2}}]$, where we denote by $\delta^i$ a symmetry transformation preserving the classical worldsheet action. These classical OPEs encode geometrical information about the $\mathrm{G}$-structure, and in particular its torsion classes. Comparing the OPEs of the chiral currents $\mathcal{J}^i$ with those of their classical counterparts $\mathcal{J}_{\text{cl.}}^i$, the geometry of $\mathcal{M}$ constrains the couplings of the $\mathcal{SW}$-algebra. Since we are comparing a quantum algebra with its classical version, the identification is only effective up to the lowest order in the string length parameter $\ell_s$, and we pay close attention to this throughout our work. Remarkably, even at this order we are able to effectively identify the contribution from scalar torsion classes. 

We study different choices of $\mathrm{G}$-structures, more precisely we take the group $\mathrm{G}$ to be equal to $\mathrm{O}(d-n)$, $\mathrm{Spin}(7)$, $\mathrm{G}_2$, $\mathrm{SU}(2)$ and $\mathrm{SU}(3)$. In the absence of torsion we recover all the special holonomy algebras, explicitly verifying the widespread lore in the literature. For $\mathrm{Spin}(7)$- and $\mathrm{SU}(2)$-structures, the effect of torsion is not explicit at the order of $\ell_s$ we are working on.
Nevertheless, for $\mathrm{G}_2$-structures with torsion we recover a one-parameter algebra---first studied in \cite{Fiset:2021azq}---which can be understood as a deformation of the Shatashvili--Vafa $\mathrm{G}_2$ algebra associated with special holonomy. We find that the scalar torsion class of the $\mathrm{G}_2$-structure is tied to the parameter of the family through the relation reported in Table \ref{tab:summary}.

The case of $\mathrm{SU}(3)$-structures poses an additional challenge since the classification of the underlying $\mathcal{SW}$-algebras is not available. Therefore, we perform a perturbative study around the Od$(3)$ algebra corresponding to the special holonomy \emph{locus}. We find an infinitesimal deformation parametrised again by the scalar torsion classes: this strongly suggests the existence of an honest $\mathcal{SW}$-algebra associated with $\mathrm{SU}(3)$-structures with torsion.
\subsection*{Outline of the paper}
 \noindent The paper is organised as follows: \begin{itemize}[label=$\diamond$]
     \item In \Cref{intro gstr} and in \Cref{sec: Walg} we review the notions of $\mathrm{G}$-structure, torsion, and the construction of $\mathcal{SW}$-algebras, providing at the same time our superspace conventions.   
     \item In \Cref{sec:Wsymm} we review the (1,0) non-linear $\sigma$-model and the $\mathcal{W}$-symmetries. We present the computation of the commutators of the superconformal and $\mathcal{W}$-symmetries and reformulate them as classical OPEs.
     \item In \Cref{sec: connection} we introduce our procedure to interpolate the classical OPEs with their corresponding quantum version; we study explicitly the $\mathcal{SW}$-algebras associated with $\mathrm{O}(d-n)$-structures as a warm-up, moving then to the study of $\mathcal{SW}$-algebras associated with manifolds endowed with $\mathrm{Spin}(7), \mathrm{G}_2, \mathrm{SU}(2)$ and $\mathrm{SU}(3)$-structures.
     \item In \Cref{conc: W} we present our conclusions and outlook for future work.
     \item In \Cref{app:nullfields} we provide more details regarding the definition of null fields in the context of $\mathcal{SW}$-algebras, adopting a more formal point of view. We also comment on how to test null fields through their OPEs with the generators of a given $\mathcal{SW}$-algebra. 
     \item In \Cref{app: spechol} we present a \emph{compendium} of explicit OPEs associated with different $\mathcal{SW}$-algebras with or without torsion.
 \end{itemize}
We present a brief summary of the algebras that we identify for each of the $\mathrm{G}$-structures---as well as the constraints imposed by the identification with the classical algebras---in \Cref{tab:summary}.
\begin{table}[h]
\centering
{\footnotesize
\setlength\tabcolsep{5.2pt}
\renewcommand\arraystretch{1.5}
    \begin{tabular}{|c||c||c|c|c|}
    \hline
        $\mathrm{G}$-struct. & $H=0$  & $H\neq0$ & Par. & Constraints found \\
        \hline
        \hline
        $\mathrm{O}(d-n)$ & \multirow{2}{*}{$\mathcal{S}\mathrm{Vir}\oplus\mathrm{Free}^n$}  & \multirow{2}{*}{$\mathcal{SW}\left( \frac{3}{2},\frac{1}{2},\dots,\frac{1}{2} \right)$} & $C_{IJ}$ & $C_{IJ} = -\sigma_I \cdot \sigma_J + O(\ell_s^2)$ \\
         $\left(\sigma_1, \dots, \sigma_n\right)$ &  &  & $f_{IJK}$ & $f_{IJK}=-\ell_s \tensor{H}{_{ijk}}\sigma^{i}_{I}\sigma^{j}_{J}\sigma^{k}_{K}+O(\ell_s^2)$ \\
        \hline
         $\mathrm{Spin}(7)$ & \multirow{2}{*}{$\mathrm{SV}^{\mathrm{Spin}(7)}$ \cite{Shatashvili:1994zw}} & \multirow{2}{*}{$\mathrm{FS}$ \cite{Figueroa-OFarrill:1990mzn}} & \multirow{2}{*}{$c$} & \multirow{2}{*}{$-$} \\
         $\left(\Psi\right)$ & & & & \\
        \hline
         $\mathrm{G}_2$ & \multirow{2}{*}{$\mathrm{SV}^{\mathrm{G}_2}$ \cite{Shatashvili:1994zw}} & \multirow{2}{*}{$\FGseven_k$ \cite{Fiset:2021azq} $\subset$ $\mathrm{Bl}$ \cite{Blumenhagen:1991nm}} & \multirow{2}{*}{$k$} & \multirow{2}{*}{$\sqrt{\frac{2}{k}}\frac{7k-4}{49k-24}=\frac{1}{6}\tau_0 \ell_s + O(\ell_s^2)$} \\
         $\left(\varphi, \psi\right)$ & & & & \\
        \hline
         $\mathrm{SU}(2)$ & \multirow{2}{*}{$\mathrm{Od}(2)$ \cite{Odake:1988bh}} & \multirow{2}{*}{$\mathcal{SW}\left( \frac{3}{2},1,1,1 \right)$} & \multirow{2}{*}{$c$} & \multirow{2}{*}{$c=6+O(\ell_s^2)$} \\
         $\left(\omega, \Omega^\pm\right)$ & & & & \\
        \hline
         $\mathrm{SU}(3)$ & \multirow{2}{*}{$\mathrm{Od}(3)$ \cite{Odake:1988bh}} & \multirow{2}{*}{$\mathrm{Od}^\varepsilon(3)\subset \mathcal{SW}(\frac{3}{2},\frac{3}{2},\frac{3}{2}, 1)$} & \multirow{2}{*}{$v_{\pm}$} & $v_{\pm}=4 \, w_0^{\pm} \ell_s+O(\ell_s^2)$  \\
          $\left(\omega, \Omega^\pm\right)$ & & & &$W_{0}^{\pm}=\veps \,  w_{0}^{\pm}+O(\veps^2, \ell_s^2) $ \\
        \hline
    \end{tabular}
    }
    \caption{\emph{Summary of the existing literature and new results. For each $\mathrm{G}$-structure we provide the characteristic forms in brackets. More details regarding $\mathrm{G}$-structures can be found in Sections \ref{intro gstr} and \ref{sec: connection}. For each one of them we list the associated special holonomy $(H=0)$ and torsionful $(H \neq 0)$ $\mathcal{SW}$-algebras studied in this paper. Up to our knowledge, we pair each algebra with the original reference where it first appeared; more details---including the explicit OPEs---can be found in Appendix \ref{app: spechol}. Finally, we indicate the real parameters of each torsionful algebra and we provide their relationship with the torsion and the characteristic forms up to order $O(\ell_s^2)$. Notice that $\tau_0$ and $W_0^{\pm}$ indicate the scalar torsion classes respectively of the $\mathrm{G}_2$ and $\mathrm{SU}(3)$-structure intrinsic torsions. The derivations are presented in Section \ref{sec: connection}. }}
    \label{tab:summary}
\end{table}

\subsection{Differential forms conventions}
We collect in this brief Section all our conventions regarding differential forms and their manipulation. Given local coordinates $\lbrace x_1,\dots,x_d\rbrace$ and a metric $G$ on a manifold $\mathcal{M}$ of dimension $d$, a generic $p$-form $\Phi$ has coefficients 
\begin{equation*}
    \Phi=\frac{1}{p!} \Phi_{i_1 \cdots i_p}(x) \de x^{i_1\cdots i_p} \, ,
\end{equation*}
where $\de x^{i_1\cdots i_p} = \de x^{i_1} \wedge \cdots \wedge \de x^{i_p}$. As usual, we define the \emph{exterior derivative} $\de \Phi$ as the $(p+1)$-form
\begin{equation*}
    \de \Phi=\frac{1}{p!} \partial_j \Phi_{i_1 \cdots i_p}(x) \de x^{j \, i_1 \cdots i_p} \, .
\end{equation*}
Let $\Psi$ be another $q$-form on $\mathcal{M}$. We define the \emph{interior product} between $\Phi$ and $\Psi$ as the $(p+q-2)$-form
\begin{equation*}
   i_{\Phi}(\Psi)=\Phi^{i} \wedge \Psi_i \, ,
\end{equation*}
where $\Phi^{i}$ is given by
\begin{equation*}
    \Phi^{i}=\frac{1}{(p-1)!} \Phi^i{}_{j_2 \cdots j_p} \de x^{j_2 \cdots j_p} \, .
\end{equation*}
Given the $p$-form $\Phi$ and the $(p+n)$-form $\Theta$, we define the \emph{hook contraction} between $\Phi$ and $\Theta$ as
\begin{equation*}
    \Phi \lrcorner \Theta = \frac{1}{p!\,n!} \Phi^{i_1 \cdots i_p} \Theta_{i_1 \cdots i_p j_1 \cdots j_{n}} \de x^{j_1\cdots j_n} \, .
\end{equation*}
The Hodge star operator $*$ acts on the $p$-form $\Phi$ as follows
\begin{equation*} 
    *\Phi=\frac{\sqrt{G}}{(d-p)! \, p!}\Phi^{i_1 \cdots i_p}\epsilon_{i_{1}\cdots i_p j_1 \cdots j_{d-p}}\de x^{j_1 \cdots j_{d-p}} \, ,
\end{equation*}
where $\epsilon_{i_{1}\dots i_d}$ is the Levi-Civita symbol. The Hodge star and the hook contraction  are tied together by the following relation 
\begin{equation} \label{eq:id hodge corner}
    \Phi \lrcorner \Psi=(-1)^{p(d-p-n)}*(\Phi \wedge *\Psi) \, .
\end{equation}
Finally, we anticipate the introduction in Section \ref{sec:Wsymm} of a set of bosonic (super)fields $X^i$ and of a fermionic (super)derivative $D$. Given a $p$-form $\Phi$ on $\mathcal{M}$, we denote by $\widehat\Phi$ the contraction
\begin{equation}
\label{eq:hatdefinition}
    \widehat\Phi = \frac{1}{p!}\, \Phi_{i_{1} \cdots i_{p}}(X)\, DX^{i_{1}} \cdots DX^{i_{p}} \, .
\end{equation}
\subsection{Review of \texorpdfstring{$\mathrm{G}$}{G}-structures with torsion} \label{intro gstr}
In a supergravity compactification on a  $d$-dimensional manifold $\mathcal{M}$, demanding that some space-time supersymmetry is preserved typically requires $\mathcal{M}$ to be endowed with a \emph{$\mathrm{G}$-structure}. We briefly review $\mathrm{G}$-structures and their most relevant properties, for more detailed discussions we refer the reader to \cite{Joyce:2007, Figueroa-OFarrill:2020gpr}.

A $\mathrm{G}$-structure is a reduction of the frame bundle of $\mathcal{M}$ to a principal subbundle with structure group $\mathrm{G}\subset \mathrm{GL}(d,\mathbb{R})$. A $\mathrm{G}$-structure on $\mathcal{M}$ naturally prescribes an action of the group $\mathrm{G}$ on tensor fields on $\mathcal{M}$, and the space of tensors decomposes into $\mathrm{G}$-representations accordingly. The singlets under this action constitute a distinguished collection of nowhere-vanishing tensor fields on $\mathcal{M}$ that are $\mathrm{G}$-invariant and receive the name of \emph{characteristic tensors}. Conversely, the existence of such tensors uniquely determines each $\mathrm{G}$-structure.

Thus, we will describe $\mathrm{G}$-structures via their characteristic tensors in what follows. For example, an $\mathrm{O}(d)$-structure on $\mathcal{M}$ describes a choice of orthonormal frames and is therefore equivalent to the manifold being Riemannian. This shows that its characteristic tensor is simply a Riemannian metric $G$ and we can equivalently work directly with $G$. 

Different types of $\mathrm{G}$-structures can be classified according to their \emph{torsion classes} \cite{Gray:1980, Fernandez:1982, Fernandez:1986}. A connection $\nabla^{\text{T}}$ with torsion tensor $\text{T}$ is said to be \emph{compatible} with a given $\mathrm{G}$-structure if all the corresponding characteristic tensors are covariantly constant under $\nabla^{\text{T}}$. We will study $\mathrm{G}$-structures for which the characteristic tensors are typically differential forms $\Phi$: it is convenient to rewrite
\begin{equation*}
    \nabla^{\text{T}} \Phi = 0 \Longrightarrow \dd \Phi = -i_{\text{T}}(\Phi) \, ,
\end{equation*}
and note that $i_{\text{T}}(\Phi)$ does not depend on the choice of compatible connection. Therefore, the exterior derivatives of the characteristic forms intrinsically characterise the $\mathrm{G}$-structure, and their decomposition into irreducible $\mathrm{G}$-representations provides a collection of forms which are known as the \emph{torsion classes} of the $\mathrm{G}$-structure. These can in turn be used to define the \emph{intrinsic torsion} of the $\mathrm{G}$-structure \cite{Figueroa-OFarrill:2020gpr}. Since torsion classes depend on the choice of group $\mathrm{G}$, we will introduce them on a case-by-case basis in \Cref{sec: connection}.

Supersymmetry requires the existence of a unique compatible connection with \emph{totally antisymmetric torsion} so that it can be identified with the NS flux $H$, see \eqref{eq:identificationHtorsion} below \cite{Friedrich:2001nh, Ivanov:2001ma, Gauntlett:2003cy}. We will always assume that we have such a connection and denote its torsion by $\text{Tor}$: this typically imposes additional constraints on the torsion classes of the $\mathrm{G}$-structure. Furthermore, $\text{Tor}$ can be uniquely expressed in term of the torsion classes. 

To illustrate the discussion above, we briefly present the example of Spin(7)-structures, which will be discussed in more detail in \Cref{sec:spin7comparison}. A Spin(7)-structure on an eight-dimensional manifold is determined by a single characteristic four-form $\Psi$ known as the \emph{Cayley form}. A Spin(7)-structure has two torsion classes $\tau_1 \in \Omega^{1}_{\bf{8}}(TM)$ and $\tau_3 \in \Omega^{3}_{\bf{48}}(TM)$, which are defined as follows
\begin{equation*}
   \de \Psi = -i_{\text{Tor}}(\Psi)=\tau_1 \wedge \Psi + *\tau_3 \, ,
\end{equation*}
where the totally antisymmetric torsion $\text{Tor}$ decomposes as
\begin{equation*}
   \text{Tor} = -\frac{1}{6}\, \tau_1\lrcorner\,\Psi - \tau_3 \, .
\end{equation*}
In this case the existence of $\text{Tor}$ does not impose any additional constraints on the torsion classes.

Finally, note that $\mathrm{G}$-structures are a natural generalisation of the concept of $\mathrm{G}$-holonomy: when all torsion classes vanish, i.e., the $\mathrm{G}$-structure is \emph{torsion-free}, we recover the usual definition of $\mathrm{G}$-holonomy in terms of closed characteristic tensors.

\subsection{Review of \texorpdfstring{$\mathcal{SW}$}{SW}-algebras}

\label{sec: Walg}

In this work we are interested in the symmetry algebras of worldsheet CFTs, where the worldsheet is spanned by a $(1,0)$ supersymmetric string. These algebras are chiral, since they split into a holomorphic and an antiholomorphic sector: we will focus on the holomorphic one, which enjoys $\mathcal{N}=1$ supersymmetry. As a consequence, we always deal with extensions of the super Virasoro algebra, generated by the holomorphic component of the stress-energy tensor $T(z)$ and its super partner $G(z)$. 

The enhancement of the Virasoro algebra via additional primary chiral currents can be described as a \emph{$\mathcal{W}$-algebra}: similarly, the enhancement of the super Virasoro algebra can be described by super $\mathcal{W}$-algebras, which will be called $\mathcal{SW}$-algebras in the following. Going into further details, in our work $\mathcal{SW}$-algebras are generated by enhancing the $\mathcal{N}=1$ Virasoro algebra through the addition of primary $\mathcal{N}=1$ (super) multiplets. 
The study of these algebras dates back to the seminal works \cite{Belavin:1984vu, Zamolodchikov:1985wn}, but we will mostly follow \cite{Blumenhagen:1990jv, Blumenhagen:1991nm, deBoer:2005pt, Figueroa-OFarrill:1990mzn, Thielemans:1994er}. Since $\mathcal{N}=(1,0)$ supersymmetry on the worldsheet will be preserved throughout this paper, it will be convenient to work in the superspace formalism \cite{Blumenhagen:1990jv, Blumenhagen:1991nm, Figueroa-OFarrill:1990mzn}. The advantage of working with manifest supersymmetry is the possibility to recast many OPEs into a single super OPE.

We start by setting up our superspace conventions. In addition to the holomorphic coordinates $(z, \bar{z})$ on the worldsheet $\Sigma$, we consider a Grassmann coordinate $\theta$ with (holomorphic) conformal weight $h_{\theta}=-\frac{1}{2}$. The coordinates on the entire \emph{super} worldsheet $\mathbb{\Sigma}$ will be called $\zeta=(\theta, z, \Bar{z})$, while the holomorphic coordinates will be called $Z=(\theta, z)$. The measure on $\mathbb{\Sigma}$ will be indicated as $\de^{2|1}\zeta=\de z \de \Bar{z} \de \theta$, whereas the holomorphic part of the measure will be denoted by $\dd Z=\dd z \dd\theta$. We introduce the super derivative
\begin{equation*}
    D=\del_{\theta}+\theta \del \, ,
\end{equation*}
and we highlight the property $D^2=\del$. Moreover, we introduce two covariant intervals on the worldsheet \cite{Blumenhagen:1991nm, Heluani:2006pk}
\begin{equation*}
    Z_{12} = z_1-z_2-\theta_1 \theta_2 \, , \qquad \theta_{12} = \theta_1-\theta_2 \, .
\end{equation*}
It should be noted that
\begin{equation*}
    \frac{1}{Z_{12}^n}=\frac{1}{(z_1-z_2)^n}\left(1 + n \frac{\theta_1 \theta_2}{z_1-z_2}\right) \, .
\end{equation*}
Working in superspace, each super multiplet can be organised into a \emph{super operator}. For example, the Virasoro multiplet $(\tfrac{1}{2}G,T)$ can be recast as the super stress-energy tensor
\begin{equation}
\label{eq: superstress}
    \ct(Z) = -\frac{1}{2}G(z)+\theta \, T(z) \, ,
\end{equation}
and the entire $\mathcal{N}=1$ Virasoro algebra can then be encoded in a single \emph{super OPE}
\begin{equation} \label{eq: superTT}
    \ct(Z_1) \ct(Z_2) \sim \frac{c}{6}\frac{1}{Z_{12}^3}+\frac{3}{2}\frac{\theta_{12}}{Z_{12}^2} \mathcal{T}(Z_2)+\frac{1}{ 2 Z_{12}} D\mathcal{T}(Z_2) +\frac{\theta_{12}}{Z_{12}} \partial\mathcal{T}(Z_2) +\cdots \, ,
\end{equation}
where the dots indicate that we are neglecting the regular terms. Note that by expanding intervals and super operators in the Grassmann coordinate $\theta$ we recover the usual Virasoro OPEs. 
Similarly, given a primary super multiplet $(P_{h},K_{h})$ of weight $h$ the four OPEs between $T$, $G$ and the operators in the multiplet can be encoded into a single super OPE between $\mathcal{T}$ and the primary super operator $\mathcal{J}_{h}$
\begin{equation} \label{eq: superTJ}
    \mathcal{T}(Z_1)\mathcal{J}_{h}(Z_2)\sim  \frac{h \, \theta_{12}}{Z_{12}^2}\mathcal{J}_{h}(Z_2) +\frac{1}{2 \, Z_{12}}D \cj_{h}(Z_2) +\frac{\theta_{12}}{Z_{12}}\partial \cj_{h}(Z_2)+\cdots \, ,
\end{equation}
where our conventions for the super operator are as follows:
\begin{equation}
\label{eq:superoperatorcomponents}
{\cal J}_{h}(Z) = (-i)^{2 h}  \big(- P_{h}(z) + \theta  \, K_{h}(z)\big)\, .
\end{equation}
We say that a super $\mathcal{W}$-algebra is \emph{generated} by a set of super operators $\Braket{\mathds{1},\mathcal{O}_1,\mathcal{O}_2, \dots}$ if any super operator in the algebra can be built out of the generators using addition, scalar multiplication, (super) derivatives and super normal ordering. Given two super operators $\mathcal{O}_i$ and $\mathcal{O}_j$, their super normal ordered product $N(\mathcal{O}_i\mathcal{O}_j)$ is simply the normal ordered product of the components, which we define in the usual way:
\begin{equation}
\label{eq:normord}
    N(AB)(z)=\oint_{z} \frac{\dd w}{2 \pi i} \, \frac{A(w) B(z)}{w-z} \, ,
\end{equation}
where $A$ and $B$ are two operators and we perform an anti-clockwise integration over a contour circling the $B(z)$ operator. In this work, our focus is on $\mathcal{SW}$-algebras $\mathcal{SW}(\frac{3}{2}, h_1, \dots, h_n)$, generated by the identity operator $\mathds{1}$, the super stress-energy tensor $\mathcal{T}$ and a \emph{finite} collection of primary super operators $\cj_{h_1}, \dots, \cj_{h_n}$ 
In this notation, $\mathcal{SW}(\frac{3}{2})=\mathcal{S}\mathrm{Vir}$ is the $\mathcal{N}=1$ super Virasoro algebra. In the following, we will often drop the prefix ``super'', leaving it implicit.

The OPEs between the generators of an $\mathcal{SW}$-algebra are constrained by superconformal symmetry. This was first described in \cite{Blumenhagen:1991nm} and we provide here a brief summary. By definition, the operator content $\mathcal{F}$ of an $\mathcal{SW}$-algebra can be split as follows 
\begin{equation*}
    \mathcal{F}=\bigoplus_{s,r=0}^{\infty} D^s \mathcal{F}_{r}^{\, \text{q.p.}} \, ,
\end{equation*}
where $\mathcal{F}_{r}^{\, \text{q.p.}}$ denotes the set of all quasi-primary super operators of conformal weight $r$. It is important to highlight that the normal ordered product of two quasi-primary operators $\mathcal{O}_i$ and $\mathcal{O}_j$ in general \emph{does not} return a quasi-primary operator. Nevertheless, $N(\mathcal{O}_i\mathcal{O}_j)$ can always be \emph{projected} onto the space $\mathcal{F}^{\, \text{q.p.}}_{h_i + h_j}$ \cite{Blumenhagen:1991nm}. We indicate by $\cn(\mathcal{O}_i\mathcal{O}_j)$ the quasi-primary projection of the normal ordering $N(\mathcal{O}_i\mathcal{O}_j)$, and refer to \cite{Blumenhagen:1991nm} for the explicit formulas.\footnote{Note however that the normal ordering convention in \cite{Blumenhagen:1991nm} differs from the one we are using \eqref{eq:normord}.} For example, the normal ordered product $N(\mathcal{T}\mathcal{T})=\frac{1}{4}\partial D\mathcal{T}$ is the descendant of a quasi-primary on the nose. On the other hand, this is not true for the operator $N(D\mathcal{T}\mathcal{T})$, so we can project it on the quasi-primary $\cn(D\mathcal{T}\mathcal{T})=N(D\mathcal{T}\mathcal{T})-\frac{3}{8}\partial \partial\mathcal{T}$.

Given a set of quasi primary operators $\left \lbrace \co_i \right \rbrace$, we introduce the two-point and the three-point function coefficients $d_{ij}$ and $C_{ijk}$ \cite{Blumenhagen:1991nm}:
\begin{align*} 
    \Braket{\mathcal{O}_i(Z_1)\mathcal{O}_j(Z_2)}&=\frac{d_{ij}}{Z_{12}^{2 h_i}} \, , \\[5pt]
      \Braket{\mathcal{O}_{i}(Z_1) \mathcal{O}_j(Z_2) \mathcal{O}_k(Z_3)}&=\left\{\begin{matrix*}[l]
\displaystyle \frac{C_{ijk}}{Z_{12}^{h_{kij}}\,Z_{23}^{h_{ijk}}\,Z_{13}^{h_{jki}}}& \qquad \text{if } h_{ijk}\in\mathbb{Z}\, ,\\[2em]
 \displaystyle (-1)^{2h_i+1}\, \Xi_{123} \, \frac{ C_{ijk}}{Z_{12}^{h_{kij}}\,Z_{23}^{h_{ijk}}\,Z_{13}^{h_{jki}}}& \qquad \text{if } h_{ijk}\in\mathbb{Z}+\frac{1}{2} \, , \\
\end{matrix*}\right.
\end{align*}
where $h_{ijk}=h_{i}+h_j-h_k$ and $\Xi_{123}=\theta_1 Z_{23}-\theta_2 Z_{13}+\theta_3 Z_{12}+\theta_1 \theta_2 \theta_3$. The coefficients $C_{ijk}$ are invariant under an even permutation of the indices, while they pick a sign $(-1)^F$ under an odd permutation where the exponent $F$ is determined by
\begin{equation} \label{eq: F}
    F=\left\{\begin{matrix*}[l]
h_{ijk}+4h_i h_j & \qquad \text{if $h_{ijk} \in \mathbb{Z}$} \, , \\[0.5em]
h_{ijk}+4(h_i+\frac{1}{2})(h_j+\frac{1}{2})+\frac{1}{2}  & \qquad  \text{if $h_{ijk} \in \mathbb{Z}+\frac{1}{2}$} \, . 
\end{matrix*}\right. 
\end{equation}
We can now write the following Ansatz for the OPE of two generators $\mathcal{J}_{h_i}$, $\mathcal{J}_{h_j}$ \cite{Blumenhagen:1991nm}:
\begin{equation} \label{eq: ANS}
    \mathcal{J}_{h_i}(Z_1) \mathcal{J}_{h_j}(Z_2)\hspace{4pt} \sim \sum_{k,r=0}^{\infty} \,  \sum_{\mathcal{O}_{k}\in \, \mathcal{F}^{\, \text{q.p.}}_{k}} \, C_{ij}^{k} \, A_{ijk}^{r} \, \frac{1}{Z_{12}^{h_{ijk}-r/2}} \, D^{r} \mathcal{O}_{k}(Z_2) \, ,
\end{equation}
where the coefficients $A_{ijk}^r$ and $C_{ij}^k$ can be computed exactly. Note that the $A_{ijk}^r$ are purely combinatorial, whereas the $C_{ij}^k$ carry physical information and are usually referred to as \emph{couplings} \cite{Figueroa-OFarrill:1990tqt, Blumenhagen:1991nm}. If $h_{ijk} \in \mathbb{Z}$, the coefficients $A_{ijk}^r$ can be found in equation (2.46) of \cite{Blumenhagen:1991nm} 
and the couplings $C_{ij}^k$ are the solutions of the linear system
\begin{equation*}
    C_{ij}^{\ell}d_{\ell k}=C_{ijk} \, .
\end{equation*}
If instead $h_{ijk} \in \mathbb{Z}+\frac{1}{2}$, the coefficients $A_{ijk}^r$ can be found in equation (2.48) of \cite{Blumenhagen:1991nm} and the couplings $C_{ij}^k$ are determined by
\begin{equation*}
     C_{ij}^{\ell}d_{\ell k}=(-1)^{2h_j +1}C_{ijk} \, .
\end{equation*}
Since $d_{ij}$ and $C_{ijk}$ enjoy symmetries under the permutation of the indices, it is natural to investigate the symmetries of the couplings $C_{ij}^{k}$. Normalising the basis of quasi-primary generators $\left\lbrace \co_i \right\rbrace$ in such a way that the $d_{ij}$ tensor is diagonal, the coefficients of the two-point functions take the form
\begin{equation*}
    d_{ij}=C_{ii}^{\mathds{1}} \, \delta_{ij} \, ,
\end{equation*}
and we obtain the following identities
\begin{equation} \label{eq: coup}
    C_{ijk}=C^{k}_{ij} C^{\mathds{1}}_{kk} \quad \text{if $h_{ijk} \in \mathbb{Z}$} \, , \qquad C_{ijk}=(-1)^{2h_j+1}C^{k}_{ij} C^{\mathds{1}}_{kk} \quad \text{if $h_{ijk} \in \mathbb{Z}+\frac{1}{2}$} \, .
\end{equation}
The symmetries of the three-point functions coefficient $C_{ijk}$ can be made explicit as follows, where the exponent $F$ is given by \eqref{eq: F}:
\begin{equation}  \label{eq: three}
   C_{ijk}=C_{jki}=C_{kij}=(-1)^F C_{kji}=(-1)^{F}C_{ikj}=(-1)^{F}C_{jik} \, .
\end{equation}
Combining the identities \eqref{eq: coup} with the identities \eqref{eq: three}, we can extract two lists of constraints, reported in Table  \ref{tab:symm}. These symmetries can be applied to further simplify the Ansatz \eqref{eq: ANS}.
\begin{table}
    \centering
    \renewcommand{\arraystretch}{2}
    \begin{tabular}{|c|r|r|}
    \hline
        Coupling & \multicolumn{1}{c|}{$h_{ijk} \in \mathbb{Z}$} & \multicolumn{1}{c|}{$h_{ijk} \in \mathbb{Z}+\frac{1}{2}$}\\
        \hline
        $C_{jk}^{i}$ & $\big(C_{kk}^{\mathds{1}}/C_{ii}^{\mathds{1}}\big)  C_{ij}^{k}$ & $(-1)^{2(h_j-h_k)}\big(C_{kk}^{\mathds{1}}/C_{ii}^{\mathds{1}}\big) C_{ij}^{k}$\\
        $C_{ki}^{j}$ & $\big(C_{kk}^{\mathds{1}}/C_{jj}^{\mathds{1}}\big) C_{ij}^{k}$ & $(-1)^{2(h_j-h_i)}\big(C_{kk}^{\mathds{1}}/C_{jj}^{\mathds{1}}\big) C_{ij}^{k}$\\
        $C_{kj}^{i}$ & $(-1)^{F}\big(C_{kk}^{\mathds{1}}/C_{ii}^{\mathds{1}}\big)C_{ij}^{k}$ & $(-1)^F\big(C_{kk}^{\mathds{1}}/C_{ii}^{\mathds{1}}\big)C_{ij}^{k}$\\
        $C_{ik}^{j}$ & $(-1)^{F}\big(C_{kk}^{\mathds{1}}/C_{jj}^{\mathds{1}}\big)C_{ij}^{k}$ & $(-1)^F(-1)^{2(h_j-h_k)}\big(C_{kk}^{\mathds{1}}/C_{jj}^{\mathds{1}}\big)C_{ij}^{k}$\\
        $C_{ji}^{k}$ & $(-1)^F C_{ij}^{k}$ & $(-1)^F (-1)^{2(h_j-h_i)} C_{ij}^{k}$\\
        \hline
    \end{tabular}
    \caption{\emph{Given three quasi-primary operators $\mathcal{O}_i$, $\mathcal{O}_j$ and $\mathcal{O}_k$, the couplings in the first column can be rewritten in terms of a single coupling $C_{ij}^{k}$, the two-point function coefficients and an overall sign which depends on the conformal weights. The couplings should be identified with the ones in the second column if $h_{ijk}$ is an integer number and with those in the third column otherwise.}}
    \label{tab:symm}
\end{table}
Finally, the Ansatz \eqref{eq: ANS} must comply with the \emph{associativity} condition, which can be encoded in a series of Jacobi identities. The outcome of imposing such condition greatly varies depending on the number and weight of the generators of the $\mathcal{SW}$-algebra: we will therefore comment on it on a case-by-case basis for the algebras we present in \Cref{sec: connection}.

Finally, it is important to recall that, in some cases, it is possible to construct \emph{null fields} (sometimes also called \emph{singular fields}) using the generators of the $\mathcal{SW}$-algebra. A null field is an operator such that every correlation function in which it is present vanishes automatically \cite{DiFrancesco:1997nk}. Null fields generate \emph{ideals} inside the original $\mathcal{SW}$-algebra. More details about null fields and their definition, following \cite{Thielemans:1994er}, can be found in \Cref{app:nullfields}.
\newline Since OPEs are only meant to hold inside correlation functions, the presence of null fields relaxes some aspects of the discussion above. For instance, the OPE consistency conditions---including the associativity of the OPE---should only really hold \emph{up to null fields}. Examples of this can be found in the literature on $\mathcal{SW}$-algebras \cite{Shatashvili:1994zw, Fiset:2021azq}. Furthermore, OPEs themselves are only well-defined up to the addition of null fields. This means the value of the couplings $C_{ij}^{k}$ can be modified by redefining the OPE using null fields: this will play a important role when giving a geometric interpretation to OPEs in \Cref{sec: connection}.

\section{Classical OPEs from \texorpdfstring{$\mathcal{W}$}{W}-symmetries}
\label{sec:Wsymm}
\noindent In this Section we work with a classical $(1,0)$ non-linear $\sigma$-model and we study the Noether currents associated with its superconformal and $\mathcal{W}$-symmetries. These currents should be seen as the classical limits of the ones appearing in the super $\mathcal{W}$-algebras introduced in the previous Section. We study their commutators and we write down a set of classical OPEs, i.e., OPEs between the classical limits of the currents. As anticipated in the Introduction, the attribute ``classical'' indicates that such OPEs find their origin in a set of commutators $[\delta^1_{\epsilon_{1}}, \delta^2_{\epsilon_{2}}]$, where $\delta^i_{\epsilon_i}$ denotes a generic symmetry transformation preserving the worldsheet classical action. 

\subsection{The (1,0) non-linear \texorpdfstring{$\sigma$}{sigma}-model}

We work with the classical $(1,0)$ non-linear $\sigma$-model, which can be employed to describe the dynamics of a heterotic superstring propagating on a general string background. Let $\mathcal{M}$ be a $d$-dimensional Riemannian manifold endowed with a metric $G$ and a two-form (Kalb--Ramond field) $B$. Let $\mathcal{V} \hookrightarrow \mathcal{M}$ be a gauge bundle over $\mathcal{M}$ of rank $q$ endowed with a connection $A$. Without loss of generality, we consider the metric on the fibres of the bundle to be constant \cite{delaOssa:2018azc}. We can make supersymmetry manifest by working in the $(1,0)$ superspace formalism (see \Cref{sec: Walg} for more details and conventions).  The $(1,0)$ $\sigma$-model has two sets of superfields\footnote{In the following, the indices $i, j, k, \dots$ run from 1 to $d$; the indices $\alpha, \beta, \gamma, \dots$ run from 1 to $q$.} 
\begin{equation*}
     X^{i}=x^{i}+i \theta \psi^{i} \, , \qquad \Lambda^{\alpha}=\lambda^{\alpha}+\theta f^{\alpha} \, .
\end{equation*}
The $X^i$ superfields are bosonic: the components $x^i$ 
describe the coordinates of the internal manifold $\mathcal{M}$, while the $\psi^i$ are a set of left-moving Majorana--Weyl fermions. The $\Lambda^{\alpha}$ superfields are fermionic: the fields $\lambda^{\alpha}$ are right-moving Majorana--Weyl fermions and represent sections of the gauge bundle $\mathcal{V}$, whereas $f^{\alpha}$ are a set of auxiliary fields. The action of the classical two-dimensional $(1,0)$ non-linear $\sigma$-model reads \cite{Hull:1986xn, delaOssa:2018azc} 
\begin{equation}  \label{eq:action}
 S[X,\Lambda] = \int_{\mathbb{\Sigma}}\frac{\de^{2|1}\zeta}{2 \, \ell_s^2}  \left[M_{ij}(X)\bar\partial X^{i} DX^{j} + \text{tr}\left(\Lambda D_{A}\Lambda\right)\right] \, ,
\end{equation}
where the trace is performed over the gauge bundle indices, $M_{ij}(X)=G_{ij}(X)+B_{ij}(X)$ and $D_{A}\Lambda^{\alpha}=D \Lambda^{\alpha} + \tensor{\widehat{A}}{^{\alpha}_{\beta}}(X)\Lambda^{\beta}$, with $\tensor{\widehat{A}}{^{\alpha}_{\beta}}(X)=\tensor{A}{_i^{\alpha}_{\beta}}(X)D X^{i}$; $\ell_{s}$ is the \emph{string length scale}, which we write in terms of the Regge slope $\alpha'$ as
\begin{equation*}
    \ell_{s}=\sqrt{2 \pi \alpha'} \, .
\end{equation*}
It is instructive to integrate out the Grassmann variable $\theta$ and the auxiliary fields $f^\alpha$ as we are then naturally led to introduce the gauge field strength $F=\de  A + A\wedge A$ and the \emph{Neveu--Schwarz three-flux} (NS flux) $H=\de B$:
\begin{equation*}
S[x,\psi,\lambda] = \int_{\Sigma} \frac{\dd z \de \Bar{z}}{2 \, \ell^2_s} \left[ M_{ij}(x)\bar\partial x^{i}\partial x^{j} 
+ G_{ij}(x)\psi^{i} \Bar{\nabla} \psi^{j} 
- \text{tr}\left(\lambda \nabla_{A} \lambda \right)
\right] \, ,
\end{equation*}
where $\Bar{\nabla}\psi^{i}=\bdel\psi^{i}+\tensor{\Gamma^{+}(x)}{^i_{k \ell}} \, \bdel x^{\ell}\psi^{k}$, $\nabla_{A} \lambda^{\alpha}=\del \lambda^{\alpha}-\tensor{\widetilde{F}(x)}{^{\alpha}_\beta}\lambda^{\beta}$ and 
\begin{equation} \label{eq: +symb}
    \Gamma^{+}(x)^{i}{}_{jk} = \mathring\Gamma(x)^{i}{}_{jk} + \frac{1}{2}\, H(x)^{i}{}_{jk} \, , \qquad \tensor{\widetilde{F}(x)}{^{\alpha}_\beta}=\frac{1}{2}\tensor{F(x)}{_{ij}^{\alpha}_\beta} \psi^{i}\psi^{j}  \, .
\end{equation}
Here $\mathring\Gamma(x)$ represents the Christoffel symbols associated with the target metric $G_{ij}(x)$. 

The NS flux $H$ will play a fundamental role in what follows. We are interested in the case where $\mathcal{M}$ is equipped with a $\mathrm{G}$-structure such that its characteristic tensors are covariantly constant with respect to the connection $\nabla^+$ with symbols \eqref{eq: +symb}. This means that $\nabla^+$ must be compatible with the G-structure: since $\nabla^+$ has totally antisymmetric torsion, this imposes constraints on the G-structure and its torsion classes \cite{Friedrich:2001nh, Ivanov:2001ma, Gauntlett:2003cy}. As a result, we identify in what follows the flux $H$ with the totally antisymmetric torsion $\text{Tor}$ we introduced in \Cref{intro gstr}:
\begin{equation}
\label{eq:identificationHtorsion}
    H= \text{Tor} \, .
\end{equation}
It should also be noted that this flux receives corrections at higher orders in $\ell_s$ in heterotic supergravity via the Green--Schwarz mechanism \cite{Green:1984sg}---which can also be reproduced at the level of the $\sigma$-model \cite{Hull:1985jv,Hull:1986xn}---of the form 
\begin{equation} \label{eq: het tors}
H = \dd B + \frac{\ell_s^2}{8 \pi} \big(\mathrm{CS}_{3}(A)- \mathrm{CS}_{3}(\Theta)\big) \, ,
\end{equation}
where  $\mathrm{CS}_{3}(\cdot)$ is the Chern-Simons three-form, $A$ is the gauge bundle connection, and $\Theta$ is an instanton connection on the tangent bundle. As we will point out later, several results for $(1,0)$ $\sigma$-models still hold (up to the appropriate order in $\ell_s$) when the $\ell_s$-corrected expression \eqref{eq: het tors} of $H$ is employed \cite{delaOssa:2018azc, Papadopoulos:2023exe}.

For later convenience, we adopt the convention where the superfields $X^i$ and $\Lambda^\alpha$ are \emph{adimensional} 
from the point of view of the string background:
\begin{equation*}
    X^i \to \ell_s X^i \, , \qquad \Lambda^\alpha \to \ell_s \Lambda^\alpha \, . 
\end{equation*}
The immediate consequence is that the coordinates of the string background $x^{i}$ are now adimensional. In this convention, the action \eqref{eq:action} reads 
\begin{equation}  \label{eq:action2}
 S[X,\Lambda] = \frac{1}{2} \int_{\mathbb{\Sigma}} \de^{2|1}\zeta  \left[M_{ij}(X)\bar\partial X^{i} DX^{j} + \text{tr}\left(\Lambda D_{A}\Lambda\right)\right] \, ,
\end{equation}
where now factors of $\ell_s$ are present in the covariant derivatives, for example
\begin{equation*}
    D_{A}\Lambda^{\alpha}=D \Lambda^{\alpha} + \ell_s \, \tensor{\widehat{A}}{^{\alpha}_{\beta}}(X)\Lambda^{\beta} \, .
\end{equation*}
In the following, we will employ the action \eqref{eq:action2}. 

\subsection{Symmetries and classical currents} \label{ssec: symm}

We now study the chiral symmetries of the action \eqref{eq:action2}. As anticipated earlier, we will focus on the holomorphic sector and largely ignore the anti-holomorphic sector. Given a continuous symmetry of the action \eqref{eq:action2}, its infinitesimal variation is given by \cite{delaOssa:2018azc}
\begin{equation*}
\delta S = \int_{\mathbb\Sigma} \de^{2|1}\zeta\, 
\Big\{
\left[- G_{ij}\big( D\bar\partial X^{j} + \ell_s \, \Gamma^{j}{}_{k \ell} DX^{k} \bar\partial X^{\ell}\big) 
+
\text{tr} (\Lambda 
\widehat{F}_{i}\Lambda ) \right] \delta X^i+  
\text{tr} (D_A \Lambda\, \delta_A\Lambda) \Big\} \, ,
\end{equation*}
where the notation $\widehat{F}_{i}$ was first defined in \eqref{eq:hatdefinition} and we introduced a gauge covariant variation of the fields $\Lambda^{\alpha}$
\begin{equation*}
    \delta_A \Lambda^{\alpha} = \delta \Lambda^{\alpha} +\ell_s \, \tensor{A}{_{i}^{\alpha}_{\beta}}\, \delta X^i\,\Lambda^{\beta} \, .
\end{equation*}
Classically, a continuous, chiral holomorphic symmetry, whose infinitesimal transformations are parametrised by the parameter $\epsilon(Z)$, allows the variation of the action to be written in the form 
\begin{equation*}
    \delta S=\int_{\mathbb \Sigma}\de^{2|1}\zeta \, \mathcal{J}_{\text{cl.}}(Z) \bdel \epsilon(Z) \, ,
\end{equation*}
where we can read off the classical current $\mathcal{J}_{\text{cl.}}(Z)$. Integrating by parts one recovers Noether's theorem and finds that, up to equations of motion, $\bdel \mathcal{J}_{\text{cl.}}(Z)=0$ classically. We can associate a charge to the chiral current $\mathcal{J}_{\text{cl.}}(Z)$, acting on worldsheet operators as follows 
\begin{equation}
\label{eq:superchargeconvention}
    \op{\mathcal{Q}}_{\mathcal{J}}\, \mathcal{O}(\zeta)=\frac{1}{2\pi i}\oint_{\zeta}\dd Z \, \epsilon(Z)  \mathcal{J}_{\text{cl.}}(Z) \mathcal{O}(\zeta) \,  ,
\end{equation}
The integral must be understood as the usual integration on the Grassmann variable $\theta$, plus an  anti-clockwise integration over a contour circling the insertion of the $\co(\zeta)$ operator.
\paragraph{Superconformal symmetry.}
The first chiral symmetry we want to consider is superconformal symmetry. The action on the fields reads in our conventions
\begin{equation*}
\delta_{\epsilon}^{\ct} X^{i} = - \epsilon \,\partial X^{i}  -  \frac{1}{2} D\epsilon \, DX^{i}\, , \qquad
\delta_{\epsilon,A}^{\ct}\Lambda =  - \epsilon \,\partial_{A} \Lambda  -  \frac{1}{2} D\epsilon \,D_{A}\Lambda\, .
\end{equation*}
The chiral current is the \emph{classical} stress-energy tensor
\begin{equation} \label{eq: class T}
{\cal T}_{\text{cl.}} = -\frac{1}{2} \left( G_{ij} \partial X^{i}DX^{j} - \ell_s \widehat H \right) \, ,
\end{equation}
where there is an explicit dependence on the torsion via $H$. The expression \eqref{eq: class T} should be thought of as a classical realisation of the stress-energy tensor operator introduced in \Cref{sec: Walg}. We recall that the stress-energy tensor can be decomposed as in equation \eqref{eq: superstress} 
\begin{equation*}
{\cal T}_{\text{cl.}}(Z) = - \frac{1}{2}  G_{\text{cl.}} (z) + \theta \, T_{\text{cl.}}(z)\, ,
\end{equation*}
and each component can be associated with a classical realisation
\begin{equation*}
 G_{\text{cl.}} = i\left(G_{ij} \partial x^{i}\psi^{j} + \ell_s \widetilde H \right) \, , \quad 
 T_{\text{cl.}} = - \frac{1}{2} G_{ij}\left[\partial x^{i}\partial x^{j} 
- (\partial \psi^{i} + \ell_s \Gamma^{i}{}_{k \ell}\, \partial x^{k}\psi^{\ell})\psi^{j}
\right] \, ,
\end{equation*}
where $\Gamma^{i}{}_{k \ell}$ was introduced in \eqref{eq: +symb} and $\widetilde H = \frac{1}{3!}\, H_{ijk}\,\psi^i \psi^j \psi^k$. In \cite{delaOssa:2018azc} the authors showed that the stress-energy tensor is still a chiral current at one-loop order.

\paragraph{$\mathcal{W}$-symmetry.}
The geometrical information provided by a $\mathrm{G}$-structure on $\mathcal{M}$ is encoded at the level of the non-linear $\sigma$-model \eqref{eq:action2} as a continuous, non-linear symmetry, which in the literature goes under the name of \emph{$\mathcal{W}$-symmetry} \cite{Bilal:1990dm, Howe:1991im, Howe:1991vs, Howe:1991ic, Stojevic:2006pq, Howe:2010az, delaOssa:2018azc, Grimanellis:2023nav, Papadopoulos:2023exe} and depends on the characteristic tensors of the $\mathrm{G}$-structure. Consider the $\sigma$-model \eqref{eq:action2}, and suppose that the ambient manifold $\mathcal{M}$ is endowed with a given $\mathrm{G}$-structure. Let $\Phi$ be a nowhere-vanishing $p$-form, and suppose it is a characteristic tensor for the $\mathrm{G}$-structure. Then, the $\mathcal{W}$-symmetry associated with $\Phi$ is encoded in the following field transformations \cite{delaOssa:2018azc}
\begin{equation}
\label{eq: trans Xlam}
    \delta_{\epsilon}^{\Phi} X^{i}=\epsilon(Z) \widehat{\Phi}^i \, , \qquad \qquad
	\delta_{\epsilon, A}^{\Phi} \Lambda^{\alpha}=2 \epsilon(Z) \widehat{\Upsilon}^{\alpha}{}_{\beta} D_A \Lambda^{\beta} \, ,
\end{equation}
where the notation $\widehat{\Phi}^i$ was introduced in \eqref{eq:hatdefinition}. The infinitesimal parameter $\epsilon(Z)$ has (holomorphic) conformal weight $h_{\epsilon}=\frac{1-p}{2}$ and it has odd/even Grassmann parity if $p$ is an even/odd integer. The second transformation in \eqref{eq: trans Xlam} requires the introduction of an $\text{End}(V)$-valued differential $(p-2)$-form $\Upsilon(X)$, which at this stage is completely arbitrary.\footnote{The role of $\Upsilon$ has been studied in relation to worldsheet supersymmetry enhancement in \cite{Howe:1988cj, Hull:1993ct}. Some comments on the $\mathcal{W}$-symmetry transformation of the gauge bundle sections have recently appeared in \cite{Grimanellis:2023nav}.} The transformations define a symmetry if and only if the following conditions are met \cite{delaOssa:2018azc}
\begin{equation}
    \nabla^{+}\Phi=0 \, , \qquad i_{F(A)}(\Phi)=0 \, , \qquad \Upsilon_{(\alpha \beta)}=0 \, , \label{eq: clas const}
\end{equation}
where $F(A)$ is the gauge field strength and $\nabla^{+}$ denotes the connection with symbols \eqref{eq: +symb}, which we identify with the unique connection---compatible with the $\mathrm{G}$-structure---with totally antisymmetric torsion \eqref{eq:identificationHtorsion}. When the conditions \eqref{eq: clas const} are satisfied we can write down the classical chiral current \cite{delaOssa:2018azc}
\begin{equation} \label{eq: current}
        \mathcal{J}^{\Phi}_{\text{cl.}}=(-1)^{p-1} \, \widehat{\Phi} \, .
\end{equation}
This current should be understood as the classical limit of a new operator extending the super Virasoro algebra on the worldsheet. We denote the components of \eqref{eq: current} following the conventions we set in \eqref{eq:superoperatorcomponents}, that is\footnote{Our conventions for \eqref{eq: current} and \eqref{eq:HPcurrentcomponents} are chosen so that when $\Phi$ is a one-form, say $\Phi=\dd x$, the associated current $\cj$ is a super free field with components $P=\psi$ and $K=i\partial x$.}
\begin{equation}
\label{eq:HPcurrentcomponents}
{\cal J}^\Phi_{\text{cl.}}(Z) = (-i)^{p}  \big(- P^\Phi_{\text{cl.}}(z) + \theta  \, K^\Phi_{\text{cl.}}(z)\big)\, .
\end{equation}
The wedge product between differential forms is mirrored at the level of the $\mathcal{W}$-symmetry classical currents. Let $\Phi$ be a $p$-form and $\Psi$ a $q$-form complying with the conditions \eqref{eq: clas const}. The $(p+q)$-form $\Phi \wedge \Psi$ automatically satisfies \eqref{eq: clas const} as well, and its current can be described in terms of the currents of each form as follows
\begin{equation*}
    {\cal J}_{\text{cl.}}^{\Phi\wedge\Psi}(Z) = (-1)^{p+q-1}\, \widehat{\Phi\wedge\Psi}(Z)= -{\cal J}_{\text{cl.}}^{\Phi}(Z){\cal J}_{\text{cl.}}^{\Psi}(Z) \, ,
\end{equation*}
where the components are
\begin{equation*}
     P_{\text{cl.}}^{\Phi\wedge\Psi}(z) = P_{\text{cl.}}^\Phi(z)P_{\text{cl.}}^\Psi(z)  \, , \qquad   K_{\text{cl.}}^{\Phi\wedge\Psi}(z) = K_{\text{cl.}}^\Phi(z)P_{\text{cl.}}^\Psi(z) + (-1)^p \, P_{\text{cl.}}^\Phi(z)K_{\text{cl.}}^\Psi(z)  \, .
\end{equation*}
Similarly to the case of the stress-energy tensor, it was shown in \cite{delaOssa:2018azc} that the $\mathcal{W}$-symmetry current \eqref{eq: current} is still chiral when we employ the $\ell_s$-corrected flux \eqref{eq: het tors}, as long as the connection $\Theta$ satisfies a \emph{quasi-instanton condition} analogous to that of $A$
 \begin{equation*} 
     i_{R(\Theta)}(\Phi)=0 \, ,
 \end{equation*}
where $R(\Theta)$ denotes the curvature of the connection $\Theta$.

It should be noted that, in the definition of the new current \eqref{eq: current}, $\Upsilon$ does not play any role. In fact, the definition looks completely independent from the structures defined on the gauge bundle: this suggests that $\Upsilon$ does not provide additional information compared to $\Phi$, even though they both appear in the $\mathcal{W}$-symmetry transformations \eqref{eq: trans Xlam}. \
In what follows, we will ignore the gauge bundle sector, leaving it for future explorations. 

\subsection{From classical commutators to OPEs}
\label{sec:commutators}

The next step is studying the classical commutators of the symmetries introduced in \Cref{ssec: symm} and rewriting them in terms of further symmetries of the $\sigma$-model \eqref{eq:action2}. As mentioned before, we will ignore the gauge bundle sector. The algebra of classical $\mathcal{W}$-symmetries when the target manifold $\mathcal{M}$ is equipped with a $\mathrm{G}$-structure has been studied case by case in terms of Poisson brackets in \cite{Howe:1991vs, Howe:1991im, Howe:1991ic, Howe:2006si, Howe:2010az}, see also \cite{Stojevic:2006pq}. We will follow a different approach to make a connection with OPEs and $\mathcal{SW}$-algebras.

\paragraph{$[\delta^{\ct}, \delta^{\ct}]$: two superconformal transformations.}
The commutator of two superconformal transformations with parameters $\epsilon_1(Z)$ and $\epsilon_2(Z)$ returns another superconformal transformation with parameter $\epsilon_{3}(Z)$
\begin{equation} \label{eq: TTcomm}
    [ \delta^{\ct}_{\epsilon_{1}}, \delta^{\ct}_{\epsilon_{2}}] X^{i} 
= \delta^{\ct}_{\epsilon_{3}}\, X^{i}\, ,
\end{equation}
where the new infinitesimal parameter is given by
\begin{equation}
\label{eq:SC-SCepsilon}
    \epsilon_{3} = \epsilon_{1} \partial \epsilon_{2}- \partial\epsilon_{1}\,\epsilon_{2} + 
    \frac{1}{2}D\epsilon_{1}D\epsilon_{2}\, .
\end{equation}
\paragraph{$[\delta^{\ct}, \delta^{\Phi}]$: superconformal and $\mathcal{W}$-transformations.} The commutator between a superconformal transformation and a $\mathcal{W}$-transformation associated with a $p$-form $\Phi$ returns a $\mathcal{W}$-transformation associated again with the $p$-form $\Phi$
\begin{equation}
    \label{eq:SC-HPcommutator1}
[\delta_{\epsilon_{1}}^{\ct}, \delta_{\epsilon_{2}}^{\Phi}] \, X^{i} = \delta_{\epsilon_{3}}^{\Phi} X^{i} \, ,
\end{equation}
where the new infinitesimal parameter is given by 
\begin{equation*}
    \epsilon_{3} = \epsilon_{1} \partial \epsilon_{2} -  \frac{1}{2} (p-1)\partial\epsilon_{1}\,\epsilon_{2} +  \frac{1}{2} D\epsilon_{1} D\epsilon_{2}\, .
\end{equation*}
\paragraph{$[\delta^{\Phi}, \delta^{\Psi}]$: two $\mathcal{W}$-transformations.}
\label{sec:commutatorHP-HP}
Finally, we consider the commutator between two $\mathcal{W}$-symmetries. This was first studied in the case of non-zero flux $H=\dd B$ in \cite{Howe:2010az}, using the $(1,1)$ $\sigma$-model. We find that the same analysis holds completely analogously for the $(1,0)$ $\sigma$-model. Furthermore, if the corrected version of the flux \eqref{eq: het tors} is used instead, the commutators still satisfy the same formulas---up to the corresponding order in the string length scale $\ell_s$. 

Given two $\mathcal{W}$-transformations corresponding to a $p$-form $\Phi$ and a $q$-form $\Psi$, their commutator is given by the sum of three different symmetry transformations:
\begin{equation}
[\delta_{\epsilon_{1}}^{\Phi}, \delta_{\epsilon_{2}}^{\Psi}] X^{i}  =
\delta_{\epsilon_U}^U X^i + \delta_{\epsilon_N}^N X^i
+ \delta^{{\cal T}V}_{\epsilon_{\mathcal{T}V}}\, X^{i}\, ,\label{eq:HPHPX}
\end{equation}
where the transformations $\delta_{\epsilon_U}^{U}$, $\delta_{\epsilon_N}^{N}$ and $\delta_{\epsilon_{\ct V}}^{\ct V}$ are described as follows:
\begin{itemize}[label=$\diamond$,leftmargin=*]
    \item The transformation $\delta_{\epsilon_U}^U$ is a $\mathcal{W}$-transformation associated with the $(p+q-2)$-form
\begin{equation} \label{eq: U}
    U = \frac{1}{c_{U}}i_{\Phi}(\Psi)\, ,
\end{equation}
where $c_{U}$ is a tunable constant parameter that we introduce for later convenience.
The infinitesimal parameter of the transformation is given by
\begin{equation*}
    \epsilon_{U }= c_U\frac{(-1)^{pq}}{p+q-2} \left[ (q-1)D\epsilon_{1}\,\epsilon_{2} 
+ (-1)^{p}\, (p-1) \epsilon_{1}\,D\epsilon_{2}\right]\, .
\end{equation*}
 \item  The transformation $\delta^{N}_{\epsilon_N}$ is a $\mathcal{W}$-transformation associated with the  $(p+q-1)$-form 
\begin{equation} \label{eq: N}
    N =\frac{\ell_s}{c_N} \left(H_{jk}\wedge \Phi^{j}\wedge\Psi^{k} +  (-1)^{p+1}\, \frac{2 \, c_V}{d - (p+q - 4)} \,  H \wedge V \right)\, , 
\end{equation}
which depends on the three-form NS flux $H$ and the $(p+q-4)$-form $V$ defined as
\begin{equation*}
    V =  \frac{1}{c_{V}} \Phi^{ij}\wedge\Psi_{ij} \, .
\end{equation*}
In the formula \eqref{eq: N}, $d$ stands for the dimension of the manifold $\mathcal{M}$ where $\Phi$ and $\Psi$ are defined. 
Analogously to the previous transformation, the constants $c_{V}$ and $c_{N}$ are introduced for later convenience. 
The infinitesimal parameter reads
\begin{equation*}
    \epsilon_N =c_N \, (-1)^{(p-1)(q-1)}  \epsilon_1\epsilon_2 \, .
\end{equation*}
\item Finally, the transformation $\delta_{\epsilon_{\ct V}}^{\ct V}$ is different from those presented in \Cref{ssec: symm} and was first introduced in \cite{Howe:2010az,Stojevic:2006pq}. It is defined as
{\small
\begingroup
\allowdisplaybreaks 
\begin{equation} \label{eq: Tphitrans}
     \delta^{{\cal T}V}_{\epsilon_{\ct V}} X^{i}
=  -\left(\epsilon_{\ct V}  \partial X^{i}+   \frac{1}{2}(-1)^{p+q} D \epsilon_{\ct V}  DX^{i} \right)\widehat V 
+ \epsilon_{\ct V} \left( -{\cal T}_{\text{cl.}}\widehat V^{i} + {\frac{1}{2}}\, DX^{i}D\widehat V \right)  \, ,
\end{equation}
\endgroup}%
where the infinitesimal parameter reads
\begin{equation*}
    \epsilon_{\mathcal{T}V} = (-1)^{(p-1)q+1} \frac{2 \, c_V}{d - (p+q - 4)}\, \epsilon_1\epsilon_2\, .
\end{equation*}
The current associated with the transformation \eqref{eq: Tphitrans} can be rewritten as a \emph{composite current}
\begin{equation} \label{eq: TVclas}
    \cj^{\ct V}_{\text{cl.}}(Z)=-\ct_{\text{cl.}}^{\phantom{V}}(Z) \, \cj^{V}_{\text{cl.}}(Z) \, .
\end{equation}
It is important to note that if $V=1$, the transformation \eqref{eq: Tphitrans} simply reduces to a superconformal transformation and---noticing that $\cj^{1}_{\text{cl.}}=-\mathds{1}$ under our conventions \eqref{eq: current}---the associated current is $\mathcal{J}^{\mathcal{T}1}_{\text{cl.}}=\mathcal{T}_{\text{cl.}}^{\phantom{\ct}}$, as expected.
\end{itemize}
Finally, we explain how to translate the commutators \eqref{eq: TTcomm}, \eqref{eq:SC-HPcommutator1}, and $\eqref{eq:HPHPX}$ into the language of OPEs. This will be used in \Cref{sec: connection} to generate a set of classical OPEs for each of the different $\mathrm{G}$-structures, which will then be compared with those of the underlying $\mathcal{SW}$-algebras. 

Consider two superconformal or $\mathcal{W}$-symmetry transformations $\delta_1$ and $\delta_2$ with chiral currents $\cj_1(Z)$ and $\cj_2(Z)$ and associated charges $\op{\mathcal{Q}}_1$ and $\op{\mathcal{Q}}_2$, defined as in \eqref{eq:superchargeconvention}.  The action of a charge on a local operator is given by the associated infinitesimal symmetry transformation \cite{DiFrancesco:1997nk}
\begin{equation*}
    \left[\op{\mathcal{Q}}_i , \mathcal{O}\right]=-\delta_i \mathcal{O} \, .
\end{equation*}
The formulas below should be understood to hold inside correlators, so $\left[\op{\mathcal{Q}} , \mathcal{O}\right]=\op{\mathcal{Q}}\,\mathcal{O}$ in what follows and we can write 
\begin{equation*}
     [\op{\mathcal{Q}}_1, \op{\mathcal{Q}}_2]\, \mathcal{O}(\zeta)= [\delta_1, \delta_2]\, \mathcal{O}(\zeta) \, .
\end{equation*}
The key idea is to take the operator $\mathcal{O}(\zeta)$ to be $X^{i}(\zeta)$:  the discussion in \Cref{sec:commutators} shows that classically the commutator can be traded for a linear combination of transformations
\begin{equation}
\label{eq:commutatorQassumofQs}
    [\op{\mathcal{Q}}_1, \op{\mathcal{Q}}_2]\, X^{i}(\zeta)= [\delta_1, \delta_2]\, X^{i}(\zeta) \overset{\text{cl.}}{=}\sum_{r} \delta_{r} X^{i}(\zeta)=-\sum_{r} \op{\mathcal{Q}}_r \, X^{i}(\zeta) \, ,
\end{equation}
where we highlighted that this statement works at the level of the classical action by employing the notation ``$\overset{\text{cl.}}{=}$''. 
The commutator of the two charges can alternatively be expressed in terms of the currents as follows \cite{DiFrancesco:1997nk} 
\begin{equation}
\label{eq:commutatorQsusingJs}
    [\op{\mathcal{Q}}_1, \op{\mathcal{Q}}_2]\, \mathcal{O}(\zeta)=(-1)^{|\epsilon_1||\cj_2|}\oint_{\zeta} \frac{\de Z_2}{2 \pi i}\oint_{\zeta_2} \frac{\de Z_1}{2 \pi i} \, \epsilon_1(Z_1) \epsilon_2(Z_2) \, \cj_1(Z_1)\cj_2(Z_2) \, \mathcal{O}(\zeta) \, ,
\end{equation}
where the prefactor is sensitive to the parity of the infinitesimal parameters and of the currents. 
Let us denote by $\epsilon_r$ the infinitesimal parameter associated with the action of the charge $\op{\mathcal{Q}}_r$. Again from the results of \Cref{sec:commutators}, note that $\epsilon_r=\epsilon_r\big(\epsilon_1(Z), \epsilon_2(Z)\big)$ is actually a functional of the infinitesimal parameters $\epsilon_1(Z)$ and $\epsilon_2(Z)$. We can then write
\begin{equation}
\label{eq:sumofQsusingJs}
   \sum_{r} \op{\mathcal{Q}}_r \, X^{i}(\zeta)=\oint_{\zeta} \frac{\de Z}{2 \pi i} \sum_{r} \epsilon_r\big(\epsilon_1(Z), \epsilon_2(Z)\big) \, \cj_r(Z) \, X^{i}(\zeta) \, , 
\end{equation}
and by substituting the right-hand sides of the equations \eqref{eq:commutatorQsusingJs} and \eqref{eq:sumofQsusingJs} into the expression \eqref{eq:commutatorQassumofQs} we obtain the following master equation
\begin{multline} \label{eq: classope}
    \oint_{\zeta} \frac{\de Z_2}{2 \pi i}\oint_{\zeta_2} \frac{\de Z_1}{2 \pi i} \, \epsilon_1(Z_1) \epsilon_2(Z_2) \, \cj_1(Z_1)\cj_2(Z_2) \,X^i(\zeta) \\ \overset{\text{cl.}}{=}-(-1)^{|\epsilon_1||\cj_2|}\oint_{\zeta} \frac{\de Z_2}{2 \pi i} \sum_{r} \epsilon_r\big(\epsilon_1(Z_2), \epsilon_2(Z_2)\big) \, \cj_r(Z_2) \, X^{i}(\zeta) \, .
\end{multline}
The connection with the OPE formalism can be made explicit working with the right-hand side of the equation \eqref{eq: classope}. To this end, we need to manipulate the functional $\epsilon_r\big(\epsilon_1(Z), \epsilon_2(Z)\big)$ using integration by parts to reproduce the $\epsilon_1(Z_1) \epsilon_2(Z_2)$ structure on the left-hand side of the equation \eqref{eq: classope}. One can then read off a ``classical OPE'' for the currents $\cj_1$ and $\cj_2$ in the right-hand side of the equation.
We now present the three classical OPEs associated with the commutators studied in \Cref{sec:commutators}.

\paragraph{$\mathcal{T}_{\text{cl.}} \times \mathcal{T}_{\text{cl.}}$ OPE.} We illustrate how to use the master equation \eqref{eq: classope} working out in detail the case where $\cj_1=\ct_{\text{cl.}}$ and $\cj_2=\ct_{\text{cl.}}$. Equation \eqref{eq: classope} reads in this case
{\small
\begingroup
\allowdisplaybreaks 
\begin{multline} \label{eq: TT}
    \oint_{\zeta} \frac{\de Z_2}{2 \pi i}\oint_{\zeta_2} \frac{\de Z_1}{2 \pi i} \, \epsilon_1(Z_1) \epsilon_2(Z_2) \, \ct_{\text{cl.}}(Z_1)\ct_{\text{cl.}}(Z_2) \,X^i(\zeta) \\[3pt]
    \overset{\text{cl.}}{=}-\oint_{\zeta} \frac{\de Z_2}{2 \pi i}  \left(\epsilon_{1}(Z_2) \partial \epsilon_{2}(Z_2)- \partial\epsilon_{1}(Z_2)\epsilon_{2}(Z_2) + \frac{1}{2}D\epsilon_{1}(Z_2)D\epsilon_{2}(Z_2)\right) \, \ct_{\text{cl.}}(Z_2) \, X^{i}(\zeta) \, ,
\end{multline}
\endgroup}%
where we plugged the expression \eqref{eq:SC-SCepsilon} in place of the functional $\epsilon_{\ct}\big(\epsilon_1(Z_2), \epsilon_2(Z_2)\big)$. The double-integral structure appearing in the left-hand side of equation \eqref{eq: TT} can be reproduced in the right-hand side by manipulating the infinitesimal parameter $\epsilon_1(Z_2)$ and its derivatives. We introduce the Dirac $\delta$-distribution on the super worldsheet \cite{Blumenhagen:1990jv, Heluani:2006pk}
\begin{equation*}
    \delta(Z_1-Z_2)=\frac{1}{2 \pi i}\frac{\theta_{12}}{Z_{12}} \, ,
\end{equation*}
and we compute the following derivatives
\begin{equation} \label{eq: derivative}
    \partial_{2} \, \delta(Z_1-Z_2)=\frac{1}{2 \pi i}\frac{\theta_{12}}{Z_{12}^2} \, , \quad D_2 \, \delta(Z_1-Z_2)=-\frac{1}{2\pi i}\frac{1}{Z_{12}}\, .
\end{equation}
The infinitesimal parameter $\epsilon_{1}(Z_2)$ can be rewritten as 
\begin{equation} \label{eq: epsZ2}
     \epsilon_1(Z_2) = \oint_{\zeta_2}\dd Z_1 \, \delta(Z_1-Z_2) \, \epsilon_1(Z_1)=\oint_{\zeta_2}\frac{\dd Z_1}{2 \pi i} \, \frac{\theta_{12}}{Z_{12}} \, \epsilon_1(Z_1) \, , 
\end{equation}
and its derivatives as follows, using the equations \eqref{eq: derivative}
\begin{equation} \label{eq: epsZ2 der}
     \partial_2  \epsilon_1(Z_2) = \oint_{\zeta_2}\frac{\dd Z_1}{2\pi i} \, \frac{\theta_{12}}{Z_{12}^2} \,  \epsilon_1(Z_1)  \, , \quad   D_2  \epsilon_1(Z_2)  = \oint_{\zeta_2}\frac{\dd Z_1}{2\pi i} \,  \frac{1}{Z_{12}} \,  \epsilon_1(Z_1) \, .
\end{equation}
We have all the ingredients needed to reshape the right-hand side of the equation \eqref{eq: TT}. The first step is integrating by parts to remove all the derivatives from the parameter $\epsilon_{2}(\zeta_2)$. Focusing on the right-hand side, we have
\begin{multline*}
    \cdots \overset{\text{cl.}}{=}\oint_{\zeta} \frac{\de Z_2}{2 \pi i}  \bigg( \frac{3}{2}\partial \epsilon_{1}(Z_2) \epsilon_{2}(Z_2) \mathcal{T}_{\text{cl.}}(Z_2) +\epsilon_{1}(Z_2) \epsilon_{2}(Z_2) \partial\mathcal{T}_{\text{cl.}}(Z_2) \\+\frac{1}{2}D\epsilon_{1}(Z_2)\epsilon_{2}(Z_2)  D\mathcal{T}_{\text{cl.}}(Z_2) \bigg)  X^{i}(\zeta) \, .
\end{multline*}
We now substitute $\epsilon_1(Z_2)$ with the expression \eqref{eq: epsZ2} and its derivatives with the expressions
{\small
\begingroup
\allowdisplaybreaks 
\eqref{eq: epsZ2 der}
\begin{multline} \label{eq: rightside}
 {\textstyle    \cdots \overset{\text{cl.}}{=}}\oint_{\zeta} \frac{\de Z_2}{2 \pi i} \oint_{\zeta_2}\frac{\dd Z_1}{2\pi i} \epsilon_{1}(Z_1) \epsilon_{2}(Z_2) \bigg( \frac{3}{2}\frac{\theta_{12}}{Z_{12}^2} \mathcal{T}_{\text{cl.}}(Z_2) +\frac{\theta_{12}}{Z_{12}} \partial\mathcal{T}_{\text{cl.}}(Z_2) +\frac{D\mathcal{T}_{\text{cl.}}(Z_2)}{ 2 Z_{12}}   \bigg)  X^{i}(\zeta) \, .
\end{multline}
\endgroup}%
By comparing the expression \eqref{eq: rightside} with the left-hand side of the expression \eqref{eq: TT}, we extract the classical OPE between the Noether current $\ct_{\text{cl.}}$ and itself
\begin{equation} \label{eq: TTOPE}
    \ct_{\text{cl.}} (Z_1) \ct_{\text{cl.}}(Z_2) \sim \frac{3}{2}\frac{\theta_{12}}{Z_{12}^2} \mathcal{T}_{\text{cl.}}(Z_2) +\frac{1}{ 2 Z_{12}} D\mathcal{T}_{\text{cl.}}(Z_2) +\frac{\theta_{12}}{Z_{12}} \partial\mathcal{T}_{\text{cl.}}(Z_2)+\cdots \, .
\end{equation}
We immediately notice that this classical OPE reproduces correctly every term in the Virasoro algebra \eqref{eq: superTT}, excluding the $\frac{1}{Z_{12}^3}$ pole proportional to the central charge $c$. This was to be expected since the information encoded in \eqref{eq: TTOPE} is the same as the one encoded in the classical commutator \eqref{eq: TTcomm}, hence this procedure only recovers the classical Witt algebra and no information about its central extension is available.

\paragraph{$\mathcal{T}_{\text{cl.}}^{\phantom{\Phi}} \times \mathcal{J}_{\text{cl.}}^{\Phi}$ OPE.}
\label{sec:SuperconformalHPOPEs} Following the same procedure used to derive the classical OPE $\ct_{\text{cl.}}\times \ct_{\text{cl.}}$, we convert the classical commutator \eqref{eq:SC-HPcommutator1} into a classical OPE. If we consider $\Phi$ to be a $p$-form, we find 
\begin{equation}
    \mathcal{T}_{\text{cl.}}^{\phantom{\Phi}}(Z_1)\mathcal{J}_{\text{cl.}}^\Phi(Z_2)\sim \frac{p}{2} \frac{\theta_{12}}{Z_{12}^2}\mathcal{J}_{\text{cl.}}^\Phi(Z_2) +\frac{1}{2 \, Z_{12}}D \cj_{\text{cl.}}^{\Phi}(Z_2) +\frac{\theta_{12}}{Z_{12}}\partial \cj_{\text{cl.}}^{\Phi}(Z_2)+\cdots \, .
    \label{eq:SC-HPSUSYOPE}
\end{equation}
This classical OPE should be compared with the quantum OPE \eqref{eq: superTJ}, where we made the assumption that the current $\mathcal{J}_{h}$ was a primary. The associated classical OPE \eqref{eq:SC-HPSUSYOPE} is compatible with such choice. Furthermore, the conformal weight of the operator $h_{\cj_{\Phi}}=\frac{p}{2}$ is determined by the degree of the form $\Phi$.

\paragraph{$\mathcal{J}_{\text{cl.}}^{\Phi} \times \mathcal{J}_{\text{cl.}}^{\Psi}$ OPE.} 
Finally, we consider the commutator \eqref{eq:HPHPX} of two $\mathcal{W}$-symmetries associated with a $p$-form $\Phi$ and a $q$-form $\Psi$. The same procedure as above then returns the classical OPE
\begin{multline}
    \mathcal{J}_{\text{cl.}}^\Phi(Z_1)\mathcal{J}_{\text{cl.}}^\Psi(Z_2) \sim (-1)^{p+1} \, c_{U} \, \frac{\mathcal{J}_{\text{cl.}}^U(Z_2)}{Z_{12}}+(-1)^{p+1}\, c_{U} \left(\frac{p-1}{p+q-2}\right)\frac{\theta_{12}}{Z_{12}}D\mathcal{J}_{\text{cl.}}^U(Z_2)\\[6pt]+(-1)^{p}\, c_{N}\, \frac{\theta_{12}}{Z_{12}}\mathcal{J}_{\text{cl.}}^N(Z_2)
    +c_{V}\left({\frac{2 }{d - (p+q - 4)}}\right) \frac{\theta_{12}}{Z_{12}} \, \ct^{\phantom{V}}_{\text{cl.}}(Z_2) \cj^{V}_{\text{cl.}}(Z_2)+\cdots  \, .
    \label{eq:HP-HPSUSYOPE}
\end{multline}
Notice that the last term in the right-hand side of the classical OPE \eqref{eq:HP-HPSUSYOPE} is simply a product of $\ct^{\phantom{V}}_{\text{cl.}}$ and $\cj^{V}_{\text{cl.}}$ evaluated at the same point, as prescribed by equation \eqref{eq: TVclas}. As it will be more evident after studying a few specific examples in Section \ref{sec: connection}, combined classical currents can be identified as the classical limits of the quasi-primary normal ordered product of the associated chiral currents
\begin{equation*}
    \mathcal{N}(\ct \cj^\Phi) \overset{\text{cl.}}{\longrightarrow} \ct_{\text{cl.}}^{\phantom{\Phi}} \cj^\Phi_{\text{cl.}} \, , \quad \mathcal{N}(\cj^\Phi \cj^\Psi) \overset{\text{cl.}}{\longrightarrow} \cj^\Phi_{\text{cl.}} \cj^\Psi_{\text{cl.}} \, .
\end{equation*}
This classical OPE is very interesting: although it is not able to capture all the terms appearing in the quantum OPE, geometrical information about the manifold $\mathcal{M}$ and the NS flux $H$ is encoded into the OPE via the currents associated with the forms $U$, $V$, $N$ as well as the classical stress-energy tensor $\ct_{\text{cl.}}$. In particular, when $\Phi$ and $\Psi$ are the characteristic forms of a $\mathrm{G}$-structure on $\mathcal{M}$, components of the intrinsic torsion of the $\mathrm{G}$-structure manifestly appear in the OPE.
The classical OPE $\mathcal{J}_{\text{cl.}}^{\Phi} \times \mathcal{J}_{\text{cl.}}^{\Psi}$ is meant to be compared with the outcome of the Ansatz \eqref{eq: ANS}---this comparison will be the focus of \Cref{sec: connection}.

\section{\texorpdfstring{$\mathrm{G}$}{G}-structures with torsion and \texorpdfstring{$\mathcal{SW}$}{SW}-algebras} \label{sec: connection}
Throughout this Section we consider a superstring background compactification on a $d$-dimensional Riemannian manifold $\mathcal{M}$ described at the classical level by the $\sigma$-model \eqref{eq:action2} with target space $\mathcal{M}$.
Suppose that $\mathcal{M}$ is endowed with a $\mathrm{G}$-structure with characteristic forms  $\Phi_1, \Phi_2, \dots, \Phi_n$ of degrees $p_1, \dots, p_n$, respectively. We can consider two different approaches:
\begin{itemize}[label=$\diamond$,leftmargin=*]
    \item The symmetries of the worldsheet CFT are described by an $\mathcal{SW}$-algebra, denoted by $\mathcal{SW}(\frac{3}{2}, h_1, \dots, h_n)$ and generated by the identity, the stress-energy tensor and a set of primary chiral currents
    \begin{equation} \label{eq: quantcurr}
        \Braket{\mathds{1}, \ct, \cj_{h_1}, \dots, \cj_{h_n}} \, .
    \end{equation}
    The $\mathcal{SW}$-algebra is completely determined by the Ansatz \eqref{eq: ANS}
    \begin{equation}  \label{eq: quaOPE}
    \mathcal{J}_{h_i}(Z_1) \mathcal{J}_{h_j}(Z_2)\hspace{4pt} \sim \sum_{k,r=0}^{\infty} \,  \sum_{\mathcal{O}_{k}\in \, \mathcal{F}^{\, \text{q.p.}}_{k}} \, C_{ij}^{k} \, A_{ijk}^{r} \, \frac{1}{Z_{12}^{h_{ijk}-r/2}} \, D^{r} \mathcal{O}_{k}(Z_2) \, ,
    \end{equation}
    and the consistency conditions of the OPEs (cf. Appendix \ref{app:nullfields}).
    \item From the results of \Cref{sec:Wsymm}, the $\sigma$-model enjoys superconformal and $\mathcal{W}$-symmetries associated with a set of classical Noether currents 
    \begin{equation} \label{eq: clascurr}
        \ct_\text{cl.}^{\phantom{\Phi}} , \ \cj^{\Phi_1}_\text{cl.} , \ \dots \  , \ \cj^{\Phi_n}_\text{cl.} \, .
    \end{equation}
    These currents carry information about the $\mathrm{G}$-structure and its intrinsic torsion,
    and they satisfy a set of classical OPEs \eqref{eq:HP-HPSUSYOPE}
    {\small
\begingroup
\allowdisplaybreaks 
    \begin{multline} \label{eq: clasOPE}
    \mathcal{J}_{\text{cl.}}^{\Phi_{1}}(Z_1)\mathcal{J}_{\text{cl.}}^{\Phi_{2}}(Z_2) \sim (-1)^{p_1+1} \, c_{U} \, \frac{\mathcal{J}_{\text{cl.}}^U(Z_2)}{Z_{12}}+(-1)^{p_1+1}\, c_{U} \left(\frac{p_1-1}{p_1+p_2-2}\right)\frac{\theta_{12}}{Z_{12}}D\mathcal{J}_{\text{cl.}}^U(Z_2)\\[6pt]
    +(-1)^{p_1}\, c_{N}\, \frac{\theta_{12}}{Z_{12}}\mathcal{J}_{\text{cl.}}^N(Z_2)
    +c_{V}\left({\frac{2 }{d - (p_1+p_2 - 4)}}\right) \frac{\theta_{12}}{Z_{12}} \, \ct^{\phantom{V}}_{\text{cl.}}(Z_2) \cj^{V}_{\text{cl.}}(Z_2)+\cdots  \, .
\end{multline}
\endgroup}
\end{itemize}
The goal of this Section is to combine these two perspectives. To this end, we will assume that the currents \eqref{eq: clascurr} are the classical limits of the generators \eqref{eq: quantcurr} \cite{Figueroa-OFarrill:1996tnk, delaOssa:2018azc}: under this assumption, the generators of the $\mathcal{SW}$-algebra read 
\begin{equation} \label{eq: gen presc}
      \Braket{\mathds{1}, \ct, \cj_{\frac{p_1}{2}}^{\Phi_1}, \dots, \cj_{\frac{p_n}{2}}^{\Phi_n}} \, .
\end{equation}
In the following, we will omit the weight of the generator $\cj_{\frac{p_i}{2}}^{\Phi_i} \to \cj^{\Phi_i}$ to make the notation less cluttered. Since symmetry arguments (see Table \ref{tab:symm}) and the associativity condition \eqref{eq: associ} will fix the OPE coefficients $C_{ij}^{k}$ up to a small number of free parameters, the idea is to compare the two sets of OPEs \eqref{eq: quaOPE} and \eqref{eq: clasOPE} in order to write those free parameters at the lowest order in the string length $\ell_s$ as functions of the torsion classes of the underlying $\mathrm{G}$-structure. In particular, the classical OPEs allow to match coefficients up to order $O(\ell_s^2)$: then, it is natural to interpret them as first order limits of the OPEs of the full $\mathcal{SW}$-algebras in a small $\ell_s$ expansion.

It should be noted that when all torsion classes vanish our manifold $\mathcal{M}$ has special holonomy and the corresponding $\mathcal{SW}$-algebras are well-understood (cf. Appendix \ref{app: spechol}). 

In what follows, when studying an OPE with the structure \eqref{eq: quaOPE} or a classical OPE \eqref{eq: clasOPE}, we will always drop the $Z_2$ dependence of the chiral currents in the right hand side of the expressions. This choice is meant to lighten the notation and the dependence should always be considered to be implicitly stated. 

\subsection{\texorpdfstring{$\mathcal{SW}\left(\frac{3}{2},\frac{1}{2}, \dots, \frac{1}{2}\right)$}{SW(3/2, 1/2, ..., 1/2)} algebra and \texorpdfstring{$\mathrm{O}(d-n)$-structures}{O(d-n)-structures}} \label{ssec: kac moody}

We illustrate the procedure described at the beginning of this Section by studying $\mathrm{O}(d-n)$-structures as a warm-up. The characteristic tensors are a set of globally defined, nowhere-vanishing vector fields on the background manifold. 

\paragraph{$\mathrm{O}(d-n)$-structures.} 
Suppose our $d$-dimensional Riemannian manifold $\mathcal{M}$ is endowed with a set of linearly independent, globally defined and nowhere-vanishing vector fields $\xi_1, \dots, \xi_{n}$. These vectors are naturally associated with a set of one-forms $\sigma_1, \dots, \sigma_n$, defined via the isomorphism $\sigma_I= G(\xi_I,\cdot)$ for $I=1, \dots, n$, where $G$ is the metric on $\mathcal{M}$. These one-forms are the characteristic tensors of an $\mathrm{O}(d-n)$-structure \cite{Tomasiello:2022dwe}. We are interested in the particular case where these forms are associated with classical chiral currents \cite{Rocek:1991ps}, which we denote by $\cj^{\sigma}_{\text{cl.}, I}= \cj^{\sigma_I}_{\text{cl.}}$ for $I=1, \dots, n$. Therefore, we must impose the condition \eqref{eq: clas const}, which reads
\begin{equation} \label{eq: nablasigma}
    \nabla_{i}^{+}\sigma^{\phantom{+}}_{I,j}=0 \, .
\end{equation}
Recalling that the torsion tensor $\mathrm{Tor}$ involved in the equation \eqref{eq: nablasigma} is totally antisymmetric, it can be deduced that the vector fields $\xi_{I}$ must satisfy the Killing equation and be the infinitesimal generators of a set of isometries \cite{Rocek:1991ps}. Thus, the $\mathrm{O}(d-n)$-structures we are studying are generated by $n$ linearly independent Killing vector fields.

\paragraph{The $\mathcal{SW}(\frac{3}{2}, \frac{1}{2}, \dots, \frac{1}{2})$ algebra.} 
The construction of the associated $\mathcal{SW}$-algebra is relatively simple, due to the low weights of the currents involved in the Virasoro extension. Following the prescription \eqref{eq: gen presc}, the generators of the algebra are
\begin{equation*}
   \Braket{\mathds{1}, \mathcal{T}, \mathcal{J}_1^{\sigma}, \dots,  \mathcal{J}_n^{\sigma}} \, ,
\end{equation*}
where each current $\mathcal{J}_I^{\sigma}$ has a weight $h_{I}=\frac{1}{2}$ and is associated with a one-form $\sigma_I$. Virasoro extends to the algebra $\mathcal{SW}(\frac{3}{2}, \frac{1}{2}, \dots , \frac{1}{2})$.
By dimensional analysis, the Ansatz \eqref{eq: quaOPE} specialises to the $\mathcal{SW}(\frac{3}{2}, \frac{1}{2}, \dots , \frac{1}{2})$ case as follows
\begin{equation} \label{eq: 1formsalg}
    \cj_I^{\sigma}(Z_1)\cj_J^{\sigma}(Z_2) \sim \frac{C_{IJ}}{Z_{12}}+C^{K}_{IJ} \frac{\theta_{12}}{Z_{12}}\cj_K^{\sigma}+\cdots \, .
\end{equation}

\paragraph{Geometrical interpretation of the couplings.}
In order to produce an Ansatz for the classical OPE \eqref{eq: clasOPE}, we first need to compute the forms $U, N$ and $V$ introduced in \Cref{sec:commutators}, being careful with the conventions:
\begin{itemize}[label=$\diamond$]
    \item $U$ is a function given by the scalar product between one-forms
    \begin{equation*}
         U_{IJ}=\frac{1}{c_{U}} 
         \sigma_{I} \cdot \sigma_{J} \, .
    \end{equation*}
    When the scalar product is non-zero, we choose $c_U=\sigma_{I} \cdot \sigma_{J}$. In this way, $U=1$ when $\sigma_{I} \cdot \sigma_{J}\neq 0$ and $U=0$ otherwise.
    \item $V$ cannot be defined, since it can be constructed only for forms of degree 2 or higher. Whenever $V$ appears, we will set the term equal to zero.
    \item $N$ is a one-form given by 
    \begin{equation*}
        N_{IJ}=\frac{\ell_s}{c_{N}} 
        \left( H_{ijk} \sigma^{i}_{I} \sigma^{j}_{J} \right) \de x^k = \frac{\ell_s}{c_N} 
         i_{\sigma_{J}} i_{\sigma_{I}}\left(H \right)  \, .
    \end{equation*}
    We will tune $c_N=\ell_s$ in the following. 
\end{itemize}
Plugging the forms above in the classical OPE \eqref{eq: clasOPE}, we obtain\footnote{By recalling our convention \eqref{eq: current}, note that $\cj^1=-\mathds{1}$.} 
\begin{equation} \label{eq: clnabe}
    \mathcal{J}_{\text{cl.},I}^{\sigma}(Z_1)\mathcal{J}_{\text{cl.},J}^{\sigma}(Z_2) \sim -\frac{\sigma_{I}\cdot \sigma_{J}}{Z_{12}}-\ell_s \frac{\theta_{12}}{Z_{12}}\mathcal{J}_{\text{cl.}}^{i_{\sigma_{J}} i_{\sigma_{I}}\left(H \right)} + \cdots \, .
\end{equation}
We can now compare the OPEs \eqref{eq: 1formsalg} and \eqref{eq: clnabe} to give a geometrical interpretation to the couplings $C_{IJ}$ and $C_{IJ}^{K}$. Comparing the poles $\frac{1}{Z_{12}}$, we find that
\begin{equation} \label{eq: kappa}
    C_{IJ} = -\sigma_I \cdot \sigma_J + O(\ell_s^2) \, ,
\end{equation}
where we took into account possible perturbative corrections in $\ell_s$ to the classical OPE \eqref{eq: clnabe}, obtained by performing loop computations.\footnote{See \cite{Bedoya:2010av, Marchioro:2013wza} for efforts in this direction in a similar context.} We can now compare the $\frac{\theta_{12}}{Z_{12}}$ poles and we find
\begin{equation} \label{eq: gamma}
  C_{IJ}^{K} \, \sigma_{K}^{k}=-\ell_s \tensor{H}{_{ij}^k}\sigma^{i}_{I}\sigma^{j}_{J}+O(\ell_s^2) \, .
\end{equation} 
Multiplying both sides of the equation \eqref{eq: gamma} by $G_{k \ell} \, \sigma_{L}^{\ell}$ and applying the result \eqref{eq: kappa}, we obtain
\begin{equation}
\label{eq:couplingcartan3form}
    f_{IJK}= -C_{IJ}^{L}C^{\phantom{K}}_{LK}=-\ell_s \tensor{H}{_{ij\ell}}\sigma^{i}_{I}\sigma^{j}_{J}\sigma^{\ell}_{K}+O(\ell_s^2) \, ,
\end{equation}
where we introduced the \emph{Cartan three-form} $f_{IJK}$. In summary, we found an interpretation of the couplings $C_{IJ}$ and $C_{IJ}^{K}$ of the $\mathcal{SW}(\frac{3}{2}, \frac{1}{2}, \dots , \frac{1}{2})$ algebra in terms of the characteristic one-forms $\sigma_I$ of the $\mathrm{O}(d-n)$-structure and the flux $H$---or equivalently the torsion $\text{Tor}$.

To better understand these OPEs, note that the forms being covariantly constant under a metric connection implies that their scalar product is constant:
\begin{equation*}
    \partial_i(\sigma_I\cdot\sigma_J)=\nabla^+_i(\sigma_I\cdot\sigma_J)= \nabla^+_i G(\sigma_I,\sigma_J)= G(\nabla_i^+\sigma_I,\sigma_J)+G(\sigma_I,\nabla_i^+\sigma_J)=0 \, .
\end{equation*}
Since the metric $G$ is Riemannian, this means that the forms $\sigma_1, \dots, \sigma_n$ can always be normalised so that their scalar product is a multiple of the identity, $\sigma_I\cdot\sigma_J=k\,\delta_{IJ}$ for $k>0$. We will choose this normalisation in what follows.

It is illustrative to consider first the case of vanishing flux $H=0$. Taking $k=1$ the OPEs read
\begin{equation*}
    \mathcal{J}^{\sigma}_I(Z_1)\mathcal{J}^{\sigma}_J(Z_2) \sim -\frac{\delta_{IJ}}{Z_{12}} + \cdots \, ,
\end{equation*}
which in our conventions are the OPEs of $n$ independent free superfields, generating $n$ copies of the $\mathrm{Free}$ algebra. This is indeed consistent: the manifold $\mathcal{M}$ has $\mathrm{O}(d-n)$ holonomy and the structure group is trivial for the dual vector fields $\xi_1, \dots, \xi_{n}$. This essentially means that the currents $\widehat\sigma_1, \dots, \widehat\sigma_n$ decouple in the $\sigma$-model and can therefore be represented by free superfields.

When $H\neq 0$, the resulting algebra is more complicated. However, there is a particularly important case: suppose that we normalise the one-forms as above for some integer $k$ and we find that the coupling $f_{IJL}$ computed in \eqref{eq:couplingcartan3form} actually corresponds to the Cartan three-form of a semisimple Lie algebra $\mathfrak{g}$. Then, the OPEs \eqref{eq: 1formsalg} with the couplings \eqref{eq: kappa} and \eqref{eq:couplingcartan3form} reproduce precisely the OPEs of an \emph{affine (super) Kac--Moody algebra} for $\mathfrak{g}$ at level $k$. 

\subsection{\texorpdfstring{$\mathcal{SW}\left(\frac{3}{2},2\right)$}{SW(3/2,2)} algebra and \texorpdfstring{$\mathrm{Spin}(7)$}{Spin(7)}-structures}
\label{sec:spin7comparison}

We now work out in full detail the case of $\mathrm{Spin}(7)$-structures: these are defined by a single characteristic four-form $\Psi$ and only possess two torsion classes, so their associated $\mathcal{SW}$-algebra is particularly simple. This algebra first appeared in relation to $\mathrm{Spin}(7)$-holonomy in \cite{Shatashvili:1994zw}.

\paragraph{$\mathrm{Spin}(7)$-structures.}
A $\mathrm{Spin}(7)$-structure on an eight-dimensional Riemannian manifold $\mathcal{M}$ is determined by a well defined, nowhere-vanishing four-form $\Psi$ such that at any given point on $\mathcal{M}$ it can be written as \cite{Joyce:2007, Karigiannis_2008}
\begin{align*}
\Psi&=\dd x^{1234}+\dd x^{1256}+\dd x^{1278}+\dd x^{3456}+\dd x^{3478}+\dd x^{5678}
+\dd x^{1357} \nn \\&-\dd x^{1368}-\dd x^{1458}-\dd x^{1467} 
-\dd x^{2358}-\dd x^{2367}-\dd x^{2457}+\dd x^{2468} \, ,
\end{align*}
for some particular choice of local coordinates $\lbrace x_1,\dots,x_8\rbrace$ around that point. We call $\Psi$ the \emph{Cayley} form. It determines a metric and an orientation on $\mathcal{M}$, and note that $\Psi$ is \emph{self-dual} with respect to the associated Hodge star operator: $*\Psi=\Psi$.

There are two torsion classes associated with a $\mathrm{Spin}(7)$-structure \cite{Fernandez:1986}, which we denote by $\tau_1 \in \Omega^{1}_{\bf{8}}(TM)$ and $\tau_3 \in \Omega^{3}_{\bf{48}}(TM)$. They are determined by the exterior derivative of the characteristic form  $\Psi$:
\begin{equation*}
    \dd \Psi =\tau_1 \wedge \Psi + *\tau_3 \, .
\end{equation*}
Remarkably, on a manifold with a $\mathrm{Spin}(7)$-structure there {\it always} exists a unique compatible connection admitting a totally antisymmetric torsion \cite{Ivanov:2001ma},  given by 
\begin{equation*}
    \text{Tor} = -\frac{1}{6}\, \tau_1\lrcorner\,\Psi - \tau_3 \, .
\end{equation*}

\paragraph{The $\mathcal{SW}\left(\frac{3}{2},2\right)$ algebra.} Following the prescription \eqref{eq: gen presc}, we study the algebra generated by
\begin{equation*}
    \Braket{\mathds{1}, \ct, \cj^{\Psi}} \, , 
\end{equation*}
where the primary current $\cj^{\Psi}$ has weight $h_{\Psi}=2$. Virasoro thus extends to the algebra $\mathcal{SW}\left(\frac{3}{2},2\right)$, derived for the first time in \cite{Figueroa-OFarrill:1990mzn} and further studied in \cite{Gepner:2001px, Benjamin:2014kna, Robbins:2024htz}. To formulate an Ansatz for the OPE $\cj^{\Psi} \times \cj^{\Psi}$, we need to introduce the following quasi-primary projections of the normal ordered operators $N(D \ct \ct)$ and $N(\ct\cj^{\Psi})$
\begin{align*}
    \mathcal{N}(D \mathcal{T}\mathcal{T}) &= N(D \mathcal{T}\mathcal{T})-\frac{3}{8}  \del \del \ct \, , \\[3pt]
    \mathcal{N}(\mathcal{T}\mathcal{J}^{\psi}) &= N(\mathcal{T}\mathcal{J}^{\psi})-\frac{1}{5}\partial D \mathcal{J}^{\psi} \, .
\end{align*}
From the general statement \eqref{eq: quaOPE}, the most general Ansatz for the OPE $\cj^{\Psi} \times \cj^{\Psi}$  is
\begingroup
\allowdisplaybreaks
\begin{align} 
\label{eq: spin7ope}
    \mathcal{J}^{\Psi}(Z_1)\mathcal{J}^{\Psi}(Z_2) &\sim \frac{C_{\Psi \Psi}^{\mathds{1}}}{Z_{12}^4} + C_{\Psi \Psi}^{\Psi}\left( \frac{1}{Z_{12}^2}\mathcal{J}^{\Psi}+ \frac{\theta_{12}}{2 \,Z_{12}^2}   D \mathcal{J}^{\Psi}+ \frac{1}{2 \, Z_{12}}   \del \mathcal{J}^{\Psi}+\frac{3}{10}\frac{\theta_{12}}{Z_{12}}  D \del  \mathcal{J}^{\Psi}\right) \nn \\[8pt]
     &+C_{\Psi \Psi}^{\ct} \left( \frac{\theta_{12}}{Z_{12}^3}  \mathcal{T}+ \frac{1}{Z_{12}^2}D \mathcal{T}+ \frac{2}{3}\frac{\theta_{12}}{Z_{12}^2} \del\mathcal{T}+\frac{1}{4 \, Z_{12}}   D \del \mathcal{T}  +\frac{\theta_{12}}{4 \, Z_{12}} \del \del \mathcal{T}\right) \nn
     \\[8pt]
     &+C_{\Psi \Psi}^{D \mathcal{T}\mathcal{T}}\frac{\theta_{12}}{Z_{12}} \mathcal{N}(D \mathcal{T}\mathcal{T})+C_{\Psi \Psi}^{\mathcal{T}\Psi} \frac{\theta_{12}}{Z_{12}}   \mathcal{N}(\mathcal{T}\mathcal{J}^{\Psi})+\cdots \, .
\end{align}
\endgroup
The coupling symmetries reported in Table \ref{tab:symm} do not provide any further constraints. The OPE consistency conditions fix the couplings to be functions of the central charge $c$ \cite{Blumenhagen:1991nm}, up to a normalisation $C^{\mathds{1}}_{\Psi \Psi}$ of the operator $\mathcal{J}^{\Psi}$
\begin{equation} \label{eq: constraints1}
    C_{\Psi \Psi}^{\ct}=\frac{12}{c} C^{\mathds{1}}_{\Psi \Psi} \, , \quad  C_{\Psi \Psi}^{D \mathcal{T}\mathcal{T}}=\frac{216}{c\, (21+4c)} C^{\mathds{1}}_{\Psi \Psi} \, , \quad  C_{\Psi \Psi}^{\mathcal{T}\Psi}=\frac{54}{6+5c}C^{\Psi}_{\Psi \Psi} \, ,
\end{equation}
and 
\begin{equation} \label{eq: constraints2}
    \left(C^{\Psi}_{\Psi \Psi}\right)^{2}=-\frac{8 (5 c+6)^2 }{c \, (c-15) (4 c+21)}C^{\mathds{1}}_{\Psi \Psi} \, .
\end{equation}
We conclude that we have a \emph{family} of algebras parametrised by the central charge $c$, bounded by unitarity as $0<c<15$.\footnote{We take the opportunity to highlight that this bound could be refined if, instead of only imposing the consistency of the symmetry algebra, we do the same for all the operators in the worldsheet CFT (by imposing crossing symmetry, for example); a similar statement can be made for all the examples provided in this Section. This goes beyond the scope of this paper, and we reserve it for future work.}
The special \emph{locus} $c=12$ corresponds to the $\SVeight$ algebra \cite{Shatashvili:1994zw}, describing string backgrounds with special holonomy equal to $\mathrm{Spin}(7)$. The OPEs for the $\mathcal{SW}(\frac{3}{2},2)$ algebra and its \emph{locus} $\SVeight$ can be found in \Cref{app:FSSpin7OPEs,app:SVSpin7OPEs}.
\paragraph{Computation of the $U$, $V$ and $N$ forms.} The simplicity of this algebra allows us to present an instructive example of how to derive the classical OPEs. In fact, there is a single classical OPE $\mathcal{J}^{\Psi}_{\text{cl.}}\times\mathcal{J}^{\Psi}_{\text{cl.}}$ to study. The associated form $U$ trivially vanishes:
\begin{equation*}
    U=\frac{1}{c_U} \Psi^{i} \wedge \Psi_{i}=0 \, ,
\end{equation*}
while the form $V$ can be computed using the identity \cite{Gauntlett:2003cy, Karigiannis_2008}
\begin{equation} \label{eq:psiid1}
    \Psi^{ijk \ell}\Psi_{mn k \ell}=12 \delta^{[i}_{[m}\delta^{j]}_{n]}+4\tensor{\Psi}{^{ij}_{mn}} \, ,
\end{equation}
leading to 
\begin{equation*}
    V=\frac{1}{c_V} \Psi^{ij}\wedge \Psi_{ij}=\frac{4!}{c_V} \Psi \, .
\end{equation*}
The choice $c_V=4!$ results in $V=\Psi$. Finally, we prove that the form $N$ vanishes. Let us consider the Hodge dual form of $N$
\begin{equation*}
    *N=\frac{\ell_s}{c_{N}}*\left(H_{jk}\wedge \Psi^{j}\wedge\Psi^{k} - 12 \,   H \wedge \Psi  \right) \, ,
\end{equation*}
thus, it is enough to show that
\begin{equation*}
    *\left(H_{jk}\wedge \Psi^{j}\wedge\Psi^{k}\right)= 12 \, *\left( H \wedge \Psi  \right) \, .
\end{equation*}
We start writing down explicitly the Hodge dual of the first term
\begin{align*}
    *\left(H_{jk}\wedge \Psi^{j}\wedge\Psi^{k}\right)&=\frac{\sqrt{G}}{7!}\frac{7!}{3!\, 3!}\tensor{H}{^{i_1}_{jk}}\Psi^{j {i_2}{i_3}{i_4}}\Psi^{k {i_5}{i_6}{i_7}} \epsilon_{i_1 i_2 i_3 i_4 i_5 i_6 i_7 i_8} \de x^{i_8} \nn \\[4pt]
    &=-\frac{5}{3!}\tensor{H}{^{i_1}_{jk}}\Psi^{j {i_2}{i_3}{i_4}}\delta^{k}_{[i_{1}}\Psi_{i_2 i_3 i_4 i_8]}^{\phantom{k}} \de x^{i_8} \, ,
\end{align*}
where we have used the identity
\begin{equation*}
    \frac{\sqrt{G}}{3!}\Psi^{k {i_5}{i_6}{i_7}} \epsilon_{i_1 i_2 i_3 i_4 i_5 i_6 i_7 i_8}=-5 \, \delta^{k}_{[i_{1}}\Psi_{i_2 i_3 i_4 i_8]}^{\phantom{k}} \, .
\end{equation*}
Using again \eqref{eq:psiid1} together with the identity 
\begin{equation*}
    \Psi^{ijk \ell}\Psi_{mj k \ell}=42 \, \delta^{i}_{m} \, ,
\end{equation*}
we reach the desired result
\begin{equation*}
    *\left(H_{jk}\wedge \Psi^{j}\wedge\Psi^{k}\right)=12 \, 
    H\lrcorner \Psi=12 \, *\left( H \wedge \Psi\right) \, ,
\end{equation*}
where we used the generic identity \eqref{eq:id hodge corner} and the self-duality of the Cayley form $\Psi=*\Psi$. Note that the only non-zero form is $V=\Psi$, which is again the Cayley form, and no torsion classes play a role in the computation. We collect this outcome in \Cref{tab:Spin7}. In subsequent Sections, we will provide similar Tables with the final results for each $\mathrm{G}$-structure.

\paragraph{Geometrical interpretation of the couplings.}

\begin{table}
\renewcommand{\arraystretch}{2}
    \centering
    \begin{tabular}{|c|c||c|c|c||c|c|c|}
    \hline
       $\cj^1_{\text{cl.}}$ & $\cj^2_{\text{cl.}}$ & $\cj^{U}_{\text{cl.}}$ & $\cj^{N}_{\text{cl.}}$ & $\cj^{V}_{\text{cl.}}$ & $c_{U}$ & $c_{N}$ & $c_{V}$ \\
        \hline
        $\cj^{\Psi}_{\text{cl.}}$&$\cj^{\Psi}_{\text{cl.}}$ & - & - & $\cj^{\Psi}_{\text{cl.}}$ & - & - & $4!$  \\
        \hline
    \end{tabular}
    \caption{\emph{Currents appearing in the classical OPE $\cj^{1}_{\mathrm{cl.}} \times \cj^2_{\mathrm{cl.}}$, in the case of a $\mathrm{Spin}(7)$-structure. The torsion classes $\tau_1$ and $\tau_3$ do not explicitly appear in the coefficients $c_U, c_N$ and $c_V$.}}
    \label{tab:Spin7}
\end{table}
The $\mathrm{Spin}(7)$ case effectively illustrates the difference between the classical and the quantum OPE: in fact, the former is significantly simpler than the latter, and reads 
\begin{equation} \label{eq: class spin7}
    \mathcal{J}^\Psi_{\text{cl.}}(Z_1)\mathcal{J}^\Psi_{\text{cl.}}(Z_2) \sim
    12 \, \frac{\theta_{12}}{Z_{12}} \, \ct_{\text{cl.}}^{\phantom{\Phi}}\cj^{\Psi}_{\text{cl.}}+\cdots  \, .
\end{equation}
The currents $\cj^{U}_{\text{cl.}}, \cj^{V}_{\text{cl.}}$ and $\cj^{N}_{\text{cl.}}$ are listed in \Cref{tab:Spin7}, together with our choices of coefficients $c_U, c_V$ and $c_N$, as computed above. Note that in this case the torsion classes do not explicitly appear in the coefficients, suggesting that they do not play a role at the order in $\ell_s$ we are considering. If we identify the product of classical currents $\ct_{\text{cl.}}^{\phantom{\Phi}}\cj^{\Psi}_{\text{cl.}}$ as the classical limit of the quasi-primary normal ordered operator $\cn(\ct\cj^{\Psi})$, the comparison between \eqref{eq: spin7ope} and \eqref{eq: class spin7} is straightforward and leads to
\begin{equation*}
    C_{\Psi \Psi}^{\mathcal{T}\Psi}= 12 + O(\ell_s^2) \, .
\end{equation*}
Since the consistency of the OPE requires the constraints \eqref{eq: constraints1} and \eqref{eq: constraints2} to be satisfied, we have
{
\allowdisplaybreaks
\begin{align*}
    C_{\Psi \Psi}^{\mathds{1}}&=-\frac{c\, (c-15)(4c+21)}{162}+O(\ell_s^2) \, , & 
    C_{\Psi \Psi}^{\Psi}&=\frac{2}{9}(6+5c)+O(\ell_s^2) \, , \\[4pt]
    C_{\Psi \Psi}^{\ct}&=-\frac{2}{27} (c-15)(4c+21)+O(\ell_s^2) \, , &
    C_{\Psi \Psi}^{D \mathcal{T}\mathcal{T}}&=-\frac{4}{3}(c-15)+O(\ell_s^2) \, .
\end{align*}
}%
In this case, the comparison between the classical and the quantum OPE was only useful to fix the normalisation $C_{\Psi \Psi}^{\mathds{1}}$. 
The central charge $c$ retains its role of free parameter of the family of algebras, bounded by the same unitarity bound: the existence of a generic Spin(7)-structure does not constrain the value of $c$ at first order in the string length scale $\ell_s$.
In this case, the role of the torsion is relegated to loop contributions at higher orders in $\ell_s$. Computing these terms would shed some light into the relationship between the central charge and torsion classes, see for example \cite{Guadagnini:1986un, Metsaev:1987zx}. As a consistency check, note that setting $c=12$ (and taking the $H\to 0$ limit) returns the special holonomy algebra: the predicted couplings are 
\begin{equation*}
    C_{\Psi \Psi}^{\mathcal{T}\Psi}=12 \, , \quad C^{\Psi}_{\Psi \Psi}= \frac{44}{3} \, , \quad C^{\mathds{1}}_{\Psi \Psi}=\frac{46}{3} \, , \quad C^{\ct}_{\Psi \Psi}=\frac{46}{3} \, , \quad C_{\Psi \Psi}^{D \mathcal{T}\mathcal{T}}=4 \, ,
\end{equation*}
and we find a perfect match with the $\SVeight$ algebra, reported explicitly in \Cref{app:SVSpin7OPEs}.

\subsection{\texorpdfstring{$\mathcal{SW}\left(\frac{3}{2},\frac{3}{2},2\right)$}{SW(3/2,3/2,2)} algebra and \texorpdfstring{$\text{G}_2$}{G2}-structures}
\label{sec:G2SW}

Next, we proceed to the case of $\mathrm{G}_2$-structures. Since they possess two characteristic forms, the associated $\mathcal{SW}$-algebra is slightly more complicated than the Spin(7) case. Nevertheless, the richer geometric structure gives additional control over the algebra, having in particular a non-trivial contribution from the scalar torsion class of the $\mathrm{G}_2$-structure. The case of $\mathrm{G}_2$-holonomy was first addressed in \cite{Shatashvili:1994zw}, whereas the case with torsion has recently received attention for the first time in \cite{Fiset:2021azq}.

\paragraph{$\mathrm{G}_2$-structures.}

A $\mathrm{G}_2$-structure on a seven-dimensional Riemannian manifold $\mathcal{M}$ is determined by a well defined, nowhere-vanishing three-form $\varphi$ such that at any given point on $\mathcal{M}$ it can be written as \cite{Joyce:2007, Bryant:2005mz, Karigiannis:2020}
\begin{equation*}
\varphi=\dd x^{246}-\dd x^{235}-\dd x^{145}- \dd x^{136}+  \dd x^{127}+\dd x^{347}+\dd x^{567} \, ,
\end{equation*}
for some particular choice of local coordinates $\lbrace x_1,\dots,x_7\rbrace$ around that point. We call $\varphi$ the \emph{associative} form. It determines a metric and an orientation on $\mathcal{M}$ that we can use to construct the Hodge dual four-form $\psi= * \varphi$, called the \emph{coassociative} form.

There are four torsion classes associated with a G$_2$-structure, which we denote by $\tau_0 \in \Omega^{0}_{\mathbf{1}}(T M)$, $\tau_1 \in \Omega^{1}_{\mathbf{7}}(T M)$, $\tau_2 \in \Omega^{2}_{\mathbf{14}}(TM)$ and $\tau_3 \in \Omega^{3}_{\mathbf{27}}(T M)$. They are determined by the exterior derivative of the characteristic tensors $\varphi$ and $\psi$:
\begin{equation*}
\dd\varphi = \tau_0\, \psi + 3 \, \tau_1\wedge\varphi + *\tau_3 \, , \qquad
\dd\psi  = 4 \, \tau_1\wedge\psi + *\tau _2 \, .
\end{equation*}
On a manifold with a $\mathrm{G}_2$-structure, the torsion class $\tau_2$ represents an obstruction to the existence of a (unique) compatible connection with totally antisymmetric torsion \cite{Friedrich:2001nh}. We set $\tau_2= 0$ in what follows, and the torsion of that connection is given by 
\begin{equation*}
\text{Tor} =  \frac{1}{6}\, \tau_0\, \varphi - \tau_1\lrcorner\,\psi - \tau_3 \, .
\end{equation*}

\paragraph{The $\mathcal{SW}(\frac{3}{2}, \frac{3}{2}, 2)$ algebra.}
Since the characteristic tensors of a manifold with a G$_2$-structure are the three-form $\varphi$ and the four-form $\psi$, the $\mathcal{SW}$-algebra that describes the corresponding worldsheet dynamics must be $\mathcal{SW}(\frac{3}{2}, \frac{3}{2}, 2)$, with generators
\begin{equation*}
   \Braket{\mathds{1}, \mathcal{T}, \mathcal{J}^{\varphi}, \mathcal{J}^{\psi}} \, .
\end{equation*}
This algebra was first obtained in \cite{Blumenhagen:1991nm}, and further studied in \cite{Noyvert:2002mc}. The algebra depends on two free parameters which we denote by $c$ and $\lambda$ following \cite{Noyvert:2002mc}: $c$ is the central charge and $\lambda$ participates in the self-coupling $C_{\phu \phu}^{\phu}$.\footnote{Up to a rescaling of the fields the algebra does not depend on the sign of $\lambda$: the true parameters of the algebra actually are $c$ and $\lambda^2$.} We will indicate the parameters as subscripts $\mathcal{SW}_{[\lambda^2,c]}(\frac{3}{2}, \frac{3}{2}, 2)$ following \cite{Fiset:2021azq}. The complete list of OPEs following our conventions---which are the ones required for comparison with the classical algebra and thus differ slightly from those of \cite{Blumenhagen:1991nm, Noyvert:2002mc}---can be found in \Cref{app:SVG2OPEs}.

There are some particular choices of $[\lambda^2,c]$ that we want to highlight since they have proven to be especially relevant for string backgrounds with a G$_2$-structure. First of all, when the parameter $\lambda$ vanishes and the central charge is fixed to $c=\frac{21}{2}$, we recover the  \emph{Shatashvili--Vafa algebra} first introduced in \cite{Shatashvili:1994zw}, which we denote by 
\begin{equation*}
    \SVseven=\mathcal{SW}_{[0,\frac{21}{2}]}\left(\tfrac{3}{2}, \tfrac{3}{2}, 2\right) \, .
\end{equation*}
It captures the worldsheet algebra enhancement for string backgrounds with G$_2$ holonomy and it has been widely studied \cite{Gaberdiel:2004vx, deBoer:2005pt, Braun:2019lnn, Fiset:2021ruv, Roiban:2001cp, Eguchi:2001xa, Blumenhagen:2001jb, Eguchi:2001ip, Noyvert:2002mc, Braun:2023zcm, Sugiyama:2001qh, Sugiyama:2002ag, Eguchi:2003yy, Heluani:2014uaa}. More recently, it was argued by the authors of \cite{Fiset:2021azq} that the \emph{locus}
\begin{equation*}
    \lambda^2=\frac{4(3-2c)^2(21-2c)}{27(10c-7)} 
\end{equation*}
defines a one-parameter family of worldsheet algebras describing string backgrounds given by a direct product of an  $\mathrm{AdS}_3$ spacetime and a manifold with a G$_2$-structure, both of them supporting a non-zero NS flux. We call this the \emph{Fiset--Gaberdiel family of algebras} and denote it by 
\begin{equation*}
    \mathrm{FG}_k=\mathcal{SW}_{\left[\frac{32(3k-2)^2}{k^2(49k-30)},\frac{21}{2}-\frac{6}{k}\right]}\left(\tfrac{3}{2}, \tfrac{3}{2}, 2\right) \, ,
\end{equation*}
where the integer parameter $k>0$ is associated with the NS flux in the background \cite{Fiset:2021azq}.\footnote{Even though the $\mathrm{FG}_k$ algebra was obtained from pure NS-NS type II backgrounds in \cite{Fiset:2021azq}, it should be equally valid for heterotic backgrounds. This is because the chiral currents in the holomorphic sector of the $(1,0)$ $\sigma$-model agree with those of the $(1,1)$ $\sigma$-model, since gauge fields do not contribute---c.f. \eqref{eq: class T} and \eqref{eq: current}.} One can also interpret $\sqrt{k}$ as the measure of the $\mathrm{AdS}_3$ radius in string units, and note that the limit $k\rightarrow\infty$ leads to a Minkowski spacetime, corresponding to a $\mathrm{G}_2$ special holonomy compactification.

Both $\mathrm{FG}_k$ and $\SVseven$ share a key property: they contain a tricritical Ising model as a subalgebra. A very important consequence is the existence of a null field at level $\frac{7}{2}$ that for $\mathrm{FG}_k$ takes the form
\begin{multline}
    \mathfrak{N}^{\mathrm{FG}}_k =8\left(3-\frac{4 c^2}{7 \mu^2}\right)\mathcal{N}(\mathcal{T}\mathcal{J}^{\psi})-\frac{2}{3}\left(1+\frac{4}{21}c\right)\mathcal{N}(D\mathcal{J}^{\varphi}\mathcal{J}^{\varphi})+\frac{8c}{\mu^2}\mathcal{N}(D\mathcal{T}\mathcal{T}) \\
    \label{eq:FGnullfield}
    +\frac{1}{\sqrt{k}}\frac{8\sqrt{2}}{7} \left( \mathcal{N}(\mathcal{J}^{\varphi}\mathcal{J}^{\psi})-\left(3-\frac{4 c^2}{3 \mu^2}\right)\mathcal{N}(D\mathcal{T}\mathcal{J}^{\varphi}) \right)  \, ,
\end{multline}
where $c=\frac{21}{2}-\frac{6}{k}$ as above and we are using, following \cite{Noyvert:2002mc},
\begin{equation}
\label{eq:muSWG2}
    \mu=\sqrt{\frac{9c\,(4 + \lambda^2)}{2(27 - 2 c)}}=\sqrt{\frac{9(7k-4)^3}{2 k^2(49k-30)}} \, .
\end{equation}
The normal ordered products appearing in the OPEs can be projected onto quasi-primary operators as follows
\begin{align*}
    \mathcal{N}(\mathcal{T}\mathcal{J}^{\psi}) &= N(\mathcal{T}\mathcal{J}^{\psi})-\frac{1}{5}\partial D \mathcal{J}^{\psi} \, , \\
    \mathcal{N}(D\mathcal{J}^{\varphi}\mathcal{J}^{\varphi}) &= N(D\mathcal{J}^{\varphi}\mathcal{J}^{\varphi}) - \frac{3k(49k-30)}{2(7k-4)^2}\partial\partial\mathcal{T}-3\frac{3k-2}{7k-4}\sqrt{\frac{2}{k}}\partial\partial\mathcal{J}^{\varphi} - \frac{9}{5}\partial D \mathcal{J}^{\psi} \, , \\
    \mathcal{N}(D\mathcal{T}\mathcal{T}) &= N(D\mathcal{T}\mathcal{T})-\frac{3}{8}\partial \partial\mathcal{T}  \, , \\
    \mathcal{N}(\mathcal{J}^{\varphi}\mathcal{J}^{\psi}) &= N(\mathcal{J}^{\varphi}\mathcal{J}^{\psi}) - \frac{7k-4}{12k}\partial\partial\mathcal{J}^{\varphi} -\frac{4(3k-2)}{5(7k-4)}\sqrt{\frac{2}{k}}\partial D\mathcal{J}^{\psi} - D \mathcal{N}(\mathcal{T}\mathcal{J}^{\varphi}) \, , \\
    \mathcal{N}(\mathcal{T}\mathcal{J}^{\varphi}) &= N(\mathcal{T}\mathcal{J}^{\varphi})-\frac{1}{4}\partial D \mathcal{J}^{\varphi}  \, , \\
    \mathcal{N}(D\mathcal{T}\mathcal{J}^{\varphi})  &= \frac{1}{2}N(D\mathcal{T}\mathcal{J}^{\varphi}) +\frac{1}{2}N(D\mathcal{J}^{\varphi}\mathcal{T}) -\frac{3}{8}\partial \partial\mathcal{J}^{\varphi}  \, .
\end{align*}
Note that for $k\rightarrow\infty$ the terms in $\mathfrak{N}^{\mathrm{FG}}_k$ proportional to $\frac{1}{\sqrt{k}}$ vanish and we recover the familiar null field up to which the $\SVseven$ algebra closes
\begin{equation*}
    \mathfrak{N}^{\mathrm{SV}}=8\,\mathcal{N}(\mathcal{T}\mathcal{J}^{\psi})-2\,\mathcal{N}(D\mathcal{J}^{\varphi}\mathcal{J}^{\varphi})+\frac{8}{3}\mathcal{N}(D\mathcal{T}\mathcal{T}) \, .
\end{equation*}
These null fields will play a fundamental role in the geometric interpretation of the $\mathrm{FG}_k$ and $\SVseven$ algebras below. This is because the different couplings of the $\mathcal{J}^\psi\times\mathcal{J}^\psi$ OPE are well-defined only up to the addition of null fields, and this must be taken into account when comparing with the classical OPE.

\paragraph{Geometrical interpretation of the couplings.}
{
\begin{table}
\renewcommand{\arraystretch}{2}
    \centering
    \begin{tabular}{|c|c||c|c|c||c|c|c|}
    \hline
       $\cj^1_{\text{cl.}}$ & $\cj^2_{\text{cl.}}$ & $\cj^{U}_{\text{cl.}}$ & $\cj^{N}_{\text{cl.}}$ & $\cj^{V}_{\text{cl.}}$ & $c_{U}$ & $c_{N}$ & $c_{V}$ \\
        \hline
        $\cj^{\phu}_{\text{cl.}}$&$\cj^{\phu}_{\text{cl.}}$ & $\cj^{\psi}_{\text{cl.}}$ & - & - & $6$ & - & -  \\
        $\cj^{\phu}_{\text{cl.}}$&$\cj^{\psi}_{\text{cl.}}$ & - & - & $\cj^{\phu}_{\text{cl.}}$ & - & - & $12$  \\
        $\cj^{\psi}_{\text{cl.}}$&$\cj^{\psi}_{\text{cl.}}$ & - & $-\cj^{\phu}_{\text{cl.}} \cj^{\psi}_{\text{cl.}}$ & $\cj^{\psi}_{\text{cl.}}$ & - & $\frac{2}{3}\, \tau_0\ell_s$ & $12$  \\
        \hline
    \end{tabular}
    \caption{\emph{Currents appearing in the classical OPE $\cj^{1}_{\mathrm{cl.}} \times \cj^2_{\mathrm{cl.}}$, in the case of a $\mathrm{G}_2$-structure.}}
    \label{tab:G2}
\end{table}}
The classical currents and their parameters for a G$_2$-structure are presented in \Cref{tab:G2}. From this Table, we study the following classical OPEs:
\begin{align}
\label{eq:classicalG2OPE1}
    \mathcal{J}^\phu_{\text{cl.}}(Z_1)\mathcal{J}^\phu_{\text{cl.}}(Z_2) &\sim 6 \left( \frac{1}{Z_{12}}\mathcal{J}_{\text{cl.}}^\psi+ \frac{\theta_{12}}{2 \, Z_{12}}D\mathcal{J}_{\text{cl.}}^\psi\right)+\cdots  \, , \\[8pt]
\label{eq:classicalG2OPE2}
     \mathcal{J}_{\text{cl.}}^\phu(Z_1)\mathcal{J}_{\text{cl.}}^\psi(Z_2) &\sim
     6 \, \frac{ \theta_{12}}{Z_{12}} \, \ct^{\phantom{\phu}}_{\text{cl.}} \cj^{\phu}_{\text{cl.}}+\cdots  \, .
\end{align}
First, we need to fix the normalisations of the operators $\mathcal{J}^\varphi$ and $\mathcal{J}^\psi$ to match those of our classical currents \eqref{eq:HPcurrentcomponents}. To do so, we impose the agreement (at first order in $\ell_s$) between the classical OPE coefficients appearing in \eqref{eq:classicalG2OPE1} and \eqref{eq:classicalG2OPE2} and the coefficients of the $\mathcal{SW}(\frac{3}{2}, \frac{3}{2}, 2)$ algebra, that is
\begin{equation*}
    C_{\varphi \varphi}^{\psi}= 6 + O(\ell_s^2) \, , \qquad    C_{\varphi \psi}^{\mathcal{T}\varphi}= 6 + O(\ell_s^2) \, ,
\end{equation*}
and we find that the right normalisation is
\begin{equation*}
    C_{\varphi \varphi}^{\mathds{1}}= \frac{2 c^2 }{\mu^2} + O(\ell_s^2) \, , \qquad    C_{\psi \psi}^{\mathds{1}}= \frac{2 c^3}{9\mu^2} + O(\ell_s^2) \, ,
\end{equation*}
where $\mu$ is again given by \eqref{eq:muSWG2}. The full list of OPEs following this normalisation can be found in \Cref{app:SVG2OPEs}, and it is immediate to check the agreement of the coefficients with the OPEs \eqref{eq:SWG2SOPEphiphi} and \eqref{eq:SWG2SOPEphipsi}.

There is one more classical OPE left to study: 
\begin{equation}
\label{eq:classicalG2OPE3}
    \mathcal{J}_{\text{cl.}}^\psi (Z_1)\mathcal{J}_{\text{cl.}}^\psi(Z_2) \sim -\frac{2}{3}\ell_s \tau_0 \frac{\theta_{12}}{Z_{12}}\mathcal{J}_{\text{cl.}}^{\phu}\mathcal{J}_{\text{cl.}}^{\psi}
    + 8 \, \frac{\theta_{12}}{Z_{12}} \, \ct^{\phantom{\psi}}_{\text{cl.}} \cj^{\psi}_{\text{cl.}}+\cdots  \, . 
\end{equation}
When comparing with the OPE \eqref{eq:SWG2SOPEpsipsi}, we seem to incur a contradiction since the coefficient $C_{\psi \psi}^{\mathcal{T}\psi}$ takes the value of $12$, which does not correspond to the classical value of $8$ in \eqref{eq:classicalG2OPE3} above. Nevertheless, this \emph{puzzle} can be solved by recalling that OPEs close up to the addition of null fields. 
Therefore, we can interpret the mismatch in the parameters as an indication that not all values of the parameters $[\lambda^2,c]$ describe superstring backgrounds endowed with a G$_2$-structure. The existence of null fields is then essential in order to have a geometric interpretation of the $\mathcal{SW}$-algebra we are studying.

As a working example, we can consider the algebra $\mathrm{FG}_{k}$. It was shown in \cite{Fiset:2021azq} that the $\mathrm{FG}_k$ algebra does describe backgrounds endowed with a G$_2$-structure, although the torsion classes were not specified. We want to show that the comparison with the classical OPE \eqref{eq:classicalG2OPE3} gives us information regarding the scalar torsion class $\tau_0$.
As we explained above, for the $\mathrm{FG}_k$ algebra the coefficient $C_{\psi \psi}^{\mathcal{T}\psi}$ is only well-defined up to addition of null fields. Therefore, specializing to the $\mathrm{FG}_k$ algebra we can modify the OPE \eqref{eq:SWG2SOPEpsipsi} by adding a convenient multiple of the null field \eqref{eq:FGnullfield} so that the condition
\begin{equation*}
    C_{\psi \psi}^{\mathcal{T}\psi}=8 + O(\ell_s^2)
\end{equation*}
is satisfied and there is an honest match between the $\mathcal{SW}$-algebra and the classical algebra. The final OPE can be found in \eqref{eq:FGG2SOPEpsipsi}. We can then read off the $C_{\psi \psi}^{\varphi\psi}$ coupling and compare with the classical prediction: 
\begin{equation}
\label{eq: class pred }
    C_{\psi \psi}^{\varphi\psi}=-4\sqrt{\frac{2}{k}}\frac{7k-4}{49k-24}=-\frac{2}{3}\tau_0 \ell_s + O(\ell_s^2) \, .
\end{equation}
This shows that the coupling $C_{\psi \psi}^{\varphi\psi}$ is directly related to the scalar torsion $\tau_0$. We conclude that the $\mathrm{FG}_k$ algebra captures the worldsheet symmetry of a string background involving a manifold endowed with a G$_2$-structure such that $\tau_0\neq 0$. It is illustrative to consider the limit $k\rightarrow\infty$ corresponding to the $\SVseven$ algebra. In this case, the coefficient $C_{\psi \psi}^{\varphi\psi}$ vanishes identically and we obtain that $\tau_0=0$, consistent with the fact that $\SVseven$ is valid when the compact manifold has G$_2$ holonomy and all the torsion classes are zero. Our analysis thus recovers the expected prediction for $\SVseven$.

Now we would like to identify the value of $\tau_0$. As shown in \cite{Gaberdiel:2021jrv}, the parameter $\sqrt{k}$ is tied to the radius $R_{\mathrm{AdS}}$ of the $\mathrm{AdS}_3$ spacetime component of the string background
\begin{equation}
\label{eq: k exp}
    k=2 \pi \left(\frac{R_{\mathrm{AdS}}}{\ell_s}\right)^2 \, .
\end{equation}
We believe the classical OPE \eqref{eq:classicalG2OPE3} to hold in the large radius limit, i.e., when the length scale of the string background is large compared to the string length scale. Therefore, to read off the value of $\tau_0$ we need to expand $C_{\psi \psi}^{\varphi\psi}$ in powers of $\frac{1}{\sqrt{k}}$. We do so around the flat limit $k=\infty$: after plugging the expression \eqref{eq: k exp} in the expansion, we obtain
\begin{equation*}
    C_{\psi \psi}^{\varphi\psi}=-\frac{1}{\sqrt{\pi}}\frac{4}{7} \frac{\ell_s} {R_{AdS}}  +  \frac{1}{\sqrt{\pi}}\frac{8}{343\pi} \left(\frac{\ell_s}{R_{AdS}}\right)^3  +\cdots \, ,
\end{equation*}
and we can plug the result in the identification \eqref{eq: class pred } to find 
\begin{equation*}
    \tau_0=  \frac{1}{\sqrt{\pi}}\frac{6}{7} \frac{1}{R_{\mathrm{AdS}}} + O(\ell_s^2) \, .
\end{equation*}
This recovers the well-known fact that in supergravity the torsion class $\tau_0$ is proportional to the inverse of the $\mathrm{AdS}$ radius, see for example \cite{Kunitomo:2009mx, delaOssa:2019cci}. Trading the parameter $k$ for $\tau_0\, \ell_s$, we can rewrite the central charge of the $\mathrm{FG}_k$ algebra as
\begin{equation*}
    c=\frac{21}{2}-\frac{49}{12}\tau_0^2 \ell_s^2 + O(\ell_s^3) \, ,
\end{equation*}
which is compatible with the notion of the central charge being corrected at order $\ell_s^2$ by the norm of the torsion squared
\cite{Guadagnini:1986un, Metsaev:1987zx}.

Finally, we point out that there are other algebras in the $\mathcal{SW}(\frac{3}{2}, \frac{3}{2}, 2)$ family in which null fields at spin $\frac{7}{2}$ arise. These could also be amenable to an interpretation along the lines we have provided here, and we leave that exploration for future work.

\subsection{\texorpdfstring{$\mathcal{SW}\left(\frac{3}{2},1,1,1\right)$}{SW(3/2,1,1,1)} algebra and \texorpdfstring{$\mathrm{SU}(2)$}{SU(2)}-structures}

We now turn our attention to generic $\mathrm{SU}(2)$-structures---note that this includes the case of $\mathrm{SU}(2)$-holonomy corresponding to $\mathrm{K}3$ surfaces. The underlying $\mathcal{SW}$-algebra has three additional currents with the same (holomorphic) weight $h=1$, so the symmetries of the couplings greatly simplify its construction.

\paragraph{$\mathrm{SU}(2)$-structures.} 
An $\mathrm{SU}(2)$-structure on a four-dimensional Riemannian manifold $\mathcal{M}$ is determined by a pair of well defined, nowhere-vanishing two-forms $(\omega,\Omega)$, where $\omega$ is real and $\Omega$ is complex, satisfying the relations \cite{Joyce:2007,Gauntlett:2003cy}
\begin{equation}
\label{eq:SU2relations}
\omega \wedge\Omega = 0 \,, \qquad \frac{1}{2}\, \omega\wedge\omega = \frac{1}{4}\, \Omega\wedge\overline\Omega\, .
\end{equation}
We call $\omega$ the \emph{Hermitian} form and $\Omega$ the \emph{holomorphic volume form}. We will denote the real and imaginary parts of $\Omega$ by
\begin{equation*}
    \Omega_{+}=\text{Re}\,\Omega \, , \qquad \Omega_{-}=\text{Im}\,\Omega \, , 
\end{equation*}
and we will often work with the triplet of real forms $(\omega,\Omega_{+},\Omega_{-})$. 
Note that the holomorphic volume form defines an almost complex structure $J$ on $\mathcal{M}$ as follows \cite{Hitchin:2000jd}:
\begin{equation} \label{eq: acsSU2}
    \tensor{J}{^i_j}=\frac{I\indices{^i_j}}{\sqrt{-\frac{1}{4}\tr(I^2)}}\, , \qquad I\indices{^i_j}=(\Omega_+)_{jk}(\Omega_-)_{lm}\epsilon^{iklm}\, ,
\end{equation}
where $\epsilon$ is the Levi-Civita symbol. We can then use $J$ to decompose (complex) forms into \emph{$(p,q)$ types}. We have that $\omega$ is of $(1,1)$ type and $\Omega$ is of $(2,0)$ type, justifying their nomenclature. Finally, note that $(\omega,\Omega)$ also define an orientation via \eqref{eq:SU2relations} and a metric $G$ through the formula $G_{ij}=\omega_{ik}J\indices{^k_j}$. 

There are three torsion classes associated with an $\mathrm{SU}(2)$-structure \cite{Gray:1980,Falcitelli:1994,Gauntlett:2003cy}: $W$ is a complex one-form, and $\vartheta$, $\vartheta'$ are complex one-forms of $(1,0)$ type. They are determined by the exterior derivative of the characteristic forms:
\begin{equation}
\dd\omega =W\wedge\omega\, , \qquad
\dd\Omega = \Bar\vartheta\wedge\Omega + \vartheta'\wedge\overline\Omega\, ,\label{eq:structSU2}
\end{equation}
On a manifold with an $\mathrm{SU}(2)$-structure there exists a compatible connection with totally antisymmetric torsion if and only if the \emph{Nijenhuis tensor} vanishes. Recall that the Nijenhuis tensor is the $(2,1)$-tensor field defined by
\begin{equation}
\label{eq: nijenhuis}
    N_{J}(X,Y)=[X,Y]+J\left([JX,Y]+[X,JY]\right)-[JX,JY] \, ,
\end{equation}
where $[\cdot, \cdot]$ is the Lie-bracket between two vector fields, $J(\cdot)$ is the endomorphism defined by the almost complex structure \eqref{eq: acsSU2} and $X$, $Y$ are vector fields on $\mathcal{M}$. We can regard the Nijenhuis tensor as a $(3,0)$-tensor using the metric $G$ on $\mathcal{M}$
\begin{equation}
\label{eq: nij2}
    N'_{J}(X,Y,Z)= G(N_{J}[X,Y],Z) \, ,
\end{equation}
where $X$, $Y$, $Z$ are vector fields on $\mathcal{M}$. The requirement that this tensor identically vanishes is equivalent to the almost complex structure being integrable, and from \eqref{eq:structSU2} one deduces that the torsion classes must satisfy
\begin{equation*}
    \vartheta'=0 \, , \qquad  \Re\vartheta=W \, ,
\end{equation*}
and the totally antisymmetric torsion reads 
\begin{equation*}
    \text{Tor}=-J(W)\wedge \omega  \, .
\end{equation*}
As a final remark, an $\mathrm{SU}(2)$-structure can be equivalently described as an $\mathrm{Sp}(1)$-structure---also known as an \emph{almost hyper-Hermitian structure}---due to the accidental isomorphism $\mathrm{SU}(2)\simeq \mathrm{Sp}(1)$. Such a structure is described by a triple of Hermitian forms $(\omega_1,\omega_2,\omega_3)$ which correspond to the forms $(\omega,\Omega_{+},\Omega_{-})$ in our notation.

\paragraph{The $\mathcal{SW}(\frac{3}{2},1,1,1)$ algebra.}
The algebra describing the dynamics of superstrings compactified on a manifold with an $\mathrm{SU}(2)$-structure is $\mathcal{SW}(\frac{3}{2},1,1,1)$. In the torsionless case, we know that this algebra should reduce to the Odake algebra $\mathrm{Od}(2)$ \cite{Odake:1988bh}, which is actually isomorphic to the (little) $\mathcal{N}=4$ Virasoro algebra \cite{Fiset:2019ecu}. The algebra is generated by the operators
\begin{equation*}
    \langle \mathds{1}, \mathcal{T}, \mathcal{J}^{\omega_1}, \mathcal{J}^{\omega_2}, \mathcal{J}^{\omega_3}   \rangle \, .
\end{equation*}
The Ansatz \eqref{eq: quaOPE} for the most general OPE $\mathcal{J}^{\omega_I} \times \mathcal{J}^{\omega_J}$, where the indices $I,J,K$ run from 1 to 3, reduces to
{\small
\begingroup
\allowdisplaybreaks
\begin{equation} \label{eq: su2 ope}
    \cj^{\omega_{I}}(Z_1)\cj^{\omega_{J}}(Z_2)\sim \frac{C_{IJ}}{Z_{12}^2}+\frac{C_{IJ}^{K}}{Z_{12}}\cj^{\omega_{K}}(Z_2)+\frac{\theta_{12}}{Z_{12}}\left(C_{IJ}^{\ct}\ct(Z_2)+ \frac{1}{2} C_{IJ}^{K} D \cj^{\omega_{K}}(Z_2) \right)+\cdots \, .
\end{equation}
\endgroup}%
The associativity of the algebra constrains the OPE \eqref{eq: su2 ope}.
To simplify the computation, we first implement the constraints coming from the symmetry of the couplings and adopt---without loss of generality---the following two-point function normalisations:
\begin{equation*}
    C_{11}=C_{22}=C_{33}=-2 \, , \qquad C_{12}=C_{23}=C_{13}=0 \, ,
\end{equation*}
so $C_{IJ}=-2 \, \delta_{IJ}$. We will soon see this is a convenient normalisation for the comparison with the underlying geometry.  
This constrains the coupling $C_{IJ}^{K}$ to be 
\begin{equation*}
    C_{IJ}^{K}= C \, \tensor{\epsilon}{_{IJ}^{K}} \, ,
\end{equation*}
where $C$ is a constant factor. 
Under this hypothesis, the OPE \eqref{eq: su2 ope} reads 
\begin{equation} \label{eq: su2 ope fin}
\begin{split}
     \cj^{\omega_{I}}(Z_1)\cj^{\omega_{J}}(Z_2)\sim &-\frac{2\,  \delta_{IJ}}{Z_{12}^2}-\delta_{IJ}\frac{12}{c} \frac{\theta_{12}}{Z_{12}} \ct(Z_2)\\[6pt]
     &\quad \quad + C \, \tensor{\epsilon}{_{IJ}^{K}}\left(\frac{1}{Z_{12}}\cj^{\omega_{K}}(Z_2)+\frac{\theta_{12}}{2 \, Z_{12}}  D \cj^{\omega_{K}}(Z_2) \right)+\cdots \, ,
     \end{split}
\end{equation}
with 
\begin{equation*}
    C^2=\frac{24}{c} \, .
\end{equation*}
The couplings of the algebra depend on a single parameter, the central charge $c$. We thus have a one-parameter family of algebras $\mathcal{SW}(\frac{3}{2},1,1,1)$. The special \emph{locus} $c=6$ corresponds to the $\mathrm{Od}(2)$ algebra.
\paragraph{Geometrical interpretation of the OPE data.}
{\renewcommand{\arraystretch}{2}
\begin{table}
    \centering
    \begin{tabular}{|c|c||c|c|c||c|c|c|}
    \hline
       $\cj^1_{\text{cl.}}$ & $\cj^2_{\text{cl.}}$ & $\cj^{U}_{\text{cl.}}$ & $\cj^{N}_{\text{cl.}}$ & $\cj^{\ct V}_{\text{cl.}}$ & $c_{U}$ & $c_{N}$ & $c_{V}$ \\
        \hline
        $\cj^{\omega}_{\text{cl.}}$&$\cj^{\omega}_{\text{cl.}}$ & - & - & $\ct$ & - & - & $4$  \\
       $\cj^{\omega}_{\text{cl.}}$&$\cj^{\Omega_{\pm}}_{\text{cl.}}$  & $\cj^{\Omega_{\mp}}_{\text{cl.}}$ & - & - & $\pm 2$ & - & - \\
        $\cj^{\Omega_{\pm}}_{\text{cl.}}$&$\cj^{\Omega_{\pm}}_{\text{cl.}}$ & - & - & $\ct$ & - & - & $4$
        \\
        $\cj^{\Omega_{+}}_{\text{cl.}}$&$\cj^{\Omega_{-}}_{\text{cl.}}$ & $\cj^{\omega}_{\text{cl.}}$ & - & - & $2$
        & - & -  \\
        \hline
    \end{tabular}
    \caption{\emph{Currents appearing in the classical OPE $\cj^{1}_{\mathrm{cl.}} \times \cj^2_{\mathrm{cl.}}$, associated with an $\mathrm{SU}(2)$-structure. Torsion classes do not explicitly appear. 
    }}
    \label{tab:SU(2)}
\end{table}}
The classical currents and their parameters for an $\mathrm{SU}(2)$-structure are presented in Table \ref{tab:SU(2)}. We obtain the following classical OPEs:
\begingroup
\allowdisplaybreaks
\begin{align*}
    \mathcal{J}^\omega_{\text{cl.}}(Z_1)\mathcal{J}^\omega_{\text{cl.}}(Z_2) &\sim- \frac{2 \, \theta_{12}}{Z_{12}} \, \ct_{\text{cl.}}(Z_2)+\cdots  \, ,   \\[8pt]
     \mathcal{J}^\omega_{\text{cl.}}(Z_1)\mathcal{J}^{\Omega_{\pm}}_{\text{cl.}}(Z_2) &\sim\mp \frac{2}{Z_{12}}\mathcal{J}^{\Omega_{\mp}}_{\text{cl.}}(Z_2) \mp \frac{\theta_{12}}{Z_{12}}D\mathcal{J}^{\Omega_{\mp}}_{\text{cl.}}(Z_2)+\cdots \, , \\[8pt]
     \mathcal{J}^{\Omega_{\pm}}_{\text{cl.}}(Z_1)\mathcal{J}^{\Omega_{\pm}}_{\text{cl.}}(Z_2) &\sim
    - \frac{2 \, \theta_{12}}{Z_{12}} \, \ct_{\text{cl.}}(Z_2)+\cdots  \, , \\[8pt]
     \mathcal{J}^{\Omega_{+}}_{\text{cl.}}(Z_1)\mathcal{J}^{\Omega_{-}}_{\text{cl.}}(Z_2) &\sim- \frac{2}{Z_{12}}\mathcal{J}^{\omega}_{\text{cl.}}(Z_2) -  \frac{\theta_{12}}{Z_{12}}D\mathcal{J}^{\omega}_{\text{cl.}}(Z_2)+\cdots   \, .
\end{align*}
\endgroup
We can rephrase these OPEs more compactly by switching to the notation introduced previously, $( \omega, \Omega_{+}, \Omega_{-})=(\omega_1, \omega_2, \omega_{3})$:
{\small
\begingroup
\allowdisplaybreaks
\begin{equation} \label{eq: su2 ope cl}
     \cj^{\omega_{I}}_{\text{cl.}}(Z_1)\cj^{\omega_{J}}_{\text{cl.}}(Z_2)\sim -\delta_{IJ} \frac{2 \, \theta_{12}}{Z_{12}}\ct_{\text{cl.}}(Z_2)-2 \, \tensor{\epsilon}{_{IJ}^{K}}\left( \frac{1}{Z_{12}}\cj^{\omega_{K}}_{\text{cl.}}(Z_2)+\frac{\theta_{12}}{2 \, Z_{12}}  D \cj^{\omega_{K}}_{\text{cl.}}(Z_2) \right)+\cdots \, ,
\end{equation}
\endgroup}%
where the indices $I,J,K$ run from 1 to 3. Since the normalisation of the currents generating the $\mathcal{SW}(\frac{3}{2},1,1,1)$ algebra has been chosen to match that of the characteristic forms associated with $\mathrm{SU}(2)$-structure, the comparison between the OPEs \eqref{eq: su2 ope fin} and \eqref{eq: su2 ope cl}  simply produces the following relation
\begin{equation*}
    C=-\sqrt{\frac{24}{c}}=-2+O(\ell_s^2) \, , 
\end{equation*}
consequently leading to 
\begin{equation*}
    c=6+O(\ell_s^2)  \, .
\end{equation*}
Therefore, deviations due to torsion away from the Odake algebra \emph{locus} would only become manifest through higher orders in the string length scale. 

\subsection{\texorpdfstring{$\mathrm{Od}^{\veps}(3)$}{Od-epsilon(3)} algebra and \texorpdfstring{$\mathrm{SU}(3)$}{SU(3)}-structures}
\label{sec: su3}

Finally, we discuss the case of $\mathrm{SU}(3)$-structures with torsion, which is interesting due to its connection with Calabi--Yau manifolds in the torsionless limit. Unfortunately, this case poses a challenge: the underlying algebra should be of the form $\mathcal{SW}(\frac{3}{2}, \frac{3}{2}, \frac{3}{2}, 1)$ and---to the best of our knowledge---a full classification of this algebra is missing in the literature.\footnote{Note two special \emph{loci} with central charges $c=9$ and $c=1$ are known and have been studied \cite{Fiset:2020lmg}.} We reserve a full description of this family of algebras for future endeavours.

We thus focus instead on an \emph{infinitesimal deformation} of the $\mathrm{Od}(3)$ algebra which we call $\mathrm{Od}^{\veps}(3)$, where $\veps$ is the infinitesimal parameter of the deformation. This algebra should be physically interpreted as the symmetry algebra of a worldsheet CFT where the string background is obtained as a deformation of a Calabi--Yau background via the introduction of an infinitesimal amount of NS flux $H$.

\paragraph{$\mathrm{SU}(3)$-structures.} 
Although $\mathrm{SU}(3)$-structures present many similarities with the case of $\mathrm{SU}(2)$-structures described above, they possess additional torsion classes that will make our study richer, as we now describe.

An $\mathrm{SU}(3)$-structure on a six-dimensional Riemannian manifold $\mathcal{M}$ is determined by a pair of well defined, nowhere-vanishing forms $(\omega,\Omega)$, where $\omega$ is a real two-form and $\Omega$ is a complex three-form, satisfying the following relations: \cite{Joyce:2007,Gauntlett:2003cy, delaOssa:2014lma}
\begin{equation} \label{eq: strusu3}
\Omega\wedge\omega = 0\, , \qquad \frac{1}{6}\, \omega\wedge\omega\wedge\omega = \frac{i}{8}
\, \Omega\wedge\overline\Omega\, ,
\end{equation} 
We call $\omega$ the Hermitian form and $\Omega$ the holomorphic volume form. We will denote the real and imaginary parts of $\Omega$ by
\begin{equation*}
    \Omega_{+}=\text{Re}\,\Omega \, , \qquad \Omega_{-}=\text{Im}\,\Omega \, , 
\end{equation*}
and we will often work with the triplet of real forms $(\omega,\Omega_{+},\Omega_{-})$. The holomorphic volume form defines an almost complex structure $J$ on $\mathcal{M}$ \cite{Hitchin:2000jd} as follows:
\begin{equation*}
    \tensor{J}{^i_j}=\frac{I\indices{^i_j}}{\sqrt{-\frac{1}{6}\tr(I^2)}}\, , \qquad I\indices{^i_j}=-(\Omega_+)_{jkl}(\Omega_+)_{mnr}\epsilon^{iklmnr}\, ,
\end{equation*}
where $\epsilon$ is the Levi-Civita symbol. As for the $\mathrm{SU}(2)$ case, we can employ $J$ to decompose (complex) forms into $(p,q)$ types. We have that $\omega$ is of $(1,1)$ type and $\Omega$ is of $(3,0)$ type, justifying their nomenclature. Note that $(\omega,\Omega)$ also define an orientation via \eqref{eq: strusu3} and a metric $G$ through the formula $G_{ij}=\omega_{ik}J\indices{^k_j}$.
\smallskip

There are five torsion classes associated with an $\mathrm{SU}(3)$-structure \cite{Gray:1980, Falcitelli:1994, LopesCardoso:2002vpf, Gauntlett:2003cy, delaOssa:2014lma}: $W_0$ is a complex function, $W_1$ is a real one-form, $\vartheta$ is a complex one-form of $(1,0)$ type, $W_2$ is a complex primitive two-form of $(1,1)$ type, and $W_3$ is a real primitive three-form of $(2,1)+(1,2)$ type. They are determined by the exterior derivative of the characteristic forms:
\begin{equation*}
\de \omega =  - \frac{3}{4} 
\, \Im(W_{0}\,\overline\Omega) + W_{1}\wedge\omega + W_{3}\, , \qquad
\de \Omega = W_{0}\, \omega\wedge\omega + W_{2}\wedge\omega  + \bar\vartheta\wedge\Omega\, .
\end{equation*}
We will often decompose the complex function $W_{0}$ into two real torsion classes
\begin{equation*}
    W_{0}^{+} = \Re W_{0} \, , \quad W_{0}^{-}=\Im W_{0} \, . 
\end{equation*}
On a manifold with an $\mathrm{SU}(3)$-structure there exists a compatible connection with totally antisymmetric torsion if and only if the Nijenhuis tensor $N'_J$, defined in equation \eqref{eq: nij2}, is totally antisymmetric \cite{Friedrich:2001nh}. This is equivalent to the vanishing of the torsion class $W_2$ \cite{Friedrich:2001nh, Gray:1980}, and the totally antisymmetric torsion then reads 
\begin{equation*}
    \text{Tor}=-J(\de \omega)+N'_{J}=-J(W_1\wedge\omega+W_3)-\frac{1}{2}\Re(\overline{W}_0 \Omega) \, .
\end{equation*}

\paragraph{The $\mathcal{SW}(\frac{3}{2}, \frac{3}{2}, \frac{3}{2}, 1)$ and $\mathrm{Od}^{\veps}(3)$ algebras.} 
The $\mathcal{SW}$-algebra underlying manifolds with an $\mathrm{SU}(3)$-structure must be $\mathcal{SW}(\frac{3}{2}, \frac{3}{2}, \frac{3}{2}, 1)$. Recalling the prescription \eqref{eq: gen presc}, the generators of the algebra will be
\begin{equation*}
   \Braket{\mathds{1}, \mathcal{T}, \mathcal{J}^{\omega}, \mathcal{J}^{\Omega_{+}}, \mathcal{J}^{\Omega_{-}}} \, ,
\end{equation*}
where the current $\mathcal{J}^{\omega}$ has weight $h_{\omega}=1$ and the currents $\mathcal{J}^{\Omega_{\pm}}$ have weight $h_{\Omega_{\pm}}=\frac{3}{2}$. 
As anticipated, the high weights and the number of generators will make the explicit construction this algebra a challenging task.
To formulate the Ansatz, we need to introduce the following quasi-primary projections of normal ordered products---note how their structures depend on the couplings, most of them yet to be determined at this stage:
{\allowdisplaybreaks
\begin{align*}
    \cn(\ct\cj^{\omega})&=N( \ct\cj^{\omega} )-\frac{1}{3}  D \del \cj^{\omega} \, , \\[8pt]
    \cn(\cj^{\omega}\cj^{\omega})&= N(\cj^{\omega}\cj^{\omega})-\frac{1}{3}C_{\omega \omega}^{\Omega_{+}}  D\cj^{\Omega_{+}}-\frac{1}{3}C_{\omega \omega}^{\Omega_{-}} D\cj^{\Omega_{-}}-\frac{1}{3}C_{\omega \omega}^{\ct}D\ct \, , \\[8pt]
    \cn(\cj^{\omega}\cj^{\Omega_{\pm}}) &=N(\cj^{\omega}\cj^{\Omega_{\pm}}) -\frac{1}{6}C_{\omega\Omega_{\pm}}^{\omega}  D \del \cj^{\omega} -\frac{1}{4}C_{\omega\Omega_{\pm}}^{\omega \omega}D \cn(\cj^{\omega} \cj^{\omega}) \nn \\[6pt]
    &-\frac{1}{3}C_{\omega\Omega_{\pm} }^{\ct}\del \ct -\frac{1}{3}C_{\omega\Omega_{\pm}}^{\Omega_{+}}\del\cj^{\Omega_{+}} -\frac{1}{3}C_{\omega\Omega_{\pm}}^{\Omega_{-}}\del\cj^{\Omega_{-}} \, .
\end{align*}}%
The symmetry arguments and the normalisation reported in \Cref{sec: Walg}  allow us to set the following couplings to zero:
\begin{align*}
    C_{\omega \omega}^{\omega}&=0 \, , & C_{\Omega_{\pm}\Omega_{\pm}}^{\omega}&=0 \, , & C_{\Omega_{\pm}\Omega_{\pm}}^{\ct\omega}&=0 \, , & C_{\Omega_{\pm}\Omega_{\pm}}^{\omega\Omega_{\pm}}&=0 \, , \nn \\
     C_{\omega \Omega_{+}}^{\Omega_{+}}&=0 \, , & C_{\omega \Omega_{-}}^{\Omega_{-}}&=0 \, , & C_{\omega \Omega_{\pm}}^{\mathds{1}}&=0 \, ,  & C_{\Omega_{+}\Omega_{-}}^{\mathds{1}}&=0 \, .
\end{align*}
Moreover, we normalise the non-zero two-point functions to be the same employed in the $\mathrm{Od}(3)$ algebra (see \Cref{app:Odakealgebras})
\begin{equation} \label{eq: normalisation}
    C_{\omega \omega}^{\mathds{1}} = -3  \, , \quad  C_{\Omega_{\pm} \Omega_{\pm}}^{\mathds{1}} = 4  \, .
\end{equation}
Additional symmetries can be employed to further constrain the couplings: adopting the normalisation \eqref{eq: normalisation} 
\begin{align*}
    C_{\omega \Omega_{\pm}}^{\omega}&= \frac{4  C_{\omega \omega}^{\Omega_{\pm}}}{3} \, , \quad C_{\Omega_{+} \Omega_{-}}^{\Omega_{+}}= C_{\Omega_{+} \Omega_{+}}^{\Omega_{-}} \, , \quad C_{\Omega_{+} \Omega_{-}}^{\Omega_{-}}= C_{\Omega_{-} \Omega_{-}}^{\Omega_{+}}\, , \\[4pt] &\hspace{1.6em} C_{\omega \Omega_{+}}^{\Omega_{-}}=-\frac{3 C_{\Omega_{+} \Omega_{-}}^{\omega}}{4}  \, , \quad 
     C_{\omega \Omega_{-}}^{\Omega_{+}}=\frac{3 C_{\Omega_{+} \Omega_{-}}^{\omega}}{4}   \, . 
\end{align*}
The most general Ansatz satisfying the symmetry constraints listed above can be found in \Cref{app: su3}.
Imposing the $\mathcal{SW}$-algebra consistency conditions onto this Ansatz leads to a large collection of non-linear constraints: we leave a full study of these equations for the future and work at perturbative level instead.

As anticipated, we will construct an \emph{infinitesimal deformation} of the $\mathrm{Od}(3)$ algebra. 
The Virasoro algebra and the OPEs with the primary generators $\mathcal{T} \times \cj^{\omega, \Omega_{\pm}}$ are assumed to take the usual forms \eqref{eq: TTOPE} and \eqref{eq:SC-HPSUSYOPE}, respectively. Consider now a generic coupling $C_{ij}^k$ appearing explicitly in the Ansatz written above, and the same coupling in the $\mathrm{Od}(3)$ algebra $C_{\text{Od},ij}^{\ \ \ k}$, where special holonomy is achieved. 
We introduce an infinitesimal parameter $\veps$: assuming that $C_{ij}^k$ is an analytic function of $\veps$, we write 
\begin{equation}
    C_{ij}^{k} =C_{\text{Od},ij}^{\ \ \ k}+\veps \, \delta C_{ij}^{k} + \frac{1}{2}\veps^2 \delta \delta C_{ij}^{k}+O(\veps^3) \, . \label{eq: deformed C}
\end{equation}
Plugging the expansion \eqref{eq: deformed C} into the OPEs of the Ansatz listed above, we produce the infinitesimal deformation of the $\mathrm{Od}(3)$ algebra that we call $\mathrm{Od}^{\veps}(3)$. 
The algebra $\mathrm{Od}^{\veps}(3)$ will be considered consistent if it satisfies the $\mathcal{SW}$-algebra consistency conditions---for example, associativity---up to terms of order $O(\veps^3)$. 

We present the final result. Interestingly, every coupling turns out to be either equal to zero (up to terms of order $O(\veps^3)$) or a function of the following two deformed couplings\footnote{The couplings $\delta\delta C_{\Omega_{+}\Omega_{-}}^{\omega\Omega_{\pm}}$ are not constrained either: however, they do not carry any physical meaning, as we argue in the following.} 
\begin{equation*}
    C_{\omega\omega}^{\Omega_{+}}=\veps \, \delta C_{\omega\omega}^{\Omega_{+}} + \frac{1}{2}\veps^2 \delta \delta C_{\omega\omega}^{\Omega_{+}}+O(\veps^3) \, , \quad C_{\omega\omega}^{\Omega_{-}}=\veps \, \delta C_{\omega\omega}^{\Omega_{-}} + \frac{1}{2}\veps^2 \delta \delta C_{\omega\omega}^{\Omega_{-}}+O(\veps^3) \, .
\end{equation*}
This is illustrated by the central charge, which exhibits a correction of order $O(\veps^2)$
\begin{equation} \label{eq: cc}
    c= 9-\frac{3}{4} \veps ^2 \left(v_{+}^{2}+v_{-}^{2}\right)+O(\veps^3) \, ,
\end{equation}
where we have introduced the following lighter notation to unclutter the formulas:
\begin{equation*}
    \delta C_{\omega\omega}^{\Omega_{+}}= v_{+} \, , \qquad  \delta \delta C_{\omega\omega}^{\Omega_{+}}=u_{+} \, , \qquad \delta C_{\omega\omega}^{\Omega_{-}}= v_{-} \, , \qquad \delta \delta C_{\omega\omega}^{\Omega_{-}}=u_{-} \, .
\end{equation*}
In analogy with the previously shown $\mathcal{SW}$-algebras, we expect $c$ to be a free parameter (subject to unitarity bounds) of the full $\mathcal{SW}(\frac{3}{2},\frac{3}{2},\frac{3}{2}, 1)$ family. Although the deformation \eqref{eq: cc} indeed shows that $c$ is allowed to take different values, it is worth stressing again that $\mathrm{Od}^{\veps}(3)$ is only valid in a small neighbourhood around $c=9$. In particular, equation \eqref{eq: cc} fails to capture the existence of the known $c=1$ \emph{locus} of the $\mathcal{SW}(\frac{3}{2},\frac{3}{2},\frac{3}{2}, 1)$ family \cite{Fiset:2020lmg}.

We list the remaining couplings, fixed by associativity up to order $O(\varepsilon^3)$
{\footnotesize
\allowdisplaybreaks
\begin{align}
    C_{\omega\omega}^{\ct}&=-2 -\frac{1}{6} \veps^2 \left(v_{+}^2+v_{-}^2\right)+O(\veps^3) \, ,  &
    C_{\omega\Omega_{+}}^{\ct}&=O(\veps^3) \, , \nn \\[4pt]
    C_{\omega\Omega_{+}}^{\omega\omega}&=-\frac{1}{2} C_{\omega\omega}^{\Omega_{-}}+O(\veps^3) \, , &
    C_{\omega\Omega_{-}}^{\ct}&=O(\veps^3) \, , \nn \\[4pt]
    C_{\omega\Omega_{-}}^{\omega\omega}&=\frac{1}{2} C_{\omega\omega}^{\Omega_{+}}+O(\veps^3)  \, , &
    C_{\Omega_{\pm}\Omega_{\pm}}^{\ct}&=4+\frac{1}{3}\veps^2(v_{+}^2+v_{-}^2)+O(\veps^3) \, , \nn \\[4pt]
    C_{\Omega_{+}\Omega_{+}}^{\Omega_{+}}&=-\frac{3}{2} C_{\omega\omega}^{\Omega_{+}}+O(\veps^3)  \, , & \label{eq: su3coup} 
    C_{\Omega_{-}\Omega_{-}}^{\Omega_{+}}&=-\frac{1}{2} C_{\omega\omega}^{\Omega_{+}}+O(\veps^3)   \, , \nn \\[4pt]
    C_{\Omega_{+}\Omega_{+}}^{\Omega_{-}}&=-\frac{1}{2} C_{\omega\omega}^{\Omega_{-}}+O(\veps^3)   \, , &
    C_{\Omega_{-}\Omega_{-}}^{\Omega_{-}}&=-\frac{3}{2} C_{\omega\omega}^{\Omega_{-}}+O(\veps^3)   \, ,  \\[4pt]
    C_{\Omega_{+}\Omega_{+}}^{\omega \omega}&=-2+\frac{1}{48}\veps^2\left(-v_{+}^2+3v_{-}^{2}\right)+O(\veps^3) \, , &
    C_{\Omega_{-}\Omega_{-}}^{\omega \omega}&=-2+\frac{1}{48}\veps^2\left(3v_{+}^2-v_{-}^{2}\right)+O(\veps^3) \, , \nn \\[4pt]
    C_{\Omega_{+}\Omega_{-}}^{\omega}&=4-\frac{1}{6}\veps^2\left(v_{+}^2 +v_{-}^2\right)+O(\veps^3) \, ,  &
    C_{\Omega_{+}\Omega_{-}}^{\ct}&=O(\veps^3) \, , \nn \\[4pt]
    C_{\Omega_{+}\Omega_{-}}^{\omega \omega}&=-\frac{1}{12}\veps^2 v_{+}v_{-}+O(\veps^3) \, , &
    C_{\Omega_{+}\Omega_{-}}^{\ct \omega}&=4+\frac{13}{24}\veps^2 \left( v_{+}^2+v_{-}^2\right)+O(\veps^3) \, , \nn \\[4pt]
    C_{\Omega_{+}\Omega_{-}}^{\omega\Omega_{+}}&=-2\veps v_{+}+\frac{1}{2}\veps^2 \delta\delta C_{\Omega_{+}\Omega_{-}}^{\omega\Omega_{+}} +O(\veps^3) \, , &
    C_{\Omega_{+}\Omega_{-}}^{\omega\Omega_{-}}&=-2\veps v_{-}+\frac{1}{2}\veps^2 \delta\delta C_{\Omega_{+}\Omega_{-}}^{\omega\Omega_{-}} +O(\veps^3) \, . \nn
\end{align}
}%
It can be checked that, with this choice of couplings, the $\mathrm{Od}^{\veps}(3)$ algebra closes up to the ideals generated by two null fields $\mathfrak{N}_1$ and $\mathfrak{N}_2$, which can be expressed as expansions in powers of $\veps$ as follows
{\allowdisplaybreaks
\begin{align*}
    \mathfrak{N}_1&= \mathcal{N}(\cj^\omega\cj^{\Omega_{-}})-\frac{\veps}{6} v_{-}\, \mathcal{N}( \ct \cj^{\omega}) -\frac{\veps^2}{12}  u_{-}\, \mathcal{N}(\ct \cj^{\omega}) +O\left(\veps ^3\right)\, ,   \\[8pt]
     \mathfrak{N}_2&= \mathcal{N}(\cj^\omega\cj^{\Omega_{+}})-\frac{\veps}{6} v_{+}\, \mathcal{N}( \ct \cj^{\omega}) -\frac{\veps^2}{12} u_{+}\, \mathcal{N}( \ct \cj^{\omega})+O\left(\veps ^3\right)  \, . 
\end{align*}}%
The null fields $\mathfrak{N}_1$ and $\mathfrak{N}_2$ satisfy the null field conditions \eqref{eq: null1} and \eqref{eq: null2} up to terms of order $O(\veps^3)$. In the limit $\veps \to 0$, they reduce to the Odake null fields \eqref{eq: odake null}
\begin{equation} \label{eq: nulllimit}
    \mathfrak{N}_1 \overset{\veps \to 0}{\longrightarrow} \mathfrak{N}_{\mathrm{Od},1} \, , \quad  \mathfrak{N}_2 \overset{\veps \to 0}{\longrightarrow} \mathfrak{N}_{\mathrm{Od},2} \, ,
\end{equation}
Following the statement \eqref{eq: nulllimit}, we realise that the parameters $\delta\delta C_{\Omega_{+}\Omega_{-}}^{\Omega_{\pm}\omega}$ in the Ansatz OPE $\cj^{\Omega_{+}} \times \cj^{\Omega_{-}}$ multiply null fields up to order $O(\veps^2)$. Thus, they are arbitrary and we can choose to set them to zero.
This completely fixes the entire $\mathrm{Od}^{\veps}(3)$ up to the free parameters $v_{\pm}, u_{\pm}$.

This infinitesimal deformation significantly alters the properties of the algebra. For instance, in the $\mathrm{Od}(3)$ algebra, the field $\cj^\omega$ is the $\mathrm{U}(1)$ supercurrent that endows the algebra with $\mathcal{N}=2$ supersymmetry. When the parameter $\varepsilon$ is turned on, the OPEs of $\cj^\omega$ are deformed away from those of a supersymmetry generator. Thus, we conclude that the deformation we have described breaks supersymmetry down to $\mathcal{N}=1$.

\paragraph{Geometrical interpretation of the couplings.}
{\begin{table}
\renewcommand{\arraystretch}{2}%
    \centering
    \begin{tabular}{|c|c||c|c|c||c|c|c|}
    \hline
       $\cj_{\text{cl.}}^1$ & $\cj_{\text{cl.}}^2$ & $\cj_{\text{cl.}}^{U}$ & $\cj_{\text{cl.}}^{N}$ & $\cj_{\text{cl.}}^{V}$ & $c_{U}$ & $c_{N}$ & $c_{V}$ \\
        \hline
        $\cj_{\text{cl.}}^{\omega}$&$\cj_{\text{cl.}}^{\omega}$ & - & $\cj_{\text{cl.}}^{\Omega_{+}} \, , \, \cj_{\text{cl.}}^{\Omega_{-}}$ & $-\mathds{1}$ & - & $4
        \, W_{0}^{+}\ell_s \, , \, 4  
        \, W_{0}^{-}\ell_s$ & $6$  \\
       $\cj_{\text{cl.}}^{\omega}$&$\cj_{\text{cl.}}^{\Omega_{\pm}}$  & $\cj_{\text{cl.}}^{\Omega_{\mp}}$ & $ -\cj_{\text{cl.}}^{\omega}\cj_{\text{cl.}}^{\omega}$ & - & $\pm 3$ & $\pm 2 \, W_{0}^{\mp}\ell_s$ & - \\
        $\cj_{\text{cl.}}^{\Omega_{\pm}}$&$\cj_{\text{cl.}}^{\Omega_{\pm}}$ & $ -\cj_{\text{cl.}}^{\omega}\cj_{\text{cl.}}^{\omega}$ & - & - & $2$ 
        & - & -  \\
        $\cj_{\text{cl.}}^{\Omega_{+}}$&$\cj_{\text{cl.}}^{\Omega_{-}}$ & - & - & $\cj_{\text{cl.}}^{ \omega}$ & - & - & $8$  
        \\
        \hline
    \end{tabular}
    \caption{\emph{Currents appearing in the classical OPE $\cj^{1}_{\mathrm{cl.}} \times \cj^2_{\mathrm{cl.}}$ for an $\mathrm{SU}(3)$-structure. Geometrical information is encoded in the scalar torsion classes $W_{0}^{+}$ and $W_{0}^{-}$.
    }}
    \label{tab:SU(3)}
\end{table}}
In Table \ref{tab:SU(3)} we report the data needed to build the classical OPEs from the $\mathrm{SU}(3)$-structure. The classical OPEs read
{\allowdisplaybreaks
\begin{align*}
    \mathcal{J}_{\text{cl.}}^\omega(Z_1)\mathcal{J}_{\text{cl.}}^\omega(Z_2) &\sim  4 \, W_{0}^{+}\,\ell_s \frac{\theta_{12}}{Z_{12}}\mathcal{J}_{\text{cl.}}^{\Omega_{+}}+ 4\,  W_{0}^{-}\,\ell_s \frac{\theta_{12}}{Z_{12}}\mathcal{J}_{\text{cl.}}^{\Omega_{-}}
    - 2 \, \frac{ \theta_{12}}{Z_{12}} \,\ct_{\text{cl.}}^{\phantom{\omega}}+\cdots  \, ,  \\[8pt]
    \mathcal{J}_{\text{cl.}}^{\omega}(Z_1)\mathcal{J}_{\text{cl.}}^{\Omega_{\pm}}(Z_2) &\sim \mp 3 \left( \frac{1}{Z_{12}} \, \mathcal{J}_{\text{cl.}}^{\Omega_{\mp}} + \frac{\theta_{12}}{3 \, Z_{12}}D\mathcal{J}_{\text{cl.}}^{\Omega_{\mp}}\right) \mp 2\,  W_{0}^{\mp}\,\ell_s \frac{\theta_{12}}{Z_{12}}\mathcal{J}_{\text{cl.}}^{\omega }\mathcal{J}_{\text{cl.}}^{\omega}+\cdots
     \,  ,  \\[8pt]
    \mathcal{J}_{\text{cl.}}^{\Omega_{\pm}}(Z_1)\mathcal{J}_{\text{cl.}}^{\Omega_{\pm}}(Z_2) &\sim -2 \left(\frac{1}{Z_{12}}\mathcal{J}_{\text{cl.}}^{\omega}\mathcal{J}_{\text{cl.}}^{\omega}+  \frac{\theta_{12}}{2 \,  Z_{12}}D\left( \mathcal{J}_{\text{cl.}}^{\omega}\mathcal{J}_{\text{cl.}}^{\omega}\right)\right)+\cdots  \, , \\[8pt]
    \mathcal{J}_{\text{cl.}}^{\Omega_{+}}(Z_1)\mathcal{J}_{\text{cl.}}^{\Omega_{-}}(Z_2) &\sim
    4 \, \frac{ \theta_{12}}{Z_{12}} \,\ct_{\text{cl.}}^{\phantom{\omega}} \mathcal{J}_{\text{cl.}}^{\omega}+\cdots  \, .
\end{align*}}
We can identify the products of the classical currents $\mathcal{J}_{\text{cl.}}^{\omega}\mathcal{J}_{\text{cl.}}^{\omega}$ and $\ct_{\text{cl.}}^{\phantom{\omega}} \mathcal{J}_{\text{cl.}}^{\omega}$ as the classical limit of the quasi-primary normal ordered operators $\cn(\cj^{\omega}\cj^{\omega})$ and $\cn(\ct \cj^{\omega})$, respectively. By comparing the couplings of the classical OPEs with the list of couplings \eqref{eq: su3coup}, we immediately realise that the comparison is effective only up to order $O(\veps)$. Then, we can extract the prediction
\begin{equation} \label{eq: su3tor}
    C_{\omega \omega}^{\Omega_{\pm}}=\veps \, v_{\pm}+O(\veps^2)=4\,  W_{0}^{\pm} \ell_s+O(\ell_s^2) \, .
\end{equation}
From the equation \eqref{eq: su3tor}, we infer that $W_0$ is of order $O(\veps)$, so the worldsheet theory described by the algebra $\mathrm{Od^{\veps}(3)}$ must be associated with an \emph{infinitesimal torsion}, ruled by the same parameter $\veps$. This scenario can be realised as follows: we introduce two fields $h$ and $b$ to slightly perturb the background metric and Kalb--Ramond field in the $\sigma$-model \eqref{eq:action}
{\small
\begingroup
\allowdisplaybreaks
\begin{equation} \label{eq: sigma def}
    S[X,\Lambda] = \int_{\mathbb\Sigma}\frac{\de^{2|1}\zeta}{2 \, \ell_s^2}\,  \left[\left(G_{ij}(X)+h_{ij}(\veps,X)\right)+\left(B_{ij}(X) + b_{ij}(\veps,X)\right)\right]\bar\partial X^{i} DX^{j} + \cdots \, ,
\end{equation}
\endgroup}%
where $B$ is a closed form and $b$ is an analytic function of $\veps$, such that
\begin{equation*}
    b(\veps, X)=\veps \, \delta b(X) + \frac{1}{2}\veps^2 \, \delta \delta b(X) + O(\veps^3) \, .
\end{equation*}
The deformation presented in the equation \eqref{eq: sigma def} can be physically interpreted as perturbing the original string background with a coherent state of gravitons, Kalb--Ramond particles \emph{etc.}; from the geometrical point of view, it corresponds to a slight deformation of the Calabi--Yau three-fold metric---which is then compensated by a deformation of the other fields to ensure the cancellation of the Weyl anomaly.
The Bianchi identity then reads
\begin{equation*}
   H = \de \left(B+b(\veps) \right) + \frac{\ell_s^2}{8\pi} \big(\mathrm{CS}_{3}(A)- \mathrm{CS}_{3}(\Theta)\big)+O(\ell_s^3) =\veps \, \de \, \delta b(X) +O(\veps^2, \ell_s^2) \, .
\end{equation*}
We can then define an adimensional $w_0$ scalar torsion class for the ``small'' torsion $ \de\,  \delta b(X)$ as follows\footnote{Perhaps readers familiar with the Hull--Strominger system \cite{Hull:1986kz,Strominger:1986uh} are surprised by the presence of a non-zero scalar torsion class $W_0$. The reason for this is that we are not assuming any particular Ansatz for the background, so our discussion applies to more general compactifications where $W_0\neq 0$. Examples of these include backgrounds where the spacetime is a domain wall \cite{Lukas:2010mf, Gray:2012md, Klaput:2012vv} or where a gaugino condensate is present \cite{LopesCardoso:2003sp,Frey:2005zz,Manousselis:2005xa,Lechtenfeld:2010dr}.}
\begin{equation} \label{eq: minitorsion}
    W_{0}^{\pm}=\veps \,  w_{0}^{\pm}+O(\veps^2, \ell_s^2) \, ,
\end{equation}
and plugging \eqref{eq: minitorsion} in the prediction \eqref{eq: su3tor}, we obtain
\begin{equation*}
    v^{\pm}=4 \, w_0^{\pm} \ell_s+O(\ell_s^2) \, .
\end{equation*}
This outcome is interesting since all the couplings \eqref{eq: su3coup} only depend on the free parameters $v_{\pm}$, without any reference to the parameters $u_{\pm}$ (which might appear at higher orders in $\veps$). For example, the central charge of the special holonomy algebra is corrected by the infinitesimal torsion as follows
\begin{equation*}
    c=9-12\, 
(\veps \, \ell_s)^2 ||w_{0}||^2+O(\veps^3, \ell_s^3) \, ,
\end{equation*}
where a new \emph{effective} string length scale $\ell^{\veps}_s=\veps \, \ell_s$ appears. By studying the list of couplings \eqref{eq: su3coup}, we realise that the $\mathrm{Od}^{\veps}(3)$ algebra should not be interpreted as an $\ell_s$ expansion of the $\mathcal{SW}(\frac{3}{2}, \frac{3}{2}, \frac{3}{2}, 1)$ algebra, but as an $\ell^{\veps}_s$ expansion around the special holonomy \emph{locus}. 

To conclude, we note that the limit $H \to 0$ implies $w_{0}^{\pm} \to 0$ in the case we are considering: the algebra flows back to the $\mathrm{Od}(3)$ algebra, and the geometry to a special holonomy manifold.

\section{Conclusions and outlook}
\label{conc: W}

In this paper we have explored the possibility of identifying the worldsheet $\mathcal{SW}$-algebras associated with string backgrounds in the presence of a non-zero NS flux $H$, focusing on backgrounds that include a manifold equipped with a $\mathrm{G}$-structure with torsion.

We have first presented families of $\mathcal{SW}$-algebras constructed from first principles that were candidates to describe each of the different $\mathrm{G}$-structure backgrounds. Next, we have considered those same backgrounds as targets of a classical $\sigma$-model and we have used the $\mathcal{W}$-symmetries originating from the $\mathrm{G}$-structures to obtain a classical limit of the OPEs. Finally, we have combined both perspectives to identify the relevant $\mathcal{SW}$-algebras and provide a geometrical interpretation of their couplings.

Our procedure recovers all well-known special holonomy algebras. In addition, for $\mathrm{G}_2$ and $\mathrm{SU}(3)$-structures we have found $\mathcal{SW}$-algebras that are explicitly parametrised by the scalar torsion classes. In the $\mathrm{G}_2$ case, we have focused on the explicit example given by the algebra studied by Fiset and Gaberdiel \cite{Fiset:2021azq}, and we correctly connected the level $k$ of their WZW worldsheet description to the scalar torsion class $\tau_0$ at the lowest order in $\ell_s$. In the $\mathrm{SU}(3)$ case, we constructed a perturbative description of the family of algebras around the special holonomy \emph{locus} $\mathrm{Od}(3)$, with central charge $c=9$. In this case, we also connected the OPE coefficients turned on by the deformation with the scalar torsion classes $W_0^{\pm}$.

Both of these examples are significative since they display the primary role of scalar torsion classes in the deformation of special holonomy algebras at the lowest order in the string length scale. Moreover, the derivation of the classical OPE is general and it only requires the geometrical information encoded into the $\mathrm{G}$-structure of the string background, so it is suited for physically interpreting not only well known string-related $\mathcal{SW}$-algebras, but also more exotic ones.

\smallskip

Our results open up several interesting future directions. 
First of all, it should be noted that, since we have restricted ourselves to the chiral holomorphic sector---and the gauge bundle does not play a role in our analysis---everything should follow analogously for $(1,1)$ non-linear $\sigma$-models and thus apply equally well to pure NS-NS backgrounds of type II superstrings. It would be interesting to verify this explicitly.

In addition, we would like to complete the study of $\mathrm{SU}(3)$-structures presented here by fully constructing the $\mathcal{SW}(\frac{3}{2},\frac{3}{2}, \frac{3}{2},1)$ algebra and finding the exact correspondence between the torsion classes and the couplings. This would be especially relevant to better understand compactifications on non-K\"ahler manifolds, see for example \cite{Hull:1986kz,Strominger:1986uh,Becker:2003yv,Becker:2003sh} for some foundational works in the heterotic setting. It would also be extremely interesting to search for a connection between the $\mathrm{Od}^{\veps}(3)$ algebra and the literature regarding tree-level marginal deformations of (2,0) non-linear $\sigma$-models \cite{Melnikov:2011ez}; a similar path could be followed by studying $\mathrm{Spin}(7)$ and $\mathrm{G}_2$-structures \cite{Fiset:2017auc} and the associated tree-level marginal deformations of the worldsheet action.

Another natural follow-up would be trying to pinpoint the role of torsion classes beyond the scalar ones. We conjecture that they should appear in loop corrections at higher orders in $\ell_s$, so a perturbative computation could be performed to better understand the precise contribution of these additional torsion classes.

Moreover, the techniques we use in this paper could be applied to backgrounds with additional geometric structures, such as the \emph{almost contact structure} always present on $\mathrm{G}_2$-manifolds \cite{delaOssa:2021cgd}. The contact form $\sigma$ would correspond to an additional chiral current of weight $h_{\sigma}=\frac{1}{2}$ in the worldsheet CFT, and in favourable circumstances it should be possible to recognise a subalgebra corresponding to the $\mathrm{SU}(3)$-structure induced on the manifold by the almost contact structure.

Another possible direction involves the gauge bundle of the $(1,0)$ non-linear $\sigma$-model: very little is known concerning the relationship between the gauge bundle and the worldsheet theory \cite{delaOssa:2018azc, Grimanellis:2023nav, Papadopoulos:2023exe}. One could study the $\mathcal{W}$-symmetry commutators applied to the superfields $\Lambda^{\alpha}$ and explore the general constraints that the existence of a $\mathrm{G}$-structure on the background imposes on the gauge bundle---analogously to our study for the superfields $X^i$.

Mirror symmetry manifests in the worldsheet as an automorphism of the underlying $\mathcal{SW}$-algebra \cite{Lerche:1989uy,Gaberdiel:2004vx,Chuang:2004th}. Studying and classifying automorphisms of the algebras $\FGseven_k$ and $\mathrm{Od}^{\veps}(3)$ corresponding to string backgrounds with nonzero NS flux would shed some light on mirror symmetry in this setting. This would complement the spacetime perspective that is already present in the literature, particularly in the case of $\mathrm{SU}(3)$-structures---see for example \cite{Gurrieri:2002wz,Fidanza:2003zi,Tomasiello:2005bp,Grana:2006hr}.

In recent years, a lot of progress has been made in the context of the AdS/CFT correspondence to relate the features of the boundary CFTs and those of the worldsheet theory in the presence of NS flux \cite{Gaberdiel:2010pz, Eberhardt:2017pty, Eberhardt:2018ouy, Eberhardt:2019ywk, Gaberdiel:2021jrv}. It would be interesting to study the possible holographic applications of the $\mathcal{SW}$-algebras found in this work and to expand the results already available for WZW models to more general non-linear $\sigma$-models.

Finally, it would also be illustrative to explore the similarities between our strategy and more mathematical approaches such as the chiral de Rham complex \cite{Malikov:1998dw}, which provides a way to quantise the classical $\sigma$-model \cite{Ekstrand:2009zd,Ekstrand:2010wu} and has recently been explored for backgrounds with nonzero NS flux \cite{Alvarez-Consul:2020hbl, Alvarez-Consul:2023zon}.

\section*{Acknowledgements}
 It is a pleasure to thank Fernando Alday, Philip Candelas, Jos\'e Figueroa-O'Farrill, Marc-Antoine Fiset, Mario Garcia-Fernandez, Alessio Miscioscia, Sujay Nair, Georgios Papadopoulos and Christopher Raymond for very useful discussions at different stages of this work. We acknowledge Kris Thielemans and the Mathematica package \texttt{OPEdefs}. MG's work is funded by the Deutsche Forschungsgemeinschaft
(DFG, German Research Foundation) under Germany’s Excellence Strategy, EXC 2121 “Quantum Universe,” 390833306. EM's work is funded by the Deutsche Forschungsgemeinschaft (DFG, German Research Foundation) – SFB 1624 – “Higher structures, moduli spaces and integrability” – 506632645. XD and MG thank the organisers of \emph{The Geometry of Moduli Spaces in String Theory} scientific programme and Johanna Knapp for hospitality at the Matrix Institute and Melbourne University during later stages of this work.
For the purpose of open access, the authors have applied a CC-BY public copyright license to any Author Accepted Manuscript (AAM) version arising from this submission.

\appendix

\section{Further details on null fields} 
\label{app:nullfields}
In this Appendix we provide further details regarding null fields, introduced in \Cref{sec: Walg}. In particular, we give a more formal definition and we discuss how to determine whether a given field is null employing only the generators of the $\mathcal{SW}$-algebra. 

Let us start by considering the two-dimensional CFT whose symmetries are encoded in a given $\mathcal{SW}$-algebra. We indicate by $\mathcal{F}$ the holomorphic sector of the operator content of such algebra. Then, we can define the following bilinear map, taking a pair of operators into a third one
\begin{equation} 
\label{eq: bilfun}
    [\cdot \ \cdot ]_{p}: \mathcal{F} \otimes \mathcal{F} \longrightarrow \mathcal{F} \, , \quad p \in \mathbb{Z} 
\end{equation}
such that for each pair $A, B \in \mathcal{F}$ there exists a $\widehat{p} \in \mathbb{Z}$ for which
\begin{equation*}
     [\cdot  \ \cdot ]_{p>\widehat{p}}: A \otimes B \longmapsto 0 \, .
\end{equation*}
For any triplet of operators $A,B,C \in \mathcal{F}$, the bilinear map satisfies three properties \cite{Thielemans:1994er}:
\begin{itemize}[label=$\diamond$]
    \item unity:
    \begin{equation} \label{eq: unity}
        [\mathds{1} A]_{0}=A \, , \quad [\mathds{1} A]_{r \neq 0}=0 \,;
    \end{equation}
    \item commutation: 
    \begin{equation} \label{eq: comm}
        [BA]_r=(-1)^{|A||B|} \sum_{s \geq r} \frac{(-1)^{s}}{(s-r)!}\del^{s-r}[AB]_s \, , \quad \forall r \in \mathbb{Z} \,;
    \end{equation}
    \item associativity: 
    \begin{equation} \label{eq: associ}
        [A[BC]_{r}]_{s}=(-1)^{|A||B|}[B[AC]_{s}]_{r}+\sum_{t>0} \binom{s-1}{t-1}[[AB]_{t}C]_{r+s-t} \, , \quad \forall r,s \in \mathbb{Z} \, .
    \end{equation}
\end{itemize}
The bilinear function \eqref{eq: bilfun} is constructed in such a way that given two operators $A$ and $B$,  $[AB]_p$ returns the $p$-th pole in the OPE between $A$ and $B$. The properties \eqref{eq: unity}, \eqref{eq: comm} and \eqref{eq: associ} encode the OPE consistency conditions. 

As already anticipated in \Cref{sec: Walg}, in some instances it is possible to relax the OPE consistency conditions \eqref{eq: unity}, \eqref{eq: comm} and \eqref{eq: associ}. Since correlation functions represent the observables of the theory, the consistency conditions should always be considered to hold up to null fields. Any correlator containing a null field insertion vanishes identically. Any null field $\mathfrak{N}\in \mathcal{F}$ generates an ideal $\mathcal{I}_{\mathfrak{N}} \in \mathcal{F}$, where every element of $\mathcal{I}_{\mathfrak{N}}$ is a null field. Physical statements are meant to hold up to these ideals. By employing the bilinear function \eqref{eq: bilfun}, we provide a necessary condition for a field $\mathfrak{N}$ to be null 
\begin{equation} \label{eq: nullfieldcond}
    [\mathfrak{N}A]_{r} \neq \mathds{1}  \quad \forall A \in \mathcal{F} \, , \forall r \in \mathbb{Z} \, .
\end{equation}
Usually, this is also regarded as a sufficient condition if all the operators apart from the identity have strictly positive conformal dimensions \cite{Thielemans:1994er}. If we consider a $\mathcal{SW}$-algebra finitely generated by the set of currents
\begin{equation*}
     \cj_{h_1}, \dots, \cj_{h_n} \, ,
\end{equation*}
we argue that testing the condition \eqref{eq: nullfieldcond} on the set of generators is sufficient. By definition, the rest of the operators of the algebra are constructed out of the generators employing addition, multiplication by scalar, action of $\del$ and normal ordering:
\begin{itemize}[label=$\diamond$, leftmargin=*]
    \item \emph{Addition and multiplication by scalar}: by linearity 
    \begin{equation} \label{eq: add}
        [\mathfrak{N}\, (\alpha \cj_{h_{i}}+ \beta \cj_{h_j})]_{r}= \alpha [\mathfrak{N} \cj_{h_{i}}]_{r}+ \beta [\mathfrak{N} \cj_{h_{j}}]_{r} \, ,
    \end{equation}
    where $\alpha$ and $\beta$ are real numbers. Since by hypothesis both terms in the right-hand side of the equation \eqref{eq: add} are null fields, we conclude that the left-hand side is a null field as well.
    \item \emph{Action of $\del$}: using the properties of the OPE \cite{Thielemans:1994er}
    \begin{equation*}
        [\mathfrak{N} \, \del \cj_{h_{i}}]_{r+1}=r [\mathfrak{N} \cj_{h_{i}}]_{r}+\del  [\mathfrak{N} \cj_{h_{i}}]_{r+1} \, .
    \end{equation*}
    The right-hand side is by hypothesis composed of a null field and the derivative of a null field, which is still a null field. Hence, the left-hand side is a null field as well.
     \item \emph{Normal ordering}: we consider the generalisation of Wick theorem for interacting fields \cite{DiFrancesco:1997nk}
     \begin{multline*}
         \mathfrak{N}(z_1) N(\cj_{h_{i}} \cj_{h_j})(z_2)=\frac{1}{2\pi i}\oint_{z_2}\frac{\de z_3}{z_3-z_2} \sum_{r>0} \bigg(\frac{[\mathfrak{N} \cj_{h_{i}} ]_{r}(z_3)  \cj_{h_{j}}(z_2)}{(z_1-z_3)^r} \\ +(-1)^{|\mathfrak{N}||\cj_{h_{i}}|}\frac{\cj_{h_{i}}(z_3) [\mathfrak{N} \cj_{h_{j}} ]_{r}(z_2)  }{(z_1-z_2)^r} \bigg)  \, .
     \end{multline*}
     Provided that 
     \begin{equation*}
        [[\mathfrak{N} \cj_{h_{i}} ]_{r}  \cj_{h_{j}}]_s \neq \mathds{1} \quad \forall r ,s >0 \, ,
     \end{equation*}
     the generalised Wick theorem ensures that the OPE $\mathfrak{N} \times N(\cj_{h_{i}} \cj_{h_j})$ contains only null fields. 
\end{itemize}
Iterating these four operations and checks leads to proving the condition \eqref{eq: nullfieldcond}.
In conclusion, the condition \eqref{eq: nullfieldcond}, which is difficult to test since the operator content is infinite, can be replaced by the conditions 
\begin{align}
     [\mathfrak{N} \cj_{h_{i}}]_{r} &\neq \mathds{1}  \quad \forall r > 0 \, , \label{eq: null1}  \\
      [[\mathfrak{N} \cj_{h_{i}} ]_{r}  \cj_{h_{j}}]_s &\neq \mathds{1} \quad \forall r ,s >0 \, , \label{eq: null2}
\end{align}
where $ \cj_{h_{i}},  \cj_{h_{j}}$ are generators of the $\mathcal{SW}$-algebra.

\section{\texorpdfstring{$\mathcal{SW}$}{SW}-algebras and their OPEs} \label{app: spechol}
This Appendix is a \emph{compendium} of explicit OPEs defining the $\mathcal{SW}$-algebras relevant for this work and already discussed in the literature. In addition, in Section \ref{app: su3} we provide an Ansatz for the most general algebra associated with an $\mathrm{SU}(3)$-structure with torsion. All the algebras are named after the authors of the works where they first appeared, up to our knowledge. 

We will only provide the non-trivial OPEs between the extra currents $\mathcal{J}^{\Phi}$. Each of the following $\mathcal{SW}$-algebras should be completed with the OPEs
\begin{align*} 
\ct(Z_1)\ct(Z_2)&\sim \frac{c}{6}\frac{1}{Z_{12}^3} +\frac{3}{2}\frac{\theta_{12}}{Z_{12}^2}\ct(Z_2) +\frac{1}{2 \, Z_{12}}\,D \ct(Z_2) +\frac{\theta_{12}}{Z_{12}}\,\partial \ct(Z_2) +\cdots \, ,\\[6pt]
\mathcal{T}(Z_1)\mathcal{J}^{\Phi}(Z_2)&\sim  \frac{p}{2}\frac{\theta_{12}}{Z_{12}^2}\mathcal{J}^{\Phi}(Z_2) +\frac{1}{2 \, Z_{12}}D \cj^{\Phi}(Z_2) +\frac{\theta_{12}}{Z_{12}}\partial \cj^{\Phi}(Z_2)+\cdots \, , 
\end{align*}
where $p$ represents the degree of the characteristic form associated with the current $\mathcal{J}^{\Phi}$ (the association is explicitly stated in Section \ref{sec: connection}). All the extra currents are considered to be primary operators. Finally, in some subcases the central charge $c$ will not be a free parameter, being instead associated with a special \emph{locus} in the corresponding family of algebras. If this is the case, its value will be specified.

\subsection{\texorpdfstring{$\mathrm{Spin}(7)$ $\mathcal{SW}$}{Spin(7) SW}-algebras}
We call $\mathrm{Spin}(7)$ $\mathcal{SW}$-algebras those algebras enhancing the super Virasoro algebra by means of a primary $\mathcal{J}^{\Psi}$ of weight $h_{\Psi}=2$.  
\subsubsection{Figueroa-O'Farrill--Schrans algebra \texorpdfstring{$\mathrm{FS}$}{FS}}
\label{app:FSSpin7OPEs}
The full family of $\mathrm{Spin}(7)$ algebras was derived for the first time by Figueroa-O'Farrill and Schrans in \cite{Figueroa-OFarrill:1990mzn}. It only depends on the central charge $c$. 
The normalisation of the $\mathcal{J}^{\Psi}$ current is encoded in the positive parameter $\mu$.
{\small
\begingroup
\allowdisplaybreaks
\begin{align*} 
    \mathcal{J}^{\Psi}(Z_1)\mathcal{J}^{\Psi}(Z_2) &\sim \frac{\mu}{Z_{12}^4} + C_{\Psi \Psi}^{\Psi}\left( \frac{1}{Z_{12}^2}\mathcal{J}^{\Psi}+ \frac{\theta_{12}}{2 \,Z_{12}^2}   D \mathcal{J}^{\Psi}+ \frac{1}{2 \, Z_{12}}   \del \mathcal{J}^{\Psi}+\frac{3}{10}\frac{\theta_{12}}{Z_{12}}  D \del  \mathcal{J}^{\Psi}\right) \nn \\[6pt]
     &+\frac{12 \mu}{c} \left( \frac{\theta_{12}}{Z_{12}^3}  \mathcal{T}+ \frac{1}{Z_{12}^2}D \mathcal{T}+ \frac{2}{3}\frac{\theta_{12}}{Z_{12}^2} \del\mathcal{T}+\frac{1}{4 \, Z_{12}}   D \del \mathcal{T}  +\frac{\theta_{12}}{4 \, Z_{12}} \del \del \mathcal{T}\right) \nn
     \\[6pt]
     &+\frac{216\mu }{c\, (21+4c)} \frac{\theta_{12}}{Z_{12}} \mathcal{N}(D \mathcal{T}\mathcal{T})+\frac{54}{6+5c} C^{\Psi}_{\Psi \Psi} \frac{\theta_{12}}{Z_{12}}   \mathcal{N}(\mathcal{T}\mathcal{J}^{\Psi})+\cdots \, ,
\end{align*}
\endgroup
}%
where the self-coupling is given by 
\begin{equation*}
    C^{\Psi}_{\Psi \Psi}=\sqrt{-\frac{8 (5 c+6)^2 \mu }{c \, (c-15) (4 c+21)} } \, .
\end{equation*}
It is immediate to derive the unitarity constraint $c<15$, which provides an upper bound to the range of the central charge.
\subsubsection{Shatashvili--Vafa algebra \texorpdfstring{$\SVeight$}{SV Spin(7)}}
\label{app:SVSpin7OPEs}
The $\SVeight$ algebra introduced by Shatashvili and Vafa in \cite{Shatashvili:1994zw} describes the symmetry algebra on the worldsheet for superstrings propagating on a background with $\mathrm{Spin}(7)$ special holonomy. It corresponds to the $\mathrm{FS}$ algebra for a particular choice of normalisation $\mu$ and central charge $c$
\begin{equation*}
    \mu=\frac{46}{3} \, , \quad c=12 \, .
\end{equation*}
{\small
\begingroup
\allowdisplaybreaks
\begin{align*} 
    \mathcal{J}^{\Psi}(Z_1)\mathcal{J}^{\Psi}(Z_2) &\sim \frac{46}{3}\frac{1}{Z_{12}^4} + \frac{44}{3}\left( \frac{1}{Z_{12}^2}\mathcal{J}^{\Psi}+ \frac{\theta_{12}}{2 \,Z_{12}^2}   D \mathcal{J}^{\Psi}+ \frac{1}{2 \, Z_{12}}   \del \mathcal{J}^{\Psi}+\frac{3}{10}\frac{\theta_{12}}{Z_{12}}  D \del  \mathcal{J}^{\Psi}\right) \nn \\[6pt]
     &+\frac{46}{3} \left( \frac{\theta_{12}}{Z_{12}^3}  \mathcal{T}+ \frac{1}{Z_{12}^2}D \mathcal{T}+ \frac{2}{3}\frac{\theta_{12}}{Z_{12}^2} \del\mathcal{T}+\frac{1}{4 \, Z_{12}}   D \del \mathcal{T}  +\frac{\theta_{12}}{4 \, Z_{12}} \del \del \mathcal{T}\right) \nn
     \\[6pt]
     &+4\, \frac{\theta_{12}}{Z_{12}} \mathcal{N}(D \mathcal{T}\mathcal{T})+12 \, \frac{\theta_{12}}{Z_{12}}   \mathcal{N}(\mathcal{T}\mathcal{J}^{\Psi})+\cdots \, .
\end{align*}
\endgroup
}%
The OPEs appearing in the original paper are given in a different basis and can be recovered by performing the field redefinition 
\begin{equation*}
    \cj^{\Psi}=-\cj^{\Psi}_{\mathrm{SV}}-\frac{1}{3}D\mathcal{T} \, ,
\end{equation*}
where $\cj^{\Psi}_{\mathrm{SV}}$ encodes the operators appearing in the original formulation of this algebra. 
\subsection{\texorpdfstring{$\mathrm{G}_2$ $\mathcal{SW}$}{G2 SW}-algebras}
We call $\mathrm{G}_2$ $\mathcal{SW}$-algebras those algebras enhancing the super Virasoro algebra by means of a primary $\mathcal{J}^{\psi}$ of weight $h_{\psi}=2$, and a primary $\mathcal{J}^{\varphi}$ of weight $h_{\varphi}=\frac{3}{2}$.
\subsubsection{Blumenhagen algebra \texorpdfstring{$\mathrm{Bl}$}{Bl}}
\label{app:SVG2OPEs} 
Although the full family of $\mathrm{G}_2$ $\mathcal{SW}$-algebras was first introduced by Blumenhagen in \cite{Blumenhagen:1991nm}, in this presentation we will follow \cite{Noyvert:2002mc}. The algebra depends on the central charge $c$ and on a positive parameter $\lambda^2$. We can pack $\lambda^2$ and $c$ into a new parameter $\mu$ 
\begin{equation*}
    \mu=\sqrt{\frac{9c\,(4 + \lambda^2)}{2(27 - 2 c)}}\, .
\end{equation*}
{\small
\allowdisplaybreaks
\begin{align} 
\nonumber
 \cj^\varphi(Z_1)\cj^\varphi(Z_2)&\sim \frac{2 c^2 }{\mu^2}\frac{1}{Z_{12}^3}+\frac{18 c }{\mu^2}\left( \frac{\theta_{12}}{Z_{12}^2}\ct+\frac{1}{3 \, Z_{12}}D\ct+\frac{2}{3}\frac{\theta_{12}}{Z_{12}}\del \ct \right) \\[6pt]
 \nn
 &+\frac{3 \lambda \sqrt{3 c}}{\mu}
 \left( \frac{\theta_{12}}{Z_{12}^2}\cj^\varphi+\frac{1}{3 \, Z_{12}}D\cj^\varphi+\frac{2}{3}\frac{\theta_{12}}{Z_{12}}\del \cj^\varphi \right) \\[6pt]
\label{eq:SWG2SOPEphiphi}
 &+ 6\left(\frac{1}{Z_{12}}\cj^\psi +\frac{\theta_{12}}{2 \, Z_{12}}D\cj^\psi\right)+\cdots \, , \\[8pt]
\nonumber
\mathcal{J}^{\varphi}(Z_1)\mathcal{J}^{\psi}(Z_2) &\sim  \frac{2c}{3}\left( \frac{1}{Z_{12}^2}\mathcal{J}^{\varphi}+ \frac{\theta_{12}}{3 \,Z_{12}^2}   D \mathcal{J}^{\varphi}+ \frac{1}{3 \, Z_{12}}   \del \mathcal{J}^{\varphi}+\frac{\theta_{12}}{6 \, Z_{12}}  D \del  \mathcal{J}^{\varphi}\right) \nn  \\[6pt]
\label{eq:SWG2SOPEphipsi}
     &+\frac{2\lambda\sqrt{3 c}}{\mu}\left( \frac{\theta_{12}}{Z_{12}^2}\mathcal{J}^{\psi}+\frac{1}{4 \, Z_{12}}D\mathcal{J}^{\psi}+\frac{\theta_{12}}{2 \, Z_{12}}\del \mathcal{J}^{\psi} \right) +6 \, \frac{\theta_{12}}{Z_{12}}   \mathcal{N}(\mathcal{T} \mathcal{J}^{\varphi})+\cdots \, , \\[8pt]
\nonumber
\mathcal{J}^{\psi}(Z_1)\mathcal{J}^{\psi}(Z_2) &\sim \frac{2 c^3}{9\mu^2}\frac{1}{Z_{12}^4}+\frac{12c}{\mu^2} \frac{\theta_{12}}{Z_{12}} \mathcal{N}(D\mathcal{T} \mathcal{T}) \nn \\[6pt]
\nonumber
&+\frac{8c^2}{3\mu^2} \left( \frac{\theta_{12}}{Z_{12}^3}  \mathcal{T}+ \frac{1}{3 \, Z_{12}^2}D \mathcal{T}+ \frac{2}{3}\frac{\theta_{12}}{Z_{12}^2} \del\mathcal{T}+\frac{1}{6 \, Z_{12}}   D \del \mathcal{T}  +\frac{\theta_{12}}{4 \, Z_{12}} \del \del \mathcal{T}\right) \nn  \\[6pt]
 &+ \frac{10c-27}{9}\left( \frac{1}{Z_{12}^2}\mathcal{J}^{\psi}+ \frac{\theta_{12}}{2 \,Z_{12}^2}   D \mathcal{J}^{\psi}+ \frac{1}{2 \, Z_{12}}   \del \mathcal{J}^{\psi}+\frac{3}{10}\frac{\theta_{12}}{Z_{12}}  D \del  \mathcal{J}^{\psi}\right) \nn \\[6pt]
\label{eq:SWG2SOPEpsipsi}
&+\frac{2\lambda c\sqrt{3c}}{9\mu} \left( \frac{\theta_{12}}{Z_{12}^3}  \mathcal{J}^{\varphi}+ \frac{1}{3 \, Z_{12}^2}D \mathcal{J}^{\varphi}+ \frac{2}{3}\frac{\theta_{12}}{Z_{12}^2} \del\mathcal{J}^{\varphi}+\frac{1}{6 \, Z_{12}}   D \del \mathcal{J}^{\varphi}  +\frac{\theta_{12}}{4 \, Z_{12}} \del \del \mathcal{J}^{\varphi}\right) \nn  \\[6pt]
     &+12 \, \frac{\theta_{12}}{Z_{12}}\mathcal{N}(\mathcal{T}\mathcal{J}^{\psi}) -\frac{\theta_{12}}{Z_{12}} \mathcal{N}(D\mathcal{J}^{\varphi} \mathcal{J}^{\varphi})+\frac{2\lambda\sqrt{3 c}}{\mu} \frac{\theta_{12}}{Z_{12}} \mathcal{N}(D\mathcal{T} \mathcal{J}^{\varphi})+\cdots \, .
\end{align}
}%
It should be noted that unitarity imposes an upper bound on the central charge, requiring $c < \frac{27}{2}$.

The OPEs of \cite{Noyvert:2002mc} can be recovered expanding the super OPEs into OPEs and performing a convenient rescaling. Our fields are related to those of \cite{Noyvert:2002mc} as follows
\begin{equation*}
    P^\varphi=i\frac{3\sqrt{3c}}{\mu} H \, , \qquad K^\varphi=i\frac{3\sqrt{3c}}{\mu} M \, , \qquad P^\psi=-\frac{2c}{3\mu}  W\, , \qquad K^\psi=-\frac{2c}{3\mu} U \, .
\end{equation*}
\subsubsection{Fiset--Gaberdiel algebra \texorpdfstring{$\mathrm{FG}_k$}{FGk}}
In \cite{Fiset:2021azq}, Fiset and Gaberdiel specialised the Blumenhagen algebra $\mathrm{Bl}$ to the backgrounds 
\begin{equation*}
    \mathrm{AdS}_3 \times S^3 \times T^4 \, , \quad \mathrm{AdS}_3  \times S^3 \times \mathrm{K}3 \, .
\end{equation*}
This family of algebras can be obtained by studying WZW models adapted to the specific backgrounds and it can be recovered from $\mathrm{Bl}$ by setting
\begin{equation}
\label{eq:candlambdaforFG}
    c=\frac{21}{2}-\frac{6}{k}\, , \quad \lambda^2=\frac{32(3k-2)^2}{k^2(49k-30)} \, ,
\end{equation}
where $k$ corresponds to the integer level of the WZW model. Choosing $\lambda>0$:
{\small
\begingroup
\allowdisplaybreaks
\begin{align} 
\nonumber
 \cj^\varphi(Z_1)\cj^\varphi(Z_2)&\sim \frac{49k-7}{7k-4}\frac{1}{Z_{12}^3}+6\frac{k(49k-30)}{(7k-4)^2}\left( \frac{\theta_{12}}{Z_{12}^2}\ct+\frac{1}{3 \, Z_{12}}D\ct+\frac{2}{3}\frac{\theta_{12}}{Z_{12}}\del \ct \right) \\[6pt]
\nonumber
 &+12\sqrt{2}\frac{1}{\sqrt{k}}\frac{3k-2}{7k-2}\left( \frac{\theta_{12}}{Z_{12}^2}\cj^\varphi+\frac{1}{3 \, Z_{12}}D\cj^\varphi+\frac{2}{3}\frac{\theta_{12}}{Z_{12}}\del \cj^\varphi \right) \\[6pt]
\nonumber
 &+ 6\left(\frac{1}{Z_{12}}\cj^\psi +\frac{\theta_{12}}{2 \, Z_{12}}D\cj^\psi\right)+\cdots \, , \\[8pt]
\nonumber
\mathcal{J}^{\varphi}(Z_1)\mathcal{J}^{\psi}(Z_2) &\sim  \left(7-\frac{4}{k}\right)\left( \frac{1}{Z_{12}^2}\mathcal{J}^{\varphi}+ \frac{\theta_{12}}{3 \,Z_{12}^2}   D \mathcal{J}^{\varphi}+ \frac{1}{3 \, Z_{12}}   \del \mathcal{J}^{\varphi}+\frac{\theta_{12}}{6 \, Z_{12}}  D \del  \mathcal{J}^{\varphi}\right) \nn  \\[6pt]
\nonumber
     &+8\sqrt{2}\frac{1}{\sqrt{k}}\frac{3k-2}{7k-4}\left( \frac{\theta_{12}}{Z_{12}^2}\mathcal{J}^{\psi}+\frac{1}{4 \, Z_{12}}D\mathcal{J}^{\psi}+\frac{\theta_{12}}{2 \, Z_{12}}\del \mathcal{J}^{\psi} \right)  \\[6pt]
     &+6 \, \frac{\theta_{12}}{Z_{12}}   \mathcal{N}(\mathcal{T} \mathcal{J}^{\varphi})+\cdots \, , \nn \\[8pt]
\nonumber
\mathcal{J}^{\psi}(Z_1)\mathcal{J}^{\psi}(Z_2) &\sim  \\[2pt]
&\hspace{-5em}\left(\frac{49}{6}-\frac{5}{k}\right)\frac{1}{Z_{12}^4} + \left(\frac{26}{3}-\frac{20}{3\, k}\right)\left( \frac{1}{Z_{12}^2}\mathcal{J}^{\psi}+ \frac{\theta_{12}}{2 \,Z_{12}^2}   D \mathcal{J}^{\psi}+ \frac{1}{2 \, Z_{12}}   \del \mathcal{J}^{\psi}+\frac{3}{10}\frac{\theta_{12}}{Z_{12}}  D \del  \mathcal{J}^{\psi}\right) \nn \\[6pt]
\nonumber
&\hspace{-5em}+\left(\frac{28}{3}-\frac{8}{3(7k-4)}\right) \left( \frac{\theta_{12}}{Z_{12}^3}  \mathcal{T}+ \frac{1}{3 \, Z_{12}^2}D \mathcal{T}+ \frac{2}{3}\frac{\theta_{12}}{Z_{12}^2} \del\mathcal{T}+\frac{1}{6 \, Z_{12}}   D \del \mathcal{T}  +\frac{\theta_{12}}{4 \, Z_{12}} \del \del \mathcal{T}\right) \nn \\[6pt]
\nonumber
&\hspace{-5em}+\sqrt{2}\frac{1}{\sqrt{k}}\frac{4(3k-2)}{3\, k}\left( \frac{\theta_{12}}{Z_{12}^3}  \mathcal{J}^{\varphi}+ \frac{1}{3 \, Z_{12}^2}D \mathcal{J}^{\varphi}+ \frac{2}{3}\frac{\theta_{12}}{Z_{12}^2} \del\mathcal{J}^{\varphi}+\frac{1}{6 \, Z_{12}}   D \del \mathcal{J}^{\varphi}  +\frac{\theta_{12}}{4 \, Z_{12}} \del \del \mathcal{J}^{\varphi}\right) \nn  \\[6pt]
\nonumber
     &\hspace{-5em}+8 \, \frac{\theta_{12}}{Z_{12}}   \mathcal{N}(\mathcal{T} \mathcal{J}^{\Psi})+\frac{8 k (49 k-30) (49 k-22)}{3 (7 k-4)^2 (49 k-24)} \frac{\theta_{12}}{Z_{12}} \mathcal{N}(D\mathcal{T} \mathcal{T})  \\[6pt]
     &\hspace{-5em}+\frac{4 \sqrt{2}}{3}\frac{1}{\sqrt{k}}\frac{k (637 k-712)+192}{(7 k-4) (49
   k-24)} \frac{\theta_{12}}{Z_{12}} \mathcal{N}(D\mathcal{T} \mathcal{J}^{\varphi}) \nn \\[6pt]
     &\hspace{-5em}-\frac{4}{9} \left(\frac{2}{49 k-24}+\frac{1}{k}\right)\frac{\theta_{12}}{Z_{12}} \mathcal{N}(D\mathcal{J}^{\varphi} \mathcal{J}^{\varphi})-4 \sqrt{2} \frac{1}{\sqrt{k}}\frac{7 k-4}{49 k-24} \frac{\theta_{12}}{Z_{12}}\mathcal{N}(\mathcal{J}^{\varphi}\mathcal{J}^{\psi})+\cdots \, . \label{eq:FGG2SOPEpsipsi}
\end{align}
\endgroup
}%
It is important to recall from Section \ref{sec:G2SW} that the $\mathrm{FG}_k$ algebra closes up to a null field, which depends on the selected level $k$
\begin{multline*}
\nonumber
    \mathfrak{N}^{\mathrm{FG}}_k =8\left(3-\frac{4 c^2}{7 \mu^2}\right)\mathcal{N}(\mathcal{T}\mathcal{J}^{\psi})-\frac{2}{3}\left(1+\frac{4}{21}c\right)\mathcal{N}(D\mathcal{J}^{\varphi}\mathcal{J}^{\varphi})+\frac{8c}{\mu^2}\mathcal{N}(D\mathcal{T}\mathcal{T}) \\
    +\frac{1}{\sqrt{k}}\frac{8\sqrt{2}}{7} \left( \mathcal{N}(\mathcal{J}^{\varphi}\mathcal{J}^{\psi})-\left(3-\frac{4 c^2}{3 \mu^2}\right)\mathcal{N}(D\mathcal{T}\mathcal{J}^{\varphi}) \right)  \, .
\end{multline*}
In fact, in our presentation substituting \eqref{eq:candlambdaforFG} into \eqref{eq:SWG2SOPEpsipsi} only yields \eqref{eq:FGG2SOPEpsipsi} up to the null field above. The OPEs found in Appendix B of \cite{Fiset:2021azq} can be recovered---again up to null fields---by changing to a non-primary basis and expanding the super OPEs into OPEs.

\subsubsection{Shatashvili--Vafa algebra \texorpdfstring{$\mathrm{SV}^{\mathrm{G}_2}$}{SV G2}}

Similarly to the $\mathrm{Spin}(7)$ case, the $\mathrm{SV}^{\mathrm{G}_2}$ algebra by Shatashvili and Vafa first appeared in \cite{Shatashvili:1994zw} and describes the symmetry algebra on the worldsheet for superstrings propagating on a background with $\mathrm{G}_2$ special holonomy. It can be obtained from the $\mathrm{Bl}$ algebra by choosing the parameters 
\begin{equation*}
    c=\frac{21}{2} \, , \quad \lambda^2=0 \, ,
\end{equation*}
or from the $\mathrm{FG}_k$ algebra in the limit $k\to \infty$. The latter method yields:
{\small
\begingroup
\allowdisplaybreaks
\begin{align*} 
 \cj^\varphi(Z_1)\cj^\varphi(Z_2)&\sim \frac{7}{Z_{12}^3}+6\left( \frac{\theta_{12}}{Z_{12}^2}\ct+\frac{1}{3 \, Z_{12}}D\ct+\frac{2}{3}\frac{\theta_{12}}{Z_{12}}\del \ct \right) \\[6pt]
 &+ 6\left(\frac{1}{Z_{12}}\cj^\psi +\frac{\theta_{12}}{2 \, Z_{12}}D\cj^\psi\right)+\cdots \, , \\[8pt]
\mathcal{J}^{\varphi}(Z_1)\mathcal{J}^{\psi}(Z_2) &\sim  7\left( \frac{1}{Z_{12}^2}\mathcal{J}^{\varphi}+ \frac{\theta_{12}}{3 \,Z_{12}^2}   D \mathcal{J}^{\varphi}+ \frac{1}{3 \, Z_{12}}   \del \mathcal{J}^{\varphi}+\frac{\theta_{12}}{6 \, Z_{12}}  D \del  \mathcal{J}^{\varphi}\right) \nn  \\[6pt]
     &+6 \, \frac{\theta_{12}}{Z_{12}}   \mathcal{N}(\mathcal{T} \mathcal{J}^{\varphi})+\cdots \, , \\[8pt]
\mathcal{J}^{\psi}(Z_1)\mathcal{J}^{\psi}(Z_2) &\sim \frac{49}{6}\frac{1}{Z_{12}^4} + \frac{26}{3}\left( \frac{1}{Z_{12}^2}\mathcal{J}^{\psi}+ \frac{\theta_{12}}{2 \,Z_{12}^2}   D \mathcal{J}^{\psi}+ \frac{1}{2 \, Z_{12}}   \del \mathcal{J}^{\psi}+\frac{3}{10}\frac{\theta_{12}}{Z_{12}}  D \del  \mathcal{J}^{\psi}\right) \nn \\[6pt]
&+\frac{28}{3} \left( \frac{\theta_{12}}{Z_{12}^3}  \mathcal{T}+ \frac{1}{3 \, Z_{12}^2}D \mathcal{T}+ \frac{2}{3}\frac{\theta_{12}}{Z_{12}^2} \del\mathcal{T}+\frac{1}{6 \, Z_{12}}   D \del \mathcal{T}  +\frac{\theta_{12}}{4 \, Z_{12}} \del \del \mathcal{T}\right) \nn
     \\[6pt]
     &+\frac{8}{3} \frac{\theta_{12}}{Z_{12}} \mathcal{N}(D\mathcal{T} \mathcal{T})+8 \, \frac{\theta_{12}}{Z_{12}}   \mathcal{N}(\mathcal{T} \mathcal{J}^{\psi})+\cdots \, .
\end{align*}
\endgroup
}%
As described in \Cref{sec:G2SW}, this algebra closes up to a null field 
\begin{equation*}
    \mathfrak{N}^{\mathrm{SV}}=8\, \mathcal{N}(\mathcal{T}\mathcal{J}^{\Psi})-2\, \mathcal{N}(D\mathcal{J}^{\varphi}\mathcal{J}^{\varphi})+\frac{8}{3}\mathcal{N}(D\mathcal{T}\mathcal{T}) \, .
\end{equation*} 
These OPEs reduce to the ones presented in \cite{Shatashvili:1994zw}. Note however that the OPEs in \cite{Shatashvili:1994zw} are expressed in a different basis, which can be recovered by performing the field redefinition
\begin{equation*}
    \cj^{\psi}=\cj^{\psi}_{\mathrm{SV}}+\frac{1}{3}D\mathcal{T} \, ,
\end{equation*}
where $\mathcal{J}^{\psi}_{\mathrm{SV}}$ encodes the operators appearing in the original formulation of this algebra.
\subsection{\texorpdfstring{$\mathrm{SU}(n)$ $\mathcal{SW}$}{SU(n) SW}-algebras}
We call $\mathrm{SU}(n)$ $\mathcal{SW}$-algebras those algebras enhancing the super Virasoro algebra by means of a primary $\mathcal{J}^{\omega}$ of weight $h_{\omega}=1$, and two primaries $\mathcal{J}^{\Omega_{\pm}}$ of weight $h_{\Omega_{\pm}}=\frac{n}{2}$. In the following, we will focus on the cases $n=2,3$.
\subsubsection{Odake algebras \texorpdfstring{$\mathrm{Od}(2)$}{Od(2)} and \texorpdfstring{$\mathrm{Od}(3)$}{Od(3)}} 
\label{app:Odakealgebras}
Although up to our knowledge the full family of $\mathrm{SU}(n)$ $\mathcal{SW}$-algebras is not known in the literature, the symmetry algebras on the worldsheet of strings propagating on manifolds with $\mathrm{SU}(n)$ special holonomy, i.e., Calabi-Yau $n$-folds, first appeared in \cite{Odake:1988bh}. 

\noindent The $\mathrm{Od}(2)$ algebra reads
{\small
\begingroup
\allowdisplaybreaks
\begin{align*}    
\cj^\omega(Z_1)\cj^\omega(Z_2)&\sim -\frac{2}{Z_{12}^2}-2\, \frac{\theta_{12}}{Z_{12}}\ct+\cdots  \, ,\\[8pt]
\cj^\omega(Z_1)\cj^{\Omega_\pm}(Z_2)&\sim \mp 2  \left(\frac{1}{Z_{12}}\cj^{\Omega_\mp}+\frac{\theta_{12}}{2 \, Z_{12}} \, D \cj^{\Omega_\mp}\right)+\cdots \, , \\[8pt]
 \cj^{\Omega_{\pm}}(Z_1)\cj^{\Omega_{\pm}}(Z_2)&\sim -\frac{2}{Z_{12}^2}-2\frac{\theta_{12}}{Z_{12}}  \ct+\cdots \, , \\[8pt]
    \cj^{\Omega_{+}}(Z_1)\cj^{\Omega_{-}}(Z_2)&\sim \mp 2  \left(\frac{1}{Z_{12}}\cj^{\omega}+\frac{\theta_{12}}{2 \, Z_{12}} \, D \cj^{\omega}\right)+\cdots \, ,
\end{align*}
\endgroup}%
and it does not require null fields to close. 

\noindent The $\mathrm{Od}(3)$ algebra reads
{\small
\begingroup
\allowdisplaybreaks
\begin{align*}    
\cj^\omega(Z_1)\cj^\omega(Z_2)&\sim -\frac{3}{Z_{12}^2}-2\, \frac{\theta_{12}}{Z_{12}}\ct+\cdots  \, ,\\[8pt]
\cj^\omega(Z_1)\cj^{\Omega_\pm}(Z_2)&\sim \mp 3 \left(\frac{1}{Z_{12}}\cj^{\Omega_\mp}+\frac{\theta_{12}}{3 \, Z_{12}} \, D \cj^{\Omega_\mp}\right)+\cdots \, , \\[8pt]
 \cj^{\Omega_{\pm}}(Z_1)\cj^{\Omega_{\pm}}(Z_2)&\sim \frac{4}{Z_{12}^3}+4\left( \frac{\theta_{12}}{Z_{12}^2}\ct+\frac{1}{3 \,Z_{12}}D\ct+\frac{2}{3}\frac{\theta_{12}}{Z_{12}}\del \ct\right) \\[6pt] &-2\left(\frac{1}{Z_{12}}\cn(\cj^{\omega} \cj^{\omega})
 +\frac{\theta_{12}}{2\, Z_{12}}D\cn(\cj^{\omega} \cj^{\omega})\right)+\cdots \, , \\[8pt]
    \cj^{\Omega_{+}}(Z_1)\cj^{\Omega_{-}}(Z_2)&\sim 4\left(\frac{1}{Z_{12}^2}\cj^{\omega}+\frac{\theta_{12}}{2 \, Z_{12}^2}D \cj^{\omega}+\frac{1}{2 \, Z_{12}} \del \cj^{\omega}+\frac{\theta_{12}}{3 \, Z_{12}}D \del \cj^{\omega}\right)\\[6pt] &+4 \frac{\theta_{12}}{Z_{12}} \, \cn(\ct \cj^{\omega})+\cdots \, .
\end{align*}
\endgroup
}%
As opposed to the $\mathrm{Od}(2)$ algebra, the $\mathrm{Od}(3)$ algebra closes up to the ideals generated by two null fields \cite{Odake:1988bh, Fiset:2018huv}
\begin{equation}
\label{eq: odake null}
    \mathfrak{N}_{\mathrm{Od},1}=\cn(\cj^{\omega}\cj^{\Omega_{-}}) \, , \quad \mathfrak{N}_{\mathrm{Od},2}=\cn(\cj^{\omega}\cj^{\Omega_{+}}) \, . 
\end{equation}
\subsubsection{Towards a torsionful \texorpdfstring{$\mathrm{SU}(3)$}{SU(3)} algebra}
\label{app: su3}
To conclude this Appendix, we provide the explicit OPEs discussed in \Cref{sec: su3}. These correspond to the \emph{unfixed} OPEs of the full family of $\mathrm{SU}(3)$ $\mathcal{SW}$-algebras. The coupling symmetries reported in Table \ref{tab:symm} have already been implemented. An important \emph{caveat} is that the chosen normalisations are the same as in the $\mathrm{Od}(3)$ algebra:
\begin{equation*}
    C_{\omega \omega}^{\mathds{1}} = -3  \, , \quad  C_{\Omega_{\pm} \Omega_{\pm}}^{\mathds{1}} = 4  \, .
\end{equation*}
{\footnotesize
\begingroup
\allowdisplaybreaks
\begin{align*}
    \cj^{\omega}(Z_1)\cj^{\omega}(Z_2)&\sim  -\frac{3}{Z_{12}^{2}} +\frac{\theta_{12}}{Z_{12}}\left(C^{\ct}_{\omega \omega} \ct+C^{\Omega_{+}}_{\omega \omega} \cj^{\Omega_{+}}+C^{\Omega_{-}}_{\omega \omega} \cj^{\Omega_{-}}\right)+\cdots \, , \\[8pt]
    \cj^{\omega}(Z_1)\cj^{\Omega_{+}}(Z_2)&\sim \frac{4}{3} \, C_{\omega \omega}^{\Omega_{+}}\left(\frac{\theta_{12}}{Z_{12}^2}\cj^{\omega}+\frac{1}{2 \, Z_{12}}D \cj^{\omega}+\frac{\theta_{12}}{2 \, Z_{12}}\del \cj^{\omega}\right) \nn \\[6pt]
    &+C^{\ct}_{\omega \Omega_{+}}\left(\frac{1}{Z_{12}}\ct+\frac{\theta_{12}}{3 \, Z_{12}}D \ct\right)-\frac{3}{4} \, C_{\Omega_{+} \Omega_{-}}^{\omega}\left(\frac{1}{Z_{12}}\cj^{\Omega_{-}}+\frac{\theta_{12}}{3 \, Z_{12}}D\cj^{\Omega_{-}} \right) \nn \\[6pt]
    &+C^{\omega \omega}_{\omega \Omega_{+}}\frac{\theta_{12}}{Z_{12}}\cn(\cj^{\omega}\cj^{\omega})+\cdots \, , \\[8pt]
    \cj^{\omega}(Z_1)\cj^{\Omega_{-}}(Z_2)&\sim \frac{4}{3} \, C_{\omega \omega}^{\Omega_{-}}\left(\frac{\theta_{12}}{Z_{12}^2}\cj^{\omega}+\frac{1}{2 \, Z_{12}}D \cj^{\omega}+\frac{\theta_{12}}{2 \, Z_{12}}\del \cj^{\omega}\right) \nn \\[6pt]
    &+C^{\ct}_{\omega \Omega_{-}}\left(\frac{1}{Z_{12}}\ct+\frac{\theta_{12}}{3 \, Z_{12}}D \ct\right)+\frac{3}{4} \, C_{\Omega_{+} \Omega_{-}}^{\omega}\left(\frac{1}{Z_{12}}\cj^{\Omega_{+}}+\frac{\theta_{12}}{3 \, Z_{12}}D\cj^{\Omega_{+}} \right) \nn \\[6pt]
    &+C^{\omega \omega}_{\omega \Omega_{-}}\frac{\theta_{12}}{Z_{12}}\cn(\cj^{\omega}\cj^{\omega})+\cdots \, , \\[8pt]
    \cj^{\Omega_{\pm}}(Z_1)\cj^{\Omega_{\pm}}(Z_2)&\sim \frac{4}{Z_{12}^3}+C_{\Omega_{\pm}\Omega_{\pm}}^{\ct}\left(\frac{\theta_{12}}{Z_{12}^2}\ct+\frac{1}{3\, Z_{12}}D\ct+\frac{2}{3}\frac{\theta_{12}}{Z_{12}}\del \ct \right) \nn \\[6pt]
    &+C_{\Omega_{\pm}\Omega_{\pm}}^{\Omega_{+}}\left(\frac{\theta_{12}}{Z_{12}^2}\cj^{\Omega_{+}}+\frac{1}{3\, Z_{12}}D\cj^{\Omega_{+}}+\frac{2}{3}\frac{\theta_{12}}{Z_{12}}\del \cj^{\Omega_{+}} \right)\nn \\[6pt]
    &+C_{\Omega_{\pm}\Omega_{\pm}}^{\Omega_{-}}\left(\frac{\theta_{12}}{Z_{12}^2}\cj^{\Omega_{-}}+\frac{1}{3\, Z_{12}}D\cj^{\Omega_{-}}+\frac{2}{3}\frac{\theta_{12}}{Z_{12}}\del \cj^{\Omega_{-}} \right) \nn \\[6pt]
    &+C_{\Omega_{\pm}\Omega_{\pm}}^{\omega \omega}\left(\frac{1}{Z_{12}}\cn(\cj^{\omega} \cj^{\omega})+\frac{\theta_{12}}{2 \, Z_{12}}D\cn(\cj^{\omega} \cj^{\omega})\right)+\cdots \, , \\[8pt]
    \cj^{\Omega_{+}}(Z_1)\cj^{\Omega_{-}}(Z_2)&\sim C_{\Omega_{+}\Omega_{-}}^{\omega}\left(\frac{1}{Z_{12}^2}\cj^{\omega}+\frac{\theta_{12}}{2 \, Z_{12}^2}D \cj^{\omega}+\frac{1}{2 \, Z_{12}} \del \cj^{\omega}+\frac{\theta_{12}}{3 \, Z_{12}}D \del \cj^{\omega}\right) \nn \\[6pt]
    &+C_{\Omega_{+}\Omega_{-}}^{\ct}\left(\frac{\theta_{12}}{Z_{12}^2}\ct+\frac{1}{3\, Z_{12}}D\ct+\frac{2}{3}\frac{\theta_{12}}{Z_{12}}\del \ct \right) \nn \\[6pt]
    &+C_{\Omega_{+} \Omega_{+}}^{\Omega_{-}}\left(\frac{\theta_{12}}{Z_{12}^2}\cj^{\Omega_{+}}+\frac{1}{3\, Z_{12}}D\cj^{\Omega_{+}}+\frac{2}{3}\frac{\theta_{12}}{Z_{12}}\del \cj^{\Omega_{+}} \right)\nn \\[6pt]
    &+C_{\Omega_{-} \Omega_{-}}^{\Omega_{+}}\left(\frac{\theta_{12}}{Z_{12}^2}\cj^{\Omega_{-}}+\frac{1}{3\, Z_{12}}D\cj^{\Omega_{-}}+\frac{2}{3}\frac{\theta_{12}}{Z_{12}}\del \cj^{\Omega_{-}} \right) \nn \\[6pt]
    &+C_{\Omega_{+}\Omega_{-}}^{\omega \omega}\left(\frac{1}{Z_{12}}\cn(\cj^{\omega} \cj^{\omega})+\frac{\theta_{12}}{2 \, Z_{12}}D\cn(\cj^{\omega} \cj^{\omega})\right) \nn \\[6pt]
    &+\frac{\theta_{12}}{Z_{12}}\left(C_{\Omega_{+}\Omega_{-}}^{\ct \omega}\cn(\ct \cj^{\omega})+ C_{\Omega_{+}\Omega_{-}}^{\omega\Omega_{+}}\cn(\cj^{\omega}\cj^{\Omega_{+}})+ C_{\Omega_{+}\Omega_{-}}^{\omega\Omega_{-}}\cn(\cj^{\omega}\cj^{\Omega_{-}})\right)+\cdots \, .
\end{align*}
\endgroup
}

\bibliography{finalbibliography.bib}

\end{document}